\documentclass[12pt,a4paper,twoside]{article}
\usepackage{latexsym,amsmath,amssymb,amsfonts,amsthm}

\newtheorem{corollary}{Corollary}
\newtheorem{theorem}{Theorem}
\def\nabla{\bigtriangledown}
\newcommand{ \R} {\mbox{\rm I$\!$R}}
\newcommand{ \C} {\mbox{\rm I$\!$C}}

\makeindex \textheight 23.5cm
\begin{document}

\title{Exact Solutions with Noncommutative \\
Symmetries in Einstein and Gauge Gravity }
\date{March 2, 2005 }
\author{Sergiu I. Vacaru \thanks{%
E-mail address:\ \ vacaru@imaff.cfmac.csic.es, ~~ sergiu$_{-}$%
vacaru@yahoo.com,\ } \\
%EndAName
{\small \textit{Instituto de Matematicas y Fisica Fundamental}}\\
{\small \textit{Consejo Superior de Investigaciones Cientificas}}\\
{\small \textit{Calle Serrano 123, Madrid 28006, Spain}}\\
}
\maketitle

\begin{abstract}
We present new classes of exact solutions with noncommutative symmetries
constructed in vacuum Einstein gravity (in general, with nonzero
cosmological constant), five dimensional (5D) gravity and (anti) de Sitter
gauge gravity. Such solutions are generated by anholonomic frame transforms
and parametrized by generic off--diagonal metrics. For certain particular
cases, the new classes of metrics have explicit limits with Killing
symmetries but, in general, they may be characterized by certain anholonomic
noncommutative matrix geometries. We argue that different classes of
noncommutative symmetries can be induced by exact solutions of the field
equations in 'commutative' gravity modeled by a corresponding moving real
and complex frame geometry. We analyze two classes of black ellipsoid
solutions (in the vacuum case and with cosmological constant) in 4D gravity
and construct the analytic extensions of metrics for certain classes of
associated frames with complex valued coefficients. The third class of
solutions describes 5D wormholes which can be extended to complex metrics in
complex gravity models defined by noncommutative geometric structures. The
anholonomic noncommutative symmetries of such objects are analyzed. We also
present a descriptive account how the Einstein gravity can be related to
gauge models of gravity and their noncommutative extensions and discuss such
constructions in relation to the Seiberg--Witten map for the gauge gravity.
Finally, we consider a formalism of vielbeins deformations subjected to
noncommutative symmetries in order to generate solutions for noncommutative
gravity models with Moyal (star) product.

\vskip15pt

Pacs:\ 02.40.Gh, 02.40.-k, 04.50.+h, 04.20.Jb

\end{abstract}

%\tableofcontents

\newpage

%\newpage\setcounter{page}1

\section{\quad Introduction}

In the last fifteen years much effort has been made to elaborate a
consistent formulation of noncommutative gravity theory generalizing the
standard Einstein theory but up to now the problem is quite difficult to
approach (see, for instance, Refs. 1-8
%\cite {con1,con2,ncg1,ncg2,ncg3,v1,ch1,madore}%
for details related to existing models). The proposed theories are for the
spaces with Euclidean signatures, and, in general, result in models of
complex gravity in noncommutative spaces provided with complex and/or
nonsymmetric metrics and anholonomic frames. There were also derived some
effective noncommutative gravity models from string/brane theory, by
considering quantum group structures and/or by proposing noncommutative
gauge like generalizations of gravity.

In this paper, we pursue the idea that noncommutative geometric structures
are present in the Einstein, five dimensional (in brief, 5D) gravity and
gauge gravity models. Such noncommutative symmetries are emphasized if the
anholonomic moving frames$^{9-12}$ %\cite{cartan1,cartan2,cartan3,cartan4}%
are introduced into consideration. This 'hidden noncommutativity' is
nontrivial for various classes of generic off--diagonal metrics admitting
effective diagonalizations by anholonomic transforms with associated
nonlinear connection structure$^{13-19}.$
%\cite{anhtv1,anhtv2,anhtv3,anhtv4,anhtv5,anhtv6,anhtv7}.%
The metrics may be subjected to the condition to define exact solutions of
the vacuum field equations with certain possible extensions to matter
sources. The noncommutative anhlonomic geometries can be derived even from
the 'commutative' general relativity theory and admit a natural embedding
into different models of complex noncommutative gravity. The metric and
frame (vielbein) coefficients corresponding to 'off--diagonal' solutions
depend on two, three or four variables and define spacetimes with associated
noncommutative symmetries. Such classes of exact solutions are very
different from the well known examples of metrics with Killing symmetry
(like the Schwarzschlild or Kerr--Newmann solutions; see a detailed analysis
in Ref. 20. %\cite{heus}).%

Our aim is to prove, by constructing and analyzing three classes of exact
solutions, that certain noncommutative geometric structures can be defined
in the framework of the Einstein (in general, with cosmological term) and 5D
gravity. We emphasize classes of anholonomic real and complex deformations
of metrics possessing associated noncommutative symmetries. Contrary to
other approaches to noncommutative gravity and field interactions theory
elaborated by substituting the commutative algebras of functions with
noncommutative algebras and/or by postulating any complex noncommutative
relations for coordinates, we try to derive noncommutative structures from
associated symmetries of metrics and frames subjected to anholonomy
relations. We shall propose a classification of such spacetimes and state a
method of complexification of exact solutions preserving the noncommutative
symmetry for black hole and wormhole metrics in 'real' and 'complex' gravity.

The study of anholonomic noncommutative symmetries of gravitational field
interactions is more involved in the moving frame formalism conventionally
adapted to equivalent redefinitions of the Einstein equations as Yang--Mills
equations for nonsemisimple gauge groups like in the Poincare gauge gravity$%
^{21,22}. $ %
%\cite{pd1,pd2}.%
This construction has direct generalizations to various type of gauge
gravity models with nondegenerate metrics in the total bundle spaces, in
both 'commutative' and 'noncommutative' forms$^{23-29,6}. $ %
%\cite{ts,v41,v42,v43,v44,v45,v5,v1}.%
The connection between the general relativity theory and gauge gravity
models is emphasized in order to apply and compare with a set of results
from noncommutative gauge theory.

Among our static solutions we find geometries having a structure as have
Schwarzschild, Reissner--Nordstrem and (anti) de Sitter spaces but with the
coefficients redefined (with certain polarization constants) with respect to
anholonomic real/complex frames which make possible definition of such
objects in noncommutative models of gravity. There are equally interesting
applications to black hole physics, quantum gravity and string gravity.

Next, the emerged anholonomic noncommutative symmetries of 'off--diagonal'
metrics prescribe explicit rules of deformation the solutions on small
noncommutative parameters and connect the results to quantum deformations of
gravity and gauge models. So, even a generally accepted version of
noncommutative gravity theory has been not yet formulated, we know how to
generate particular classes of 'real' and 'complex' stable metrics with
noncommutative symmetries and possessing properties very similar to the
usual black hole and wormhole solutions. In particular, we present a
systematic procedure for constructing exact solutions both in commutative
and noncommutative gravity models, to define black hole and wormhole objects
with noncommutative symmetries and quantum corrections. We are able to
investigate the physical properties of such objects subjected to certain
classes of anholonomic and/or quantum deformations.

\vskip4pt

The paper is organized as follows:

We begin in section II with a brief introduction into the geometry of
spacetimes provided with anholonomic frame structure and associated
nonlinear connections. Such geometries are characterized by corresponding
anholonomy relations induced by nonlinear connection coefficients related to
certain off--diagonal metric components. This also induces a corresponding
noncommutative spacetime structure.

In section III, we illustrate that such noncommutative anholonomic
geometries can be associated even to real spacetimes and that a simple
realization holds within the algebra for complex matrices. We emphasize that
a corresponding noncommutative differential calculus can be derived from the
anholonomy coefficients deforming the structure constants of the related Lie
algebras.

Section IV is devoted to a rigorous analysis of two classes of static black
ellipsoid solutions (the first and second type metrics defining respectively
4D vacuum Einstein and induced by cosmological constant configurations). We
prove that such metrics can be complexified in order to admit associated
complex frame/ nonlinear connection structures inducing noncommutative
matrix geometries and show how analytic extensions of such real and
complexified spacetimes can be constructed.

In section V, a class of 5D wormhole solutions with anisotropic elliptic
polarizations is considered for the 5D gravity. We argue that such generic
off--diagonal metrics may be also complexified as to preserve the wormhole
configurations being additionally characterized by complex valued
coefficients for the associated nonlinear connection. Such objects posses
the same noncommutative symmetry for both type of real and complex solutions.

Section VI is a discussion how the Einstein gravity and its higher dimension
extensions can be incorporated naturally into 'commutative' and
'noncommutative' gauge models. A new point is that the proposed geometric
formalism is elaborated in order to include anholonomic complex vielbeins.

In section VII, we define the Seiberg--Witten map for the de Sitter gauge
gravity and state a prescription how the exact solutions possessing
anhlonomic noncommutative symmetries can be adapted to deformations via star
products with noncommutative relations for coordinates.

We conclude and discuss the results in section VIII. For convenience, we
summarize the necessary results from Refs. 13-19 and 30-32 %
% \cite{anhtv1,anhtv2,anhtv3,anhtv4,anhtv5,anhtv6,anhtv7,vmethod1,vmethod2,vnp}%
in Appendices A, B and C and state some definitions on ''star'' products and
enveloping algebras in Appendix D.

\section{\quad Off--Diagonal Metrics and Anhlonomic Fra\-mes}

We consider a $(n+m)$--dimensional spacetime manifold $V^{n+m}$ provided
with a (pseudo) Riemanni\-an metric $\mathbf{g}=\{g_{\mu \nu }\}$ and denote
the local coordinates $u=(x,y),$ or in component form, $u^{\alpha
}=(x^{i},y^{a}),$ where the Greek indices are conventionally split into two
subsets, $x=\{x^{i}\}$ and $y=\{y^{a}\},$ labelled, correspondingly, by
Latin indices of type $i,j,k,...=1,2,...,n,$ and $a,b,...=1,2,...,m.$ In
general, the geometric objects on such spacetimes may posses some nontrivial
Killing symmetries (the Killing case is emphasized by the condition $L_{X}%
\mathbf{g=}0,$ where $L_{X}$ is the Lie derivative with respect to a vector
field $X$ on $V^{n+m},$ see, for instance, Ref. 20) %\cite{heus}),%
or some deformations \ of such symmetries, for instance, by frame
transforms. The spacetimes may have some additional frame structures with
associated nonlinear connection, bundle structure and even nontrivial
torsions being adapted to the frame structure.

We shall define our constructions for a general metric ansatz of type%
\begin{equation}
\mathbf{g}=g_{\mu \nu }\delta u^{\mu }\otimes \delta u^{\nu }=g_{ij}\left(
x^{k}\right) dx^{i}\otimes dx^{j}+h_{ab}\left( x^{k},v\right) \delta
y^{a}\otimes \delta y^{b}  \label{dmetric}
\end{equation}%
with respect to a locally adapted basis $[dx^{i},\delta y^{a}],$ where the
Einstein's summation rule is applied and by $v$ we emphasize the dependence
on a so--called 'anisotropic' coordinate from the set $\{y^{a}\}.$ The local
basis
\begin{equation}
e_{[N]}^{\mu }=\delta ^{\mu }=\delta u^{\mu }=[dx^{i},\delta
y^{a}]=[dx^{i},\delta y^{a}=dy^{a}+N_{i}^{a}\left( x^{k},v\right) dx^{i}]
\label{ndif}
\end{equation}%
(called to be N--elongated; we shall provide an additional index [N] if
would be necessary to distinguish such objects) is dual to the local basis%
\begin{equation}
e_{\alpha }^{[N]}=\delta _{\alpha }=\frac{\delta }{\delta u^{\alpha }}=\left[
\frac{\delta }{\partial x^{i}}=\frac{\partial }{\partial x^{i}}%
-N_{i}^{a}\left( x^{k},v\right) \frac{\partial }{\partial y^{a}},\frac{%
\partial }{\partial y^{b}}\right] .  \label{nder}
\end{equation}%
We consider an off--diagonal metric ansatz for (\ref{dmetric}) having the
components%
\begin{equation}
\widehat{g}_{\alpha \beta }=\left[
\begin{array}{cc}
g_{ij}+N_{i}^{a}N_{j}^{b}h_{ab} & N_{j}^{e}h_{ae} \\
N_{i}^{e}h_{be} & h_{ab}%
\end{array}%
\right] .  \label{offdig}
\end{equation}%
So, we can write equivalently $\mathbf{g}=\widehat{g}_{\alpha \beta
}du^{\alpha }du^{\beta }$ if the metric is rewritten with respect to the
local dual coordinate basis $du^{\mu }=[dx^{i},dy^{a}],$ the dual to $%
\partial /\partial u^{\alpha }=\left[ \partial /\partial x^{i},\partial
/\partial y^{b}\right] $ (defined correspondingly by usual partial
derivatives and differentials).

A very surprising fact is that the off--diagonal metric ansatz (\ref{offdig}%
) for dimensions $n+m=3,4,5$ and certain imbedding of such configurations in
extra dimension (super) spaces results in completely integrable systems of
partial differential equations (see details in Refs. 13-19
% \cite{anhtv1,anhtv2,anhtv3,anhtv4,anhtv5,anhtv6,anhtv7}%
with a review of results in Refs. 30,31 %\cite{vmethod1,vmethod2}%
and Theorems 1-3 in the Appendix B). In this paper, we shall consider that
any metric (\ref{offdig}), or equivalently (\ref{dmetric}) and frames
(vielbeins) (\ref{ndif}) and (\ref{nder}), parametrizes an exact solution of
the Einstein equations in a 'commutative' gravity theory.

Let us state the main geometric properties of spacetimes provided with
off--diagonal metrics which can be effectively diagonalized with respect to
the N--elongated frames (\ref{ndif}) and (\ref{nder}):

\begin{enumerate}
\item Such spacetimes are characterized by certain anholonomic frame
relations \ (anholonomy conditions)
\begin{equation}
e_{\alpha }^{[N]}e_{\beta }^{[N]}-e_{\beta }^{[N]}e_{\alpha
}^{[N]}=w_{~\alpha \beta }^{[N]\gamma }e_{\gamma }^{[N]}  \label{anh}
\end{equation}%
with some nontrivial anholonomy coefficients $w_{~\alpha \beta }^{[N]\gamma
} $ computed as
\begin{eqnarray}
w_{~ij}^{k} &=&0,~w_{~aj}^{k}=0,~~w_{~ab}^{k}=0,~w_{~ab}^{c}=0,
\label{anhol} \\
w_{~bj}^{a} &=&-w_{~jb}^{a}=\partial _{b}N_{j}^{a},~w_{~ij}^{a}=-\Omega
_{ij}^{a}=\delta _{i}N_{j}^{a}-\delta _{j}N_{i}^{a}  \notag
\end{eqnarray}%
(we shall omit the label $[N]$ if this will not result in any confusion; as
a matter of principle, we can consider arbitrary anhlonomy coefficients not
related to any off--diagonal metric terms). If the values $w_{~\alpha \beta
}^{[N]\gamma }$ do not vanish, it is not possible to diagonalize the metric (%
\ref{offdig}) by any coordinate transforms: such spacetimes are generic
off--diagonal. The holonomic frames (in particular the coordinate ones)
consist a subclass of vielbeins with vanishing anholonomy coefficients.

\item To any frame (vielbein) transform defined by the coefficients of $%
e_{\alpha }^{[N]}$ decomposed with respect to usual coordinate frames, we
can associate a nonlinear connection structure (in brief, N--connection) $%
\mathbf{N}$ with the coefficients $\{N_{j}^{a}\}$ (in global form the
N--connection was defined in Ref. 33 %\cite{barthel}%
by developing previous ideas from Finsler geometry$^{9-12,34-36},$
%\cite{cartan1,cartan2,cartan3,cartan4,kaw1,kaw2,kaw3},%
investigated in details for vector bundle spaces in Refs. 37,38;\
% \cite{ma1,m2};%
see also Refs. 13-19, 30-32
%\cite{anhtv1,anhtv2,anhtv3,anhtv4,anhtv5,anhtv6,anhtv7,vmethod1,vmethod2,vnp}%
on definition of such objects in (pseudo) Riemannian and
Riemann--Cartan--Weyl geometry or on superspaces). Here we note that the
N--connection structure is characterized by its curvature (N--curvature) $%
\mathbf{\Omega }=\{\Omega _{ij}^{a}\}$ with the coefficients computed as in (%
\ref{anhol}). The well known class of linear connections is to be
distinguished as a particular case when $N_{j}^{a}(x,y)=\Gamma
_{jb}^{a}(x)y^{b}.$ On (pseudo) Riemannian spaces, the N--connection is a
geometric object completely defined by anholonomic frames when the vielbein
transforms $e_{\alpha }^{[N]}$ are parametrized explicitly via certain
values $\left( N_{i}^{a},\delta _{i}^{j},\delta _{b}^{a}\right) ,$ where $%
\delta _{i}^{j}$ and $\delta _{b}^{a}$ are the Kronecker symbols, like in (%
\ref{nder}).

\item The N--coefficients define a conventional global horizontal--vertical
(in brief, h--v ) splitting of spacetime $V^{n+m\text{ }}$ into
holonomic--anholonomic subsets of geometrical objects labelled by
h--components with indices $i,j,...$ and v--components with indices$%
\,a,b,....,$ see details in Refs. 13-19, 30-32. %
%\cite{anhtv1,anhtv2,anhtv3,anhtv4,anhtv5,anhtv6,anhtv7,vmethod1,vmethod2,vnp}.%
The necessary formulas for the h--v--decompositions of the curvature, Ricci
and Einstein tensors are contained in Appendix A.

\item Such generic ''off--diagonal'' spacetimes may be characterized by the
so--called canonical N--adapted linear connection $\Gamma ^{\lbrack
c]}=\{L_{\ jk}^{i},L_{\ bk}^{a},C_{\ jc}^{i},C_{\ bc}^{a}\}$ satisfying the
metricity condition $D_{\gamma }^{[c]}g_{\alpha \beta }=0$ and being adapted
to the h-v--distribution. The coefficients of $\Gamma ^{\lbrack c]}$ are
\begin{eqnarray}
L_{\ jk}^{i} &=&\frac{1}{2}g^{in}\left( \delta _{k}g_{nj}+\delta
_{j}g_{nk}-\delta _{n}g_{jk}\right) ,  \label{dcon} \\
L_{\ bk}^{a} &=&\partial _{b}N_{k}^{a}+\frac{1}{2}h^{ac}\left( \delta
_{k}h_{bc}-h_{dc}\partial _{b}N_{k}^{d}-h_{db}\partial _{c}N_{k}^{d}\right) ,
\notag \\
C_{\ jc}^{i} &=&\frac{1}{2}g^{ik}\partial _{c}g_{jk},\ C_{\ bc}^{a}=\frac{1}{%
2}h^{ad}\left( \partial _{c}h_{db}+\partial _{b}h_{dc}-\partial
_{d}h_{bc}\right) ,  \notag
\end{eqnarray}%
where $\delta _{k}=\delta /\partial x^{k}$ and $\partial _{c}=\partial
/\partial y^{a};$ they are constructed from the coefficients (and their
partial derivatives) of the metric and N--connection. This connection is an
anholonomic deformation (by \ N--coefficients) of the Levi--Civita
connection.

\item The torsion of the connection $\Gamma ^{\lbrack c]}$ is defined (for
simplicity, we omit the label $[c])$
\begin{equation}
T_{\ \beta \gamma }^{\alpha }=\Gamma _{\ \beta \gamma }^{\alpha }-\Gamma _{\
\gamma \beta }^{\alpha }+w_{\ \beta \gamma }^{\alpha },  \label{torsion}
\end{equation}%
with h--v--components
\begin{eqnarray}
T_{.jk}^{i} &=&T_{jk}^{i}=L_{jk}^{i}-L_{kj}^{i}=0%
\mbox{ in the canonical
case },\quad  \notag \\
\quad T_{.bc}^{a} &=&S_{.bc}^{a}=C_{bc}^{a}-C_{cb}^{a}=0%
\mbox{ in the
canonical case },  \label{dtors} \\
T_{.ja}^{i} &=&0,\ T_{ja}^{i}=-T_{aj}^{i}=-C_{ja}^{i},\ T_{.ij}^{a}=-\Omega
_{ij}^{a},\quad T_{.bi}^{a}=-T_{.bi}^{a}=\partial
_{b}N_{i}^{a}-L_{.bi}^{a}.\quad  \notag
\end{eqnarray}%
The nonvanishing components of torsion are induced as an anholonomic frame
effect which is obtained by vielbien transforms (\ref{ndif}) and (\ref{nder}%
) even for a (pseudo) Riemannian metric (\ref{offdig}). In this paper, we
shall also consider some nontrivial torsion structures existing in extra
dimension gravity.

\item By straightforward calculations with respect to the frames (\ref{ndif}%
) and (\ref{nder}) (see for instance, Refs. 39,40) \ %
% \cite{ste1,mtw})%
we can compute the coefficients of the Levi--Civita connection $%
\bigtriangledown ,$ i. e. $\Gamma _{\alpha \beta \gamma }^{[\bigtriangledown
]}=g\left( e_{\alpha }^{[N]},\bigtriangledown _{\gamma }e_{\beta
}^{[N]}\right) =g_{\alpha \tau }\Gamma _{\beta \gamma }^{[\bigtriangledown
]\tau },$ satisfying the metricity condition $\bigtriangledown _{\gamma
}g_{\alpha \beta }=0$ for $g_{\alpha \beta }=\left( g_{ij},h_{ab}\right) ,$%
\begin{equation*}
\Gamma _{\alpha \beta \gamma }^{[\bigtriangledown ]}=\frac{1}{2}\left(
e_{\beta }^{[N]}g_{\alpha \gamma }+e_{\gamma }^{[N]}g_{\beta \alpha
}-e_{\alpha }^{[N]}g_{\gamma \beta }+g_{\alpha \tau }w_{\ \gamma \beta
}^{\tau }+g_{\beta \tau }w_{\ \alpha \gamma }^{\tau }-g_{\beta \tau }w_{\
\beta \alpha }^{\tau }\right) .
\end{equation*}%
Using the values (\ref{anhol}) and (\ref{nder}), we can write
\begin{equation}
\Gamma _{\beta \gamma }^{[\bigtriangledown ]\tau }=\{L_{\ jk}^{i},L_{\
bk}^{a}+\frac{\partial N_{k}^{a}}{\partial y^{b}},C_{\ jc}^{i}+\frac{1}{2}%
g^{ik}\Omega _{jk}^{a}h_{ca},C_{\ bc}^{a}\}.  \label{lcc}
\end{equation}%
Comparing the coefficients of $\Gamma ^{\lbrack c]}$ and $\Gamma ^{\lbrack
\bigtriangledown ]},$ we conclude that both connections have the same
coefficients with respect to the N--adapted frames (\ref{ndif}) and (\ref%
{nder}) if and only if $\partial N_{k}^{a}/$ $\partial y^{b}=0$ and $\Omega
_{jk}^{a}=0,$ i. e. the N--connection curvature vanishes.

\item The ansatz of type (\ref{offdig}) have been largely used in
Kaluza--Klein theories (see, for instance, Refs. 41-43). %
%\cite{salst,prd,ow}).%
For the corresponding compactifications, the coefficients $N_{i}^{a}$ may be
associated to the potential of certain gauge fields but, in general, they
belong to some noncompactified metric and vielbein gravitational fields.
There were elaborated general methods for constructing exact solutions
without compactification and arbitrary $N_{i}^{a}$ in various type of
gravity models$^{13-19}.$ %
%\cite{anhtv1,anhtv2,anhtv3,anhtv4,anhtv5,anhtv6,anhtv7}.%
\end{enumerate}

Any ansatz of type (\ref{offdig}) with the components satisfying the
conditions of the Theorems 1-3 from the Appendix B define a new class of
exact solutions, vacuum and nonvacuum ones, in 3-5 dimensional gravity
parametrized by generic off--diagonal metrics with the coefficients
depending on 2,3 or even 4 variables. These solutions can be constructed in
explicit form by using corresponding boundary and symmetry conditions
following the so--called 'anholonomic frame method' elaborated and developed
in Refs. 13-19,30,31
%\cite{anhtv1,anhtv2,anhtv3,anhtv4,anhtv5,anhtv6,anhtv7,vmethod1,vmethod2}%
(for instance, they can describe black elipsoid/tori configurations, 2-3
dimensional solitonic--spinor--dilaton interactions, polarized wormhole/flux
tube solutions, locally anisotropic Taub\ NUT spacetimes and so on).

Perhaps, by using the anholonomic frame method, we can construct the most
general known class of exact solutions in Einstein gravity and its extra
dimension and string generalizations. From a formal point of view, we can
use supperpositions of anholonomic maps in order to construct integral
varieties of the Einstein equations with the metric/frame coefficients being
functions of necessary smooth class depending on arbitrary number of
variables but parametrized as products of functions depending on 1,2,3 and 4
real, or some complex, variables with real and complex valued functions. The
physical meaning of such classes of solutions should be stated following
explicit physical models. We note that the bulk of the well known black hole
and cosmological solutions (for instance, the Schwarschild, Kerr--Newman,
Reisner--Nordstrom and Friedman--Roberston--Walker solutions) are with
metrics being diagonalizable by coordinate transforms and depending only on
one variable (radial or timelike), with imposed spherical or cylindrical
symmetries and subjected to the conditions of Killing symmetry being
asymptotically flat.

In general, the solutions with anholonomic configurations do not posses
Killing symmetries (for instance, they are not restricted by ''black hole
uniqueness theorems'', proved for Killing spacetimes satisfying
corresponding asymptotic conditions, see details and references in Ref. 20)
%
% \cite{heus})%
but have new properties like the 1--7 stated above. There is a subclass of
'off--diagonal' solutions resulting in corresponding limits into the well
known asymptotically flat spacetimes, or with (anti) de\ Sitter symmetries$%
^{13-19,30,31}.$ %
% \cite {anhtv1,anhtv2,anhtv3,anhtv4,anhtv5,anhtv6,anhtv7,vmethod1,vmethod2}.%
We are interested to investigate possible symmetries of such 'non--Killing'
exact solutions.

The purpose of the next section is to prove that the spacetimes with a
nontrivial anholonomic and associated N--connec\-ti\-on structure posses a
natural noncommutative symmetry.

\section{\quad Anholonomic Noncommutative Structures}

\label{ncgg}

We shall analyze two simple realizations of noncommutative geometry of
anholonomic frames within the algebra of complex $k\times k$ matrices, $%
M_{k}(\C,u^{\alpha })$ depending on coordinates $u^{\alpha }$ on \ spacetime
$V^{n+m}$ connected to complex Lie algebras $SL\left( k,\C\right) $ and $%
SU_{k}.$ We shall consider matrix valued functions of necessary smoothly
class derived from the anholonomic frame relations (\ref{anh}) (being
similar to the Lie algebra relations) with the coefficients (\ref{anhol})
induced by off--diagonal metric terms in (\ref{offdig}) and by N--connection
coefficients $N_{i}^{a}.$ We shall use algebras of complex matrices in order
to have the possibility for some extensions to complex solutions. Usually,
for commutative gravity models, the anholonomy coefficients $w_{~\alpha
\beta }^{[N]\gamma }$ are real functions but in the section 7 we shall
consider also complex spacetimes related to noncommutative gravity$^{3-5}.$
%\cite{ncg1,ncg2,ncg3}.%

We start with the basic relations for the simplest model of noncommutative
geometry realized with the algebra of complex $\left( k\times k\right) $
noncommutative matrices$^{44}, $\ %\cite{dub},%
$M_{k}(\C).$ An element $M\in M_{k}(\C)$ can be represented as a linear
combination of the unit $\left( k\times k\right) $ matrix $I$ and $\left(
k^{2}-1\right) $ hermitian traseless matrices $q_{\underline{\alpha }}$ with
the underlined index $\underline{\alpha }$ running values $1,2,...,k^{2}-1,$
i. e.
\begin{equation*}
M=\alpha \ I+\sum \beta ^{\underline{\alpha }}q_{\underline{\alpha }}
\end{equation*}%
for some constants $\alpha $ and $\beta ^{\underline{\alpha }}.$ It is
possible to chose the basis matrices $q_{\underline{\alpha }}$ satisfying
the relations%
\begin{equation}
q_{\underline{\alpha }}q_{\underline{\beta }}=\frac{1}{k}\rho _{\underline{%
\alpha }\underline{\beta }}I+Q_{\underline{\alpha }\underline{\beta }}^{%
\underline{\gamma }}q_{\underline{\gamma }}-\frac{i}{2}f_{~\underline{\alpha
}\underline{\beta }}^{\underline{\gamma }}q_{\underline{\gamma }},
\label{gr1}
\end{equation}%
where $i^{2}=-1$ and the real coefficients satisfy the properties
\begin{equation*}
Q_{\underline{\alpha }\underline{\beta }}^{\underline{\gamma }}=Q_{%
\underline{\beta }\underline{\alpha }}^{\underline{\gamma }},\ Q_{\underline{%
\gamma }\underline{\beta }}^{\underline{\gamma }}=0,\ f_{~\underline{\alpha }%
\underline{\beta }}^{\underline{\gamma }}=-f_{~\underline{\beta }\underline{%
\alpha }}^{\underline{\gamma }},f_{~\underline{\gamma }\underline{\alpha }}^{%
\underline{\gamma }}=0
\end{equation*}%
with $f_{~\underline{\alpha }\underline{\beta }}^{\underline{\gamma }}$
being the structure constants of the Lie group $SL\left( k,\C\right) $ and
the Killing--Cartan metric tensor $\rho _{\underline{\alpha }\underline{%
\beta }}=f_{~\underline{\alpha }\underline{\gamma }}^{\underline{\tau }}f_{~%
\underline{\tau }\underline{\beta }}^{\underline{\gamma }}.$ The interior
derivatives $\widehat{\partial }_{\underline{\gamma }}$ of this algebra can
be defied as
\begin{equation}
\widehat{\partial }_{\underline{\gamma }}q_{\underline{\beta }}=ad\left( iq_{%
\underline{\gamma }}\right) q_{\underline{\beta }}=i[q_{\underline{\gamma }%
},q_{\underline{\beta }}]=f_{~\underline{\gamma }\underline{\beta }}^{%
\underline{\alpha }}q_{\underline{\alpha }}.  \label{der1a}
\end{equation}%
Following the Jacoby identity, we obtain
\begin{equation}
\widehat{\partial }_{\underline{\alpha }}\widehat{\partial }_{\underline{%
\beta }}-\widehat{\partial }_{\underline{\beta }}\widehat{\partial }_{%
\underline{\alpha }}=f_{~\underline{\alpha }\underline{\beta }}^{\underline{%
\gamma }}\widehat{\partial }_{\underline{\gamma }}.  \label{jacob}
\end{equation}

Our idea is to construct a noncommutative geometry starting from the
anholonomy relations of frames (\ref{anh}) by adding to the structure
constants $f_{~\underline{\alpha }\underline{\beta }}^{\underline{\gamma }}$
the anhlonomy coefficients $w_{\ \alpha \gamma }^{[N]\tau }$ (\ref{anhol}),
Such deformed structure constants consist from N--connection coefficients $%
N_{i}^{a}$ and their first partial derivatives, i. e. they are induced by
some off--diagonal terms in the metric (\ref{offdig}) being a solution of
the gravitational field equations . We note that there is a rough analogy
between formulas (\ref{jacob}) and (\ref{anh}) because the anholonomy
coefficients do not satisfy, in general, the condition $w_{\ \tau \alpha
}^{[N]\tau }=0.$ There is also another substantial difference: the
anholonomy relations are defined for a manifold of dimension $n+m,$ with
Greek indices $\alpha ,\beta ,...$ running values from $1$ to $\ n+m$ but
the matrix noncommutativity relations are stated for traseless matrices
labeled by underlined indices $\underline{\alpha },\underline{\beta },$
running values from $1$ to $k^{2}-1.$ It is not possible to satisfy the
condition $k^{2}-1=n+m$ by using integer numbers for arbitrary $n+m.$ We
suggest to extend the dimension of spacetime from $n+m$ to any $n^{\prime
}\geq n$ and $m^{\prime }\geq m$ when the condition $k^{2}-1=n^{\prime
}+m^{\prime }$ can be satisfied by a trivial embedding of the metric (\ref%
{offdig}) into higher dimension, for instance, by adding the necessary
number of unities on the diagonal by writing
\begin{equation*}
\widehat{g}_{\underline{\alpha }\underline{\beta }}=\left[
\begin{array}{ccccc}
1 & ... & 0 & 0 & 0 \\
... & ... & ... & ... & ... \\
0 & ... & 1 & 0 & 0 \\
0 & ... & 0 & g_{ij}+N_{i}^{a}N_{j}^{b}h_{ab} & N_{j}^{e}h_{ae} \\
0 & ... & 0 & N_{i}^{e}h_{be} & h_{ab}%
\end{array}%
\right]
\end{equation*}%
and $e_{\underline{\alpha }}^{[N]}=\delta _{\underline{\alpha }}=\left(
1,1,...,e_{\alpha }^{[N]}\right) ,$ where, for simplicity, we preserve the
same type of underlined Greek indices, $\underline{\alpha },\underline{\beta
}...=1,2,...,k^{2}-1=n^{\prime }+m^{\prime }.$ The anholonomy coefficients $%
w_{~\alpha \beta }^{[N]\gamma }$ can be extended with some trivial zero
components and for consistency we rewrite them without labeled indices, $%
w_{~\alpha \beta }^{[N]\gamma }\rightarrow W_{~\underline{\alpha }\underline{%
\beta }}^{\underline{\gamma }}.$ The set of anholonomy coefficients $%
w_{~\alpha \beta }^{[N]\gamma }$(\ref{anhol}) may result in degenerated
matrices, for instance for certain classes of exact solutions of the
Einstein equations. Nevertheless, we can consider an extension $w_{~\alpha
\beta }^{[N]\gamma }\rightarrow W_{~\underline{\alpha }\underline{\beta }}^{%
\underline{\gamma }}$ when the coefficients $w_{~\underline{\alpha }%
\underline{\beta }}^{\underline{\gamma }}(u^{\underline{\tau }})$ for any
fixed value $u^{\underline{\tau }}=u_{[0]}^{\underline{\tau }}$ would be
some deformations of the\ structure constants of the Lie algebra $SL\left( k,%
\C\right) ,$ like
\begin{equation}
W_{~\underline{\alpha }\underline{\beta }}^{\underline{\gamma }}=f_{~%
\underline{\alpha }\underline{\beta }}^{\underline{\gamma }}+w_{~\underline{%
\alpha }\underline{\beta }}^{\underline{\gamma }},  \label{anhb}
\end{equation}%
being nondegenerate. \

Instead of the matrix algebra $M_{k}(\C),$ constructed from constant complex
elements, we shall consider dependencies on coordinates $u^{\underline{%
\alpha }}=\left( 0,...,u^{\alpha }\right) ,$ for instance, like a trivial
matrix bundle on $V^{n^{\prime }+m^{\prime }},$ and denote this space $M_{k}(%
\C,u^{\underline{\alpha }}).$ Any element $B\left( u^{\underline{\alpha }%
}\right) \in M_{k}(\C,u^{\underline{\alpha }})$ with a noncommutative
structure induced by $W_{~\underline{\alpha }\underline{\beta }}^{\underline{%
\gamma }}$ is represented as a linear combination of the unit $(n^{\prime
}+m^{\prime })\times (n^{\prime }+m^{\prime })$ matrix $I$ and the $%
[(n^{\prime }+m^{\prime })^{2}-1]$ hermitian traceless matrices $q_{%
\underline{\alpha }}\left( u^{\underline{\tau }}\right) $ with the
underlined index $\underline{\alpha }$ running values $1,2,...,(n^{\prime
}+m^{\prime })^{2}-1,$%
\begin{equation*}
B\left( u^{\underline{\tau }}\right) =\alpha \left( u^{\underline{\tau }%
}\right) \ I+\sum \beta ^{\underline{\alpha }}\left( u^{\underline{\tau }%
}\right) q_{\underline{\alpha }}\left( u^{\underline{\tau }}\right)
\end{equation*}%
under condition that the following relation holds:%
\begin{equation*}
q_{\underline{\alpha }}\left( u^{\underline{\tau }}\right) q_{\underline{%
\beta }}\left( u^{\underline{\gamma }}\right) =\frac{1}{n^{\prime
}+m^{\prime }}\rho _{\underline{\alpha }\underline{\beta }}\left( u^{%
\underline{\nu }}\right) +Q_{\underline{\alpha }\underline{\beta }}^{%
\underline{\gamma }}q_{\underline{\gamma }}\left( u^{\underline{\mu }%
}\right) -\frac{i}{2}W_{~\underline{\alpha }\underline{\beta }}^{\underline{%
\gamma }}q_{\underline{\gamma }}\left( u^{\underline{\mu }}\right)
\end{equation*}%
with the same values of $Q_{\underline{\alpha }\underline{\beta }}^{%
\underline{\gamma }}$ from the Lie algebra for $SL\left( k,\C\right) $ but
with the Killing--Cartan like metric tensor defined by anholonomy
coefficients, i. e. $\rho _{\underline{\alpha }\underline{\beta }}\left( u^{%
\underline{\nu }}\right) =W_{~\underline{\alpha }\underline{\gamma }}^{%
\underline{\tau }}\left( u^{\underline{\alpha }}\right) W_{~\underline{\tau }%
\underline{\beta }}^{\underline{\gamma }}\left( u^{\underline{\alpha }%
}\right) .$ For complex spacetimes, we shall consider that the coefficients $%
N_{\underline{i}}^{\underline{a}}$ and $W_{~\underline{\alpha }\underline{%
\beta }}^{\underline{\gamma }}$ may be some complex valued functions of
necessary smooth (in general, with complex variables) class. In result, the
Killing--Cartan like metric tensor $\rho _{\underline{\alpha }\underline{%
\beta }}$ can be also complex.

We rewrite (\ref{anh}) as
\begin{equation}
e_{\underline{\alpha }}e_{\underline{\beta }}-e_{\underline{\beta }}e_{%
\underline{\alpha }}=W_{~\underline{\alpha }\underline{\beta }}^{\underline{%
\gamma }}e_{\underline{\gamma }}  \label{anh1}
\end{equation}%
being equivalent to (\ref{jacob}) and defining a noncommutative anholonomic
structure (for simplicity, we use the same symbols $e_{\underline{\alpha }}$
as for some 'N--elongated' partial derivatives, but with underlined
indices). The analogs of derivation operators (\ref{der1a}) are stated by
using $W_{~\underline{\alpha }\underline{\beta }}^{\underline{\gamma }},$%
\begin{equation}
e_{\underline{\alpha }}q_{\underline{\beta }}\left( u^{\underline{\gamma }%
}\right) =ad\left[ iq_{\underline{\alpha }}\left( u^{\underline{\gamma }%
}\right) \right] q_{\underline{\beta }}\left( u^{\underline{\gamma }}\right)
=i\left[ q_{\underline{\alpha }}\left( u^{\underline{\gamma }}\right) q_{%
\underline{\beta }}\left( u^{\underline{\gamma }}\right) \right] =W_{~%
\underline{\alpha }\underline{\beta }}^{\underline{\gamma }}q_{\underline{%
\gamma }}  \label{nder1}
\end{equation}

The operators (\ref{nder1}) define a linear space of anholonomic derivations
satisfying the conditions (\ref{anh1}), denoted $AderM_{k}(\C,u^{\underline{%
\alpha }}),$ elongated by N--connection and distinguished into irreducible
h-- and v--components, respectively, into $e_{\underline{i}}$ and $e_{%
\underline{b}},$ like $e_{\underline{\alpha }}=\left( e_{\underline{i}%
}=\partial _{\underline{i}}-N_{\underline{i}}^{\underline{a}}e_{\underline{a}%
},e_{\underline{b}}=\partial _{\underline{b}}\right) .$ The space $AderM_{k}(%
\C,u^{\underline{\alpha }})$ is not a left module over \ the algebra $M_{k}(%
\C,u^{\underline{\alpha }})$ which means that there is a a substantial
difference with respect to the usual commutative differential geometry where
a vector field multiplied on the left by a function produces a new vector
field.

The duals to operators (\ref{nder1}), $e^{\underline{\mu }},$ found from $e^{%
\underline{\mu }}\left( e_{\underline{\alpha }}\right) =\delta _{_{%
\underline{\alpha }}}^{\underline{\mu }}I,$ define a canonical basis of
1--forms $e^{\underline{\mu }}$ connected to the N--connection structure. By
using these forms, we can span a left module over $M_{k}(\C,u^{\underline{%
\alpha }})$ following $q_{\underline{\alpha }}e^{\underline{\mu }}\left( e_{%
\underline{\beta }}\right) =q_{\underline{\alpha }}\delta _{_{\underline{%
\beta }}}^{\underline{\mu }}I=q_{\underline{\alpha }}\delta _{_{\underline{%
\beta }}}^{\underline{\mu }}.$ \ For an arbitrary vector field
\begin{equation*}
Y=Y^{\alpha }e_{\alpha }\rightarrow Y^{\underline{\alpha }}e_{\underline{%
\alpha }}=Y^{\underline{i}}e_{\underline{i}}+Y^{\underline{a}}e_{\underline{a%
}},
\end{equation*}%
it is possible to define an exterior differential (in our case being
N--elongated), starting with the action on a function $\varphi $
(equivalent, a 0--form),
\begin{equation*}
\delta \ \varphi \left( Y\right) =Y\varphi =Y^{\underline{i}}\delta _{%
\underline{i}}\varphi +Y^{\underline{a}}\partial _{\underline{a}}\varphi
\end{equation*}%
when%
\begin{equation*}
\left( \delta \ I\right) \left( e_{\underline{\alpha }}\right) =e_{%
\underline{\alpha }}I=ad\left( ie_{\underline{\alpha }}\right) I=i\left[ e_{%
\underline{\alpha }},I\right] =0,\mbox{ i. e. }\delta I=0,
\end{equation*}%
and
\begin{equation}
\delta q_{\underline{\mu }}(e_{\underline{\alpha }})=e_{\underline{\alpha }%
}(e_{\underline{\mu }})=i[e_{\underline{\mu }},e_{\underline{\alpha }}]=W_{~%
\underline{\alpha }\underline{\mu }}^{\underline{\gamma }}e_{\underline{%
\gamma }}.  \label{aux1}
\end{equation}%
Considering the nondegenerated case, we can invert (\ref{aux1}) as to obtain
a similar expression with respect to $e^{\underline{\mu }},$%
\begin{equation}
\delta (e_{\underline{\alpha }})=W_{~\underline{\alpha }\underline{\mu }}^{%
\underline{\gamma }}e_{\underline{\gamma }}e^{\underline{\mu }},
\label{aux2}
\end{equation}%
from which a very important property follows by using the Jacobi identity, $%
\delta ^{2}=0,$ resulting in a possibility to define a usual Grassman
algebra of $p$--forms with the wedge product $\wedge $ stated as%
\begin{equation*}
e^{\underline{\mu }}\wedge e^{\underline{\nu }}=\frac{1}{2}\left( e^{%
\underline{\mu }}\otimes e^{\underline{\nu }}-e^{\underline{\nu }}\otimes e^{%
\underline{\mu }}\right) .
\end{equation*}%
We can write (\ref{aux2}) as
\begin{equation*}
\delta (e^{\underline{\alpha }})=-\frac{1}{2}W_{~\underline{\beta }%
\underline{\mu }}^{\underline{\alpha }}e^{\underline{\beta }}e^{\underline{%
\mu }}
\end{equation*}%
and introduce the canonical 1--form $e=q_{\underline{\alpha }}e^{\underline{%
\alpha }}$ being coordinate--independent and adapted to the N--connection
structure and satisfying the condition $\delta e+e\wedge e=0.$

In a standard manner, we can introduce the volume element induced by the
canonical Cartan--Killing metric and the corresponding star operator $\star $%
\ (Hodge duality).\ We define the volume element $\sigma $ by using the
complete antisymmetric tensor $\epsilon _{\underline{\alpha }_{1}\underline{%
\alpha }_{2}...\underline{\alpha }_{k^{2}-1}}$as
\begin{equation*}
\sigma =\frac{1}{\left[ (n^{\prime }+m^{\prime })^{2}-1\right] !}\epsilon _{%
\underline{\alpha }_{1}\underline{\alpha }_{2}...\underline{\alpha }%
_{n^{\prime }+m^{\prime }}}e^{\underline{\alpha }_{1}}\wedge e^{\underline{%
\alpha }_{2}}\wedge ...\wedge e^{\underline{\alpha }_{n^{\prime }+m^{\prime
}}}
\end{equation*}%
to which any $\left( k^{2}-1\right) $--form is proportional $\left(
k^{2}-1=n^{\prime }+m^{\prime }\right) .$ The integral of such a form is
defined as the trace of the matrix coefficient in the from $\sigma $ and the
scalar product introduced for any couple of $p$--forms $\varpi $ and $\psi $%
\begin{equation*}
\left( \varpi ,\psi \right) =\int \left( \varpi \wedge \star \psi \right) .
\end{equation*}

Let us analyze how a noncommutative differential form calculus (induced by
an anholonomic structure) can be developed and related to the Hamiltonian
classical and quantum mechanics and Poisson bracket formalism:

For a $p$--form $\varpi ^{\lbrack p]},$ the anti--derivation $i_{Y}$ with
respect to a vector field $Y\in AderM_{k}(\C,u^{\underline{\alpha }})$ can
be defined as in the usual formalism,%
\begin{equation*}
i_{Y}\varpi ^{\lbrack p]}\left( X_{1},X_{2},...,X_{p-1}\right) =\varpi
^{\lbrack p]}\left( Y,X_{1},X_{2},...,X_{p-1}\right)
\end{equation*}%
where $X_{1},X_{2},...,X_{p-1}\in AderM_{k}(\C,u^{\underline{\alpha }}).$ By
a straightforward calculus we can check that for a 2--form $\Xi =\delta $ $e$
one holds
\begin{equation*}
\delta \Xi =\delta ^{2}e=0\mbox{ and }L_{Y}\Xi =0
\end{equation*}%
where the Lie derivative of forms is defined as $L_{Y}\varpi ^{\lbrack
p]}=\left( i_{Y}\ \delta +\delta \ i_{Y}\right) \varpi ^{\lbrack p]}.$

The Hamiltonian vector field $H_{[\varphi ]}$ of an element of algebra $%
\varphi \in M_{k}(\C,u^{\underline{\alpha }})$ is introduced following the
equality $\Xi \left( H_{[\varphi ]},Y\right) =Y\varphi $ which holds for any
vector field. Then, we can define the Poisson bracket of two functions (in a
quantum variant, observables) $\varphi $ and $\chi ,$ $\{\varphi ,\chi
\}=\Xi \left( H_{[\varphi ]},H_{[\chi ]}\right) $ when
\begin{equation*}
\{e_{\underline{\alpha }},e_{\underline{\beta }}\}=\Xi \left( e_{\underline{%
\alpha }},e_{\underline{\beta }}\right) =i[e_{\underline{\alpha }},e_{%
\underline{\beta }}].
\end{equation*}%
This way, a simple version of noncommutative classical and quantum mechanics
(up to a factor like the Plank constant, $\hbar $) is proposed, being
derived by anholonomic relations for a certain class of exact
'off--diagonal' solutions in commutative gravity.

We note that by using the Lie algebra $SU\left( k,\C\right) $ we can
elaborate an alternative noncommutative calculus related to the special
unitary group $SU_{k}$ in $k$ dimensions when the anholonomic coefficients
\begin{equation}
W_{~\underline{\alpha }\underline{\beta }}^{\underline{\gamma }}=p_{~%
\underline{\alpha }\underline{\beta }}^{\underline{\gamma }}+w_{~\underline{%
\alpha }\underline{\beta }}^{\underline{\gamma }}  \label{anhcsu}
\end{equation}%
induce a linear connection in the associated noncommutative space
(noncommutative geometries with $p_{~\underline{\alpha }\underline{\beta }}^{%
\underline{\gamma }}$ being the structure constants of $SU_{k}$ were
investigated in Refs. 44,46-49,8).
%\cite{dub,dvkm1,dvkm2,madm,mmm,madore}).%

Let us state the main formulas for a such realization of anholonomic
noncommutativity: In this case, the matrix basis $q_{\underline{\alpha }}$
consists from anti--hermitian (and not hermitian) matrices and the relations
(\ref{gr1}) are stated in a different form
\begin{equation}
q_{\underline{\alpha }}q_{\underline{\beta }}=-\frac{1}{k}\rho _{\underline{%
\alpha }\underline{\beta }}I+Z_{\underline{\alpha }\underline{\beta }}^{%
\underline{\gamma }}q_{\underline{\gamma }}+\frac{1}{2}p_{~\underline{\alpha
}\underline{\beta }}^{\underline{\gamma }}q_{\underline{\gamma }},
\label{gr2}
\end{equation}%
where $\rho _{\underline{\alpha }\underline{\gamma }}Z_{\underline{\alpha }%
\underline{\beta }}^{\underline{\gamma }}$ is trace--free and symmetric in
all pairs of indices and $\rho _{\underline{\alpha }\underline{\beta }}=p_{~%
\underline{\alpha }\underline{\gamma }}^{\underline{\tau }}p_{~\underline{%
\tau }\underline{\beta }}^{\underline{\gamma }}.$ We consider dependencies
of matrix coefficients on coordinates $u^{\underline{\alpha }}=\left(
0,...,u^{\alpha }\right) ,$ i. e. we work in the space $M_{k}(\C,u^{%
\underline{\alpha }}),$ \ and introduce the 'anholonomic' derivations $e_{%
\underline{\alpha }},$
\begin{equation*}
e_{\underline{\alpha }}\varphi =[q_{\underline{\alpha }},\varphi ]
\end{equation*}%
for arbitrary matrix function $\varphi \in M_{k}(\C,u^{\underline{\alpha }})$
defining a basis for the Lie algebra of derivations $Der\left[ M_{k}(\C,u^{%
\underline{\alpha }})\right] $ of $M_{k}(\C,u^{\underline{\alpha }}).$ In
this case, we generalize (\ref{gr2}) to
\begin{equation*}
q_{\underline{\alpha }}(u^{\underline{\tau }})q_{\underline{\beta }}(u^{%
\underline{\tau }})=-\frac{1}{k}\rho _{\underline{\alpha }\underline{\beta }%
}(u^{\underline{\tau }})I+Z_{\underline{\alpha }\underline{\beta }}^{%
\underline{\gamma }}q_{\underline{\gamma }}(u^{\underline{\tau }})+\frac{1}{2%
}W_{~\underline{\alpha }\underline{\beta }}^{\underline{\gamma }}(u^{%
\underline{\tau }})q_{\underline{\gamma }}(u^{\underline{\tau }}),
\end{equation*}%
with an effective (of N--anholonomy origin) metric $\rho _{\underline{\alpha
}\underline{\beta }}(u^{\underline{\lambda }})=W_{~\underline{\alpha }%
\underline{\gamma }}^{\underline{\tau }}(u^{\underline{\lambda }})W_{~%
\underline{\tau }\underline{\beta }}^{\underline{\gamma }}(u^{\underline{%
\lambda }})$ being an anholonomic deformation of the Killing metric of $%
SU_{k}.$

In order to define the algebra of forms $\Omega ^{\ast }\left[ M_{k}(\C,u^{%
\underline{\alpha }})\right] $ over $M_{k}(\C,u^{\underline{\alpha }})$ we
put $\Omega ^{0}=$ $M_{k}$ and write
\begin{equation*}
\delta \varphi \left( e_{\underline{\alpha }}\right) =e_{\underline{\alpha }%
}(\varphi )
\end{equation*}
for every matrix function $\varphi \in M_{k}(\C,u^{\underline{\alpha }}).$
As a particular case, we have
\begin{equation*}
\delta q^{\underline{\alpha }}\left( e_{\underline{\beta }}\right) =-W_{~%
\underline{\beta }\underline{\gamma }}^{\underline{\alpha }}q^{\underline{%
\gamma }}
\end{equation*}%
where indices are raised and lowered with the anholonomically deformed
metric $\rho _{\underline{\alpha }\underline{\beta }}(u^{\underline{\lambda }%
}).$ This way, we can define the set of 1--forms $\Omega ^{1}\left[ M_{k}(\C%
,u^{\underline{\alpha }})\right] $ to be the set of all elements of the form
$\varphi \delta \beta $ with $\varphi $ and $\beta $ belonging to $M_{k}(\C%
,u^{\underline{\alpha }}).$ The set of all differential forms define a
differential algebra $\Omega ^{\ast }\left[ M_{k}(\C,u^{\underline{\alpha }})%
\right] $ with the couple $\left( \Omega ^{\ast }\left[ M_{k}(\C,u^{%
\underline{\alpha }})\right] ,\delta \right) $ said to be a differential
calculus in $M_{k}(\C,u^{\underline{\alpha }})$ induced by the anholonomy of
certain exact solutions (with off--diagonal metrics and associated
N--connections) in a gravity theory.

We can also find a set of generators $e^{\underline{\alpha }}$ of $\Omega
^{1}\left[ M_{k}(\C,u^{\underline{\alpha }})\right] ,$ as a left/ right
--module completely characterized by the duality equations $e^{\underline{%
\mu }}\left( e_{\underline{\alpha }}\right) =\delta _{_{\underline{\alpha }%
}}^{\underline{\mu }}I$ and related to $\delta q^{\underline{\alpha }},$%
\begin{equation*}
\delta q^{\underline{\alpha }}=W_{~\underline{\beta }\underline{\gamma }}^{%
\underline{\alpha }}q^{\underline{\beta }}q^{\underline{\gamma }}\mbox{ and }%
e^{\underline{\mu }}=q_{\underline{\gamma }}q^{\underline{\mu }}\delta q^{%
\underline{\gamma }}.
\end{equation*}%
Similarly to the formalism presented in details in Ref. 8, %\cite{madore},%
we can elaborate a differential calculus with derivations by introducing a
linear torsionless connection%
\begin{equation*}
\mathcal{D}e^{\underline{\mu }}=-\omega _{\ \underline{\gamma }}^{\underline{%
\mu }}\otimes e^{\underline{\gamma }}
\end{equation*}%
with the coefficients $\omega _{\ \underline{\gamma }}^{\underline{\mu }}=-%
\frac{1}{2}W_{~\underline{\gamma }\underline{\beta }}^{\underline{\mu }}e^{%
\underline{\gamma }},$ resulting in the curvature 2--form%
\begin{equation*}
\mathcal{R}_{\ \underline{\gamma }}^{\underline{\mu }}=\frac{1}{8}W_{~%
\underline{\gamma }\underline{\beta }}^{\underline{\mu }}W_{~\underline{%
\alpha }\underline{\tau }}^{\underline{\beta }}e^{\underline{\alpha }}e^{%
\underline{\tau }}.
\end{equation*}%
So, even the anholonomy coefficients (\ref{anhol}) of a solution, for
instance, in string gravity, has nontrivial torsion coefficients, (\ref%
{dtors}), the associated linear connection induced by the anholonomy
coefficients in the associated noncommutative space of symmetries of the
solution can be defined to be torsionless but to have a specific metrics and
curvature being very different from the spacetime curvature tensor. This is
a surprising fact that 'commutative' curved spacetimes provided with
off--diagonal metrics and associated anhlonomic frames and N--connections
may be characterized by a noncommutative 'shadow' space with a Lie algebra
like structure induced by the frame anolonomy. We argue that such metrics
possess anholonomic noncommutative symmetries.

Finally, in this section, we conclude that for the 'holonomic' solutions of
the Einstein equations, with vanishing $w_{~\underline{\alpha }\underline{%
\beta }}^{\underline{\gamma }},$ any associated noncommutative geometry or $%
SL\left( k,\C\right) ,$ or $SU_{k}$ type, decouples from the off--diagonal
(anholonomic) gravitational background and transforms into a trivial one
defined by the corresponding structure constants of the chosen Lie algebra.
The anholonomic noncommutativity and the related differential geometry are
induced by the anholonomy coefficients. All such structures reflect a
specific type of symmetries of generic off--diagonal metrics and associated
frame/ N--connection structures. Considering exact solutions of the
gravitational field equations, we can assert that we constructed a class of
vacuum or nonvacuum metrics possessing a specific noncommutative symmetry
instead, for instance, of any usual Killing symmetry. In general, we can
introduce a new classification of spacetimes following anholonomic
noncommutative aglebraic properties of metrics and vielbein structures.

\section{\quad Black Ellipsoids with Noncommutative Symmetry}

In this section, we shall analyze two classes of black ellipsoid solutions
of the Einstein and (anti) de Sitter gravity (with arbitrary cosmological
term) possessing hidden noncommutative symmetries. Such off--diagonal
metrics will be constructed as to generate also exact solutions in complex
gravity, with respect to complex N--elongated vielbeins (for simplicity, we
shall consider the metric coefficients to be real with respect to such
complex frames) which have to be considered if any noncommutativity of
coordinates with complex parameters and/or Wick like rotations to Euclidean
signatures are introduced. Such metrics are stable for certain
configurations with complex off--diagonal terms (a rigorous proof may be
performed by generalizing to complex spaces the results from Refs. 18,19).
%
%\cite {anhtv6,anhtv7}).%

\subsection{Anholonomic Complex Deformations of the Schwarzschild Solution}

We consider a 4D off--diagonal metric ansatz (a complex generalization of (%
\ref{offdig}), or equivalently, of (\ref{dmetric}) with complex frames
(vielbeins) (\ref{ndif}) and (\ref{nder}), \ see also the ansatz (\ref%
{ansatz4}) in the Appenix C),
\begin{eqnarray}
\delta s^{2} &=&-\left( 1-\frac{2m}{r}+\frac{\varepsilon }{r^{2}}\right)
^{-1}dr^{2}-r^{2}\gamma (r)d\theta ^{2}  \label{sch} \\
&&-\eta _{3}\left( r,\varphi \right) r^{2}\sin ^{2}\theta d\varphi ^{2}+\eta
_{4}\left( r,\varphi \right) \left( 1-\frac{2m}{r}+\frac{\varepsilon }{r^{2}}%
\right) \delta t^{2}  \notag
\end{eqnarray}%
for usual local spherical and time like coordinates $u=\{u^{\alpha }=\left(
u^{2}=r,u^{3}=\theta ,u^{4}=\varphi ,u^{5}=t\right) \}.$ In order to have
compatibility with notations from Appendices B and C, in this subsection, we
consider that 4D Greek indices run \ from 2 till 5 where the
''polarization'' functions $\eta _{3,4}$ are decomposed on a small parameter
$\varepsilon ,0<|\varepsilon |\ll 1,$
\begin{eqnarray}
\eta _{4}\left( r,\varphi \right) &=&\eta _{4[0]}\left( r,\varphi \right)
+\varepsilon \lambda _{4}\left( r,\varphi \right) +\varepsilon ^{2}\gamma
_{4}\left( r,\varphi \right) +...,  \label{decom1} \\
\eta _{5}\left( r,\varphi \right) &=&1+\varepsilon \lambda _{5}\left(
r,\varphi \right) +\varepsilon ^{2}\gamma _{5}\left( r,\varphi \right) +...,
\notag
\end{eqnarray}%
$\gamma (r)$ is a necessary smooth class function satisfying $\gamma (r)=1$
if $\varepsilon \rightarrow 0$ (it will be defined below) and
\begin{equation}
\delta t=dt+n_{2}\left( r,\varphi \right) dr  \label{anhc}
\end{equation}%
for $n_{2}=n_{2}^{[re]}+in_{2}^{[im]}\sim \varepsilon ...+\varepsilon ^{2}$
terms being, in general, a complex valued function. In the particular case,
when $n_{2}$ is real, i. e. when $n_{2}=n_{2}^{[re]}$ and $n_{2}^{[im]}=0,$
the labels $[re]$ and $[im]$ being used respectively for the real and
imaginary parts, the metric (\ref{sch}) was investigated in connection to
the definition of static and non--static black ellipsoid configurations in
Refs. 13-19,30,31). %
%\cite{anhtv1,anhtv2,anhtv3,anhtv4,anhtv5,anhtv6,anhtv7,vmethod1,vmethod2}).%
\ The functions $\eta _{4,5}\left( r,\varphi \right) $ and $n_{2}\left(
r,\varphi \right) $ will be found as the metric will define a solution of
the vacuum Einstein equations (\ref{einsteq3}) (see Appendices A, B and C
for the explicit form of field equations (\ref{ricci1a})--(\ref{ricci4a})
written for the 4D ansatz (\ref{ansatz4})). By introducing certain complex
components of metric generated by small deformations of the spherical static
symmetry on a small positive parameter $\varepsilon $ (in the limits $%
\varepsilon \rightarrow 0$ and $\eta _{4,5}\rightarrow 1$ we have just the
Schwarzschild solution for a point particle of mass $m)$ we show here that
it is possible to extend the results of Refs. 18,19 %
%\cite{anhtv6,anhtv7}%
with respect to complex anholonomic structures (\ref{ndif}) with a
nontrivial component $N_{2}^{5}=n_{2}\left( r,\varphi \right) $ given by
N--elongation (\ref{anhc}).

A more interesting class of exact solutions with an effective electric
charge $q$ induced from the complex/ noncommutative/ anholonomic gravity may
be constructed if we state that the parameter of anholonomic deformations is
of type $\varepsilon =(iq)^{2}$ for a real $q$ and imaginary $i.$ In this
case the metric (\ref{sch}) will have real coefficients in the first order
of $\varepsilon ,$ being very similar to those from the well known
Reissner--N\"{o}rdstrom metric with, in our case effective, electric charge $%
q.$ For convenience, in our further calculations we shall use both small
parameters $\varepsilon $ and/or $q.$

The set of $\eta ,\lambda $ and $\gamma $ functions from\ (\ref{decom1})
define arbitrary anholonomic (in our case with certain complexity)
deformations of the Schwarschild metric. As a particular case, we can
consider the condition of vanishing of the metric coefficient before $\delta
t^{2}$%
\begin{eqnarray}
\eta _{5}\left( r,\varphi \right) \left( 1-\frac{2m}{r}+\frac{\varepsilon }{%
r^{2}}\right) &=&1-\frac{2m}{r}+\varepsilon \frac{\Phi _{5}}{r^{2}}%
+\varepsilon ^{2}\Theta _{5}=0,  \label{hor1} \\
\Phi _{5} &=&\lambda _{4}\left( r^{2}-2mr\right) +1  \notag \\
\Theta _{5} &=&\gamma _{4}\left( 1-\frac{2m}{r}\right) +\lambda _{4},  \notag
\end{eqnarray}%
defining a rotation ellipsoid configuration when
\begin{equation*}
\lambda _{5}=\left( 1-\frac{2m}{r}\right) ^{-1}(\cos \varphi -\frac{1}{r^{2}}%
)\mbox{ and }\gamma _{5}=-\lambda _{5}\left( 1-\frac{2m}{r}\right) ^{-1}.
\end{equation*}%
Really, in the first order on $\varepsilon ,$ one follows \ a zero value for
the coefficient before $\delta t^{2}$ if
\begin{equation}
r_{+}=\frac{2m}{1-q^{2}\cos \varphi }=2m[1+q^{2}\cos \varphi ],  \label{ebh}
\end{equation}%
which is the equation for a 3D ellipsoid like hypersurface with a small
eccentricity $q^{2}.$ In general, we can consider arbitrary pairs of
functions $\lambda _{5}(r,\theta ,\varphi )$ and $\gamma _{5}(r,\theta
,\varphi )$ (for $\varphi $--anisotropies, \ $\lambda _{5}(r,\varphi )$ and $%
\gamma _{5}(r,\varphi ))$ which may be singular for some values of $r,$ or
on some hypersurfaces $r=r\left( \theta ,\varphi \right) $ ($r=r(\varphi )).$
Such a configuration may define a static black ellipsoid object (a
Scwharszschild like static solution with the horizon slightly deformed to an
ellipsoidal hypersurface$^{18,19}$). % \cite{anhtv6,anhtv7}).%

In general, not being restricted only to ellipsoidal configurations, the
simplest way to analyze the condition of vanishing of the metric coefficient
before $\delta t^{2}$ is to choose $\gamma _{5}$ and $\lambda _{5}$ as to
have $\Theta =0.$ In this case we find from \ (\ref{hor1})%
\begin{equation}
r_{\pm }=m\pm m\sqrt{1-\varepsilon \frac{\Phi }{m^{2}}}\simeq m\left[ 1\pm
\left( 1+q^{2}\frac{\Phi _{5}}{2m^{2}}\right) \right]  \label{hor1a}
\end{equation}%
where $\Phi _{5}\left( r,\varphi \right) $ is taken for $r=2m.$

Let us introduce a new radial coordinate
\begin{equation}
\xi =\int dr\sqrt{|1-\frac{2m}{r}-\frac{q^{2}}{r^{2}}|}  \label{int2}
\end{equation}%
and define
\begin{equation}
h_{4}=-\eta _{4}(\xi ,\varphi )r^{2}(\xi )\sin ^{2}\theta ,\ h_{5}=1-\frac{2m%
}{r}-q^{2}\frac{\Phi _{5}}{r^{2}}.  \label{sch1q}
\end{equation}%
For $r=r\left( \xi \right) $ found as the inverse function after integration
in (\ref{int2}), we can write the metric (\ref{sch}) as
\begin{eqnarray}
ds^{2} &=&-d\xi ^{2}-r^{2}\left( \xi \right) \gamma \left( \xi \right)
d\theta ^{2}+h_{4}\left( \xi ,\theta ,\varphi \right) \delta \varphi
^{2}+h_{5}\left( \xi ,\theta ,\varphi \right) \delta t^{2},  \label{sch1} \\
\delta t &=&dt+n_{2}\left( \xi ,\varphi \right) d\xi ,  \notag
\end{eqnarray}%
where the coefficient $n_{2}$ is taken to solve the equation (\ref{ricci4a})
and to satisfy the condition $\Omega _{jk}^{a}=0$ which states that we fix
the canonical N--adapted connection (\ref{dcon}) to coincide with the
Levi--Civita connection (\ref{lcc}), i. e. to consider a complex like
Einstein but not Einstein--Cartan theory,$\ $which together with the
condition $r^{2}\left( \xi \right) \gamma \left( \xi \right) =\xi ^{2}$ will
be transformed into the usual Schwarschild metric for $\varepsilon
\rightarrow 0.$

Let us define the conditions when the coefficients of metric (\ref{sch})
define solutions of the vacuum Einstein equations (such solutions exists in
the real case following the Theorems 1-3 from the Appendix B, in our case we
only state a generalization for certain complex valued metric coefficients).
For $g_{2}=-1,g_{3}=-r^{2}\left( \xi \right) \gamma \left( \xi \right) $ and
arbitrary $h_{4}(\xi ,\theta ,\varphi )$ and $h_{5}\left( \xi ,\theta
\right) ,$ we get solutions the equations (\ref{ricci1a})--(\ref{ricci3a}).
If $h_{5}$ depends on anisotropic variable $\varphi ,$ the equation (\ref%
{ricci2a}) may be solved if
\begin{equation}
\sqrt{|\eta _{4}|}=\eta _{0}\left( \sqrt{|\eta _{5}|}\right) ^{\ast }
\label{conda}
\end{equation}%
for $\eta _{0}=const.$ Considering decompositions of type (\ref{decom1}) we
put $\eta _{0}=\eta /|\varepsilon |,$ where the constant $\eta $ is taken as
to have $\sqrt{|\eta _{3}|}=1$ in the limits
\begin{equation}
\frac{\left( \sqrt{|\eta _{4}|}\right) ^{\ast }\rightarrow 0}{|\varepsilon
|\rightarrow 0}\rightarrow \frac{1}{\eta }=const.  \label{condb}
\end{equation}%
These conditions are satisfied if the functions $\eta _{4[0]},$ $\lambda
_{4,5}$ and $\gamma _{4,5}$ are related via relations
\begin{equation*}
\sqrt{|\eta _{4[0]}|}=\frac{\eta }{2}\lambda _{5}^{\ast },\lambda _{5}=\eta
\sqrt{|\eta _{4[0]}|}\gamma _{5}^{\ast }
\end{equation*}%
for arbitrary $\gamma _{4}\left( r,\varphi \right) .$ In this paper we
select only such solutions which satisfy the conditions (\ref{conda}) and (%
\ref{condb}) being a complex variant of the conditions (\ref{p1}), see
Appendix B. Similar classes of solutions were selected also in Refs.
[18,19],\ %
%\cite{anhtv6},%
for static black ellipsoid metrics for the (non--complex) Einstein gravity
with real $\varepsilon $\ parameter.

The next step is to construct the solution of (\ref{ricci4a}) which in
general real form is (\ref{n}). To consider linear infinitesimal extensions
on $\varepsilon $ of the Schwarzschild metric, we may write the solution of (%
\ref{ricci4a}) as
\begin{equation*}
n_{2}=\varepsilon \widehat{n}_{2}\left( \xi ,\varphi \right)
\end{equation*}%
where
\begin{eqnarray}
\widehat{n}_{2}\left( \xi ,\varphi \right) &=&n_{2[1]}\left( \xi \right)
+n_{2[2]}\left( \xi \right) \int d\varphi \ \eta _{4}\left( \xi ,\varphi
\right) /\left( \sqrt{|\eta _{5}\left( \xi ,\varphi \right) |}\right)
^{3},\eta _{5}^{\ast }\neq 0;  \label{auxf4s} \\
&=&n_{2[1]}\left( \xi \right) +n_{2[2]}\left( \xi \right) \int d\varphi \
\eta _{4}\left( \xi ,\varphi \right) ,\eta _{5}^{\ast }=0;  \notag \\
&=&n_{2[1]}\left( \xi \right) +n_{2[2]}\left( \xi \right) \int d\varphi
/\left( \sqrt{|\eta _{5}\left( \xi ,\varphi \right) |}\right) ^{3},\eta
_{4}^{\ast }=0;  \notag
\end{eqnarray}%
with the functions $n_{2[1,2]}\left( \xi \right) $ may be complex valued and
have to be stated by boundary conditions.

The data
\begin{eqnarray}
g_{1} &=&-1,g_{2}=-1,g_{3}=-r^{2}(\xi )\gamma (\xi ),  \label{data} \\
h_{4} &=&-\eta _{4}(\xi ,\varphi )r^{2}(\xi )\sin ^{2}\theta ,~h_{5}=1-\frac{%
2m}{r}+\varepsilon \frac{\Phi _{5}}{r^{2}},  \notag \\
w_{2,3} &=&0,n_{2}=\varepsilon \widehat{n}_{2}\left( \xi ,\varphi \right)
,n_{3}=0,  \notag
\end{eqnarray}%
for the metric (\ref{sch}) written in variables $\left( \xi ,\theta ,\varphi
,t\right) $ define a class of solutions of the complex vacuum Einstein
equations with non--trivial polarization function $\eta _{4}$ and extended
on parameter $\varepsilon $ up to the second order (the polarization
functions being taken as to make zero the second order coefficients). Such
solutions are generated by small complex deformations (in particular cases
of rotation ellipsoid symmetry) of the Schwarschild metric. It is possible
to consider some particular parametrizations of N--coefficients resulting in
hermitian metrics and frames, or another type complex configurations. Such
constructions do not affect the stability properties of solutions elaborated
in this paper.

We can relate our complex exact solutions (\ref{data}) with some small
deformations of the Schwar\-zschild metric to a Reissner--N\"{o}rdstrom like
metric with the ''electric'' charge induced effectively from the anholonomic
complex gravity, as well we can satisfy the asymptotically flat condition,
if we chose such functions $n_{k[1,2]}\left( x^{i}\right) $ as $%
n_{k}\rightarrow 0$ for $\varepsilon \rightarrow 0$ and $\eta
_{4}\rightarrow 1.$ These functions have also to be selected as to vanish
far away from the horizon, for instance, like $\sim 1/r^{1+\tau },\tau >0,$
for long distances $r\rightarrow \infty .$ we get a static metric with
effective ''electric'' charge induced by a small, quadratic on $\varepsilon
, $ off--diagonal metric extension. Roughly, we can say that we have
constructed a Reissner--Nordstrem like world ''living'' in a a 'slightly'
complexified frame which induced both an effective electric charge and
certain polarizations of the metric coefficients via the functions $%
h_{4[0]},\eta _{4,5}$ and $n_{5}.$ Another very important property is that
the deformed metric was stated to define a vacuum solution which differs
substantially from the usual Reissner--Nordstrem metric being an exact
static solution of the Einstein--Maxwell equations. For $\varepsilon
\rightarrow 0$ and $h_{4[0]}\rightarrow 1$ and for $\gamma =1,$ the metric (%
\ref{sch}) transforms into the usual Schwarzschild metric. A solution with
ellipsoid symmetry can be selected by a corresponding condition of vanishing
of the coefficient before $\delta t$ which defines an ellipsoidal
hypersurface like for the Kerr metric, but in our case the metric is
non--rotating.

\subsection{ Analytic Extensions of Ellipsoid Complex Metrics}

In order to understand that the constructed in this section exact solution
of vacuum complex gravity really defines black hole like objects we have to
analyze it's analytic extensions, horizon and geodesic behaviour and
stability.

The metric (\ref{sch}) has a singular behaviour for $r=r_{\pm },$ see (\ref%
{hor1a}) like the usual Reissner--Nordstrem one. Our aim is to prove that
this way we have constructed a solution of the vacuum complex Einstein
equations with a static ''anisotropic'' horizon being a small deformation on
parameter $\varepsilon $ of the Schwarzschild's solution horizon. We may
analyze the anisotropic horizon's properties for some fixed ''direction''
given in a smooth vicinity of any values $\varphi =\varphi _{0}$ and $%
r_{+}=r_{+}\left( \varphi _{0}\right) .$ $\ $The final conclusions will be
some general ones for arbitrary $\varphi $ when the explicit values of
coefficients will have a parametric dependence on angular coordinate $%
\varphi .$ Of course, in order to avoid singularities induced by integration
of the equation (\ref{ricci4a}) we have choose such solutions (\ref{auxf4s})
as the associated anhlonomic frames would be of necessary smooth class,
without singularities.

The metrics (\ref{sch}) and (\ref{sch1}) are regular in the regions I ($%
\infty >r>r_{+}^{\Phi }),$ II ($r_{+}^{\Phi }>r>r_{-}^{\Phi })$ and III$%
\;(r_{-}^{\Phi }>r>0).$ As in the Schwarzschild, Reissner--Nordstrem and
Kerr cases these singularities can be removed by introducing suitable
coordinates and extending the manifold to obtain a maximal analytic extension%
$^{50,51}. $ %
%\cite{gb,carter}.%
We have similar regions as in the Reissner--Nordstrem space--time, but with
just only one possibility $\varepsilon <1$ instead of three relations for
static electro--vacuum cases ($q^{2}<m^{2},q^{2}=m^{2},q^{2}>m^{2};$ where $%
q $ and $m$ are correspondingly the electric charge and mass of the point
particle in the Reissner--Nordstrem metric). This property holds for both
type of anholonomic deformatons, real or complex ones. So, we may consider
the usual Penrose's diagrams as for a particular case of the
Reissner--Nordstrem spacetime but keeping in mind that such diagrams and
horizons have an additional parametrization on an angular coordinate and
that the frames have some complex coefficients.

We can construct the maximally extended manifold step by steps like in the
Schwarzschild case (see details, for instance, in Ref. 52)) %
%\cite{haw}))%
by supposing that the complex valued coefficients of metrics and frame are
of necessary smooth class as real and complex valued functions (for
simplicity, we consider the simplest variant when the spacetime is provided
with a complex valued metric but admits covering by real coordinates which
after certain coordinate transform may be also complex). We introduce a new
coordinate
\begin{equation*}
r^{\Vert }=\int dr\left( 1-\frac{2m}{r}-\frac{q^{2}}{r^{2}}\right) ^{-1}
\end{equation*}%
for $r>r_{+}^{1}$ and find explicitly the coordinate
\begin{equation}
r^{\Vert }=r+\frac{(r_{+}^{1})^{2}}{r_{+}^{1}-r_{-}^{1}}\ln (r-r_{+}^{1})-%
\frac{(r_{-}^{1})^{2}}{r_{+}^{1}-r_{-}^{1}}\ln (r-r_{-}^{1}),  \label{r1}
\end{equation}%
where $r_{\pm }^{1}=r_{\pm }^{\Phi }$ with $\Phi =1.$ If $r$ is expressed as
a function on $\xi ,$ than $r^{\Vert }$ can be also expressed as a function
on $\xi $ depending additionally on some parameters.

Defining the advanced and retarded coordinates, $v=t+r^{\Vert }$ and $%
w=t-r^{\Vert },$ with corresponding elongated differentials
\begin{equation}
\delta v=\delta t+dr^{\Vert }\mbox{ and }\delta w=\delta t-dr^{\Vert },
\label{coord1}
\end{equation}%
which are N--adapted frames like (\ref{ndif}) but complex one, the metric (%
\ref{sch1}) takes the form%
\begin{equation*}
\delta s^{2}=-r^{2}(\xi )\gamma (\xi )d\theta ^{2}-\eta _{3}(\xi ,\varphi
_{0})r^{2}(\xi )\sin ^{2}\theta \delta \varphi ^{2}+(1-\frac{2m}{r(\xi )}%
-q^{2}\frac{\Phi _{4}(r,\varphi _{0})}{r^{2}(\xi )})\delta v\delta w,
\end{equation*}%
where (in general, in non--explicit form) $r(\xi )$ is a function of type $%
r(\xi )=r\left( r^{\Vert }\right) =$ $r\left( v,w\right) .$ Introducing new
coordinates $(v^{\prime \prime },w^{\prime \prime })$ by%
\begin{equation}
v^{\prime \prime }=\arctan \left[ \exp \left( \frac{r_{+}^{1}-r_{-}^{1}}{%
4(r_{+}^{1})^{2}}v\right) \right] ,w^{\prime \prime }=\arctan \left[ -\exp
\left( \frac{-r_{+}^{1}+r_{-}^{1}}{4(r_{+}^{1})^{2}}w\right) \right]
\label{coord2}
\end{equation}%
and multiplying the last term on the conformal factor we obtain
\begin{equation}
\delta s^{2}=-r^{2}\gamma (r)d\theta ^{2}-\eta _{4}(r,\varphi _{0})r^{2}\sin
^{2}\theta \delta \varphi ^{2}+64\frac{(r_{+}^{1})^{4}}{%
(r_{+}^{1}-r_{-}^{1})^{2}}[1-\frac{2m}{r(\xi )}-q^{2}\frac{\Phi
_{5}(r,\varphi _{0})}{r^{2}(\xi )}]\delta v^{\prime \prime }\delta w^{\prime
\prime },  \label{el2b}
\end{equation}%
where $r$ is defined implicitly by
\begin{equation}
\tan v^{\prime \prime }\tan w^{\prime \prime }=-\exp \left[ \frac{%
r_{+}^{1}-r_{-}^{1}}{2(r_{+}^{1})^{2}}r\right] \sqrt{\frac{r-r_{+}^{1}}{%
(r-r_{-}^{1})^{\chi }}},\chi =\left( \frac{r_{+}^{1}}{r_{-}^{1}}\right) ^{2}
\label{coord3}
\end{equation}%
where the functions $\tan $ and $\exp $ should be considered as the complex
functions. As particular cases, we may chose $\eta _{5}\left( r,\varphi
\right) $ as the condition of vanishing of the metric coefficient before $%
\delta v^{\prime \prime }\delta w^{\prime \prime }$ will describe a horizon
parametrized by a resolution ellipsoid hypersurface.

The metric (\ref{el2b}) is very similar to that analyzed in Refs. 18,19 %
%\cite{anhtv6,anhtv7}%
but the coordinate transforms defined by (\ref{coord1})--(\ref{coord3})
involve complex coordinate transforms, so $\delta v^{\prime \prime }\delta
w^{\prime \prime }$ is a product defined by complexified N--adapted frames.

The maximal \ extension of the Schwarzschild metric deformed by a small
parameter $\varepsilon ,$ i. e. \ the extension of the metric (\ref{sch}),
is defined by taking (\ref{el2b}) as the metric on the maximal manifold
(which for corresponding coordinate transforms can be considered as a real
one but with complex valued coefficients of the metric and moving frames) on
which this metric is of smoothly class $C^{2}.$ The Penrose diagram of this
static but locally anisotropic space--time, for any fixed angular value $%
\varphi _{0}$ is similar to the Reissner--N\"{o}rdstrom solution, for the
case $q^{2}\rightarrow \varepsilon $ and $q^{2}<m^{2}$ (see, for instance,
Ref. 52). %
% \cite{haw})).%
There are an infinite number of asymptotically flat regions of type I,
connected by intermediate regions II and III, where there is still an
irremovable singularity at $r=0$ for every region III. We may travel from a
region I to another ones by passing through the 'wormholes' made by
anisotropic deformations (ellipsoid off--diagonality of metrics, or
anholonomy) like in the Reissner--Nordstrom universe because $\sqrt{%
\varepsilon }\sim q$ may model an effective electric charge. One can not
turn back in a such travel because the complex frames ''do not allow us''.

It should be noted that the metric (\ref{el2b}) $\ $\ is analytic every were
except at $r=r_{-}^{1}$ (we have to use the term analytic as real functions
for the metric coefficients in the lower approximations on $\varepsilon $
and analytic as complex functions for the higher aproximations of the metric
coefficients and for all terms contained in the vielbein coefficients). We
may eliminate this coordinate degeneration by introducing another new
complex coordinates%
\begin{equation*}
v^{\prime \prime \prime }=\arctan \left[ \exp \left( \frac{%
r_{+}^{1}-r_{-}^{1}}{2n_{0}(r_{+}^{1})^{2}}v\right) \right] ,w^{\prime
\prime \prime }=\arctan \left[ -\exp \left( \frac{-r_{+}^{1}+r_{-}^{1}}{%
2n_{0}(r_{+}^{1})^{2}}w\right) \right] ,
\end{equation*}%
where the integer $n_{0}\geq (r_{+}^{1})^{2}/(r_{-}^{1})^{2}$ and complex
functions. In \ these coordinates, the metric is (in general, complex)
analytic every were except at $r=r_{+}^{1}$ where it is degenerate. This way
the space--time manifold can be covered by an analytic atlas by using
coordinate carts defined by $(v^{\prime \prime },w^{\prime \prime },\theta
,\varphi )$ and $(v^{\prime \prime \prime },w^{\prime \prime \prime },\theta
,\varphi ).$ We also note that the analytic extensions of the deformed
metrics were performed with respect to anholonomic complex frames which
distinguish such constructions from those dealing only with holonomic and/or
real coordinates, like for the usual Reissner--N\"{o}rdstrom and Kerr
metrics.

A more rigorous analysis of the metric (\ref{sch}) should involve a
computation of its curvature and investigation of singularity properties. We
omit here this cumbersome calculus by emphasizing that anholonomic
deformations of the Schwarzschild solution defined by a small real or
complex parameter $\varepsilon $ can not remove the bulk singularity of such
spacetimes; there are deformations of the horizon, frames and specific
polarizations of constants.

The metric (\ref{sch}) and its analytic extensions do not posses Killing
symmetries being deformed by anholonomic transforms. Nevertheless, we can
associate to such solutions certain noncommutative symmetries following the
procedure described in section \ref{ncgg}. Taking the data (\ref{data}) and
formulas (\ref{anhol}), we compute the corresponding nontrivial anholonomy
coefficients
\begin{equation}
w_{\ 42}^{[N]5}=-w_{\ 24}^{[N]5}=\partial n_{2}\left( \xi ,\varphi \right)
/\partial \varphi =n_{2}^{\ast }\left( \xi ,\varphi \right)  \label{auxxx1}
\end{equation}%
with $n_{2}$ defined by (\ref{auxf4s}). Our vacuum solution is for 4D, so
for $n+m=4,$ the condition $k^{2}-1=n+m$ can not satisfied in integer
numbers. We may trivially extend the dimensions like $n^{\prime }=6$ and $%
m^{\prime }=m=2$ and for $k=3$ to consider the \ Lie group $SL\left( 3,\C%
\right) $ noncommutativity with corresponding values of $Q_{\underline{%
\alpha }\underline{\beta }}^{\underline{\gamma }}$\ and structure constants $%
f_{~\underline{\alpha }\underline{\beta }}^{\underline{\gamma }},$ see (\ref%
{gr1}). An extension $w_{~\alpha \beta }^{[N]\gamma }\rightarrow W_{~%
\underline{\alpha }\underline{\beta }}^{\underline{\gamma }}$ may be
performed by stating the N--deformed ''structure'' constants (\ref{anhb}), $%
W_{~\underline{\alpha }\underline{\beta }}^{\underline{\gamma }}=f_{~%
\underline{\alpha }\underline{\beta }}^{\underline{\gamma }}+w_{~\underline{%
\alpha }\underline{\beta }}^{[N]\underline{\gamma }},$ with only two
nontrivial values of $w_{~\underline{\alpha }\underline{\beta }}^{[N]%
\underline{\gamma }}$ given by (\ref{auxxx1}).

The associated anholonomic noncommutative symmetries of the black ellipsoid
solutions can be alternatively defined as in the trivial anholonomy limit
they will result in a certan noncommutativity for the Lie group $SU_{3.}.$
In this case, we have to consider a N--deformaton of the group structure
constants $p_{~\underline{\alpha }\underline{\beta }}^{\underline{\gamma }},$
like in (\ref{anhcsu}), $W_{~\underline{\alpha }\underline{\beta }}^{%
\underline{\gamma }}=p_{~\underline{\alpha }\underline{\beta }}^{\underline{%
\gamma }}+w_{~\underline{\alpha }\underline{\beta }}^{\underline{\gamma }}.$
This variant of deformations can be related directly with the ''de Sitter
nonlinear gauge gravity model of (non) commutative gravity''$^{6}$ \ %
%\cite{v1}%
and the $SU_{k}[SO(k)]$--models of noncommutative gravity$^{7}$
%\cite{ch1}%
by considering complex vielbeins.

\subsection{Black ellipsoids and the cosmological constant}

\label{belsg}

We can generalize the vacuum equations to the gravity with cosmological
constant $\lambda $,%
\begin{equation}
R_{\mu ^{\prime }\nu ^{\prime }}=\lambda g_{\mu ^{\prime }\nu ^{\prime }},
\label{eq17}
\end{equation}%
where $R_{\mu ^{\prime }\nu ^{\prime }}$ is the Ricci tensor (\ref{dricci}),
in general with anholonomic variables and the indices take values $i^{\prime
},k^{\prime }=1,2$ and $a^{\prime },b^{\prime }=3,4.$

For an ansatz of type (\ref{dmetric4})
\begin{eqnarray}
\delta s^{2} &=&g_{1}(dx^{1})^{2}+g_{2}(dx^{2})^{2}+h_{3}\left( x^{i^{\prime
}},y^{3}\right) (\delta y^{3})^{2}+h_{4}\left( x^{i^{\prime }},y^{3}\right)
(\delta y^{4})^{2},  \label{ansatz18} \\
\delta y^{3} &=&dy^{3}+w_{i^{\prime }}\left( x^{k^{\prime }},y^{3}\right)
dx^{i^{\prime }},\quad \delta y^{4}=dy^{4}+n_{i^{\prime }}\left(
x^{k^{\prime }},y^{3}\right) dx^{i^{\prime }},  \notag
\end{eqnarray}%
the Einstein equations (\ref{eq17}) are written (see Refs. 30,31,13-19 \ %
%\cite{vmethod1,vmethod2,anhtv1,anhtv2,anhtv3,anhtv4,anhtv5,anhtv6,anhtv7}%
for details on computation; this is a particular case of source of type (\ref%
{emcond4}), see Appendix B)
\begin{eqnarray}
R_{1}^{1}=R_{2}^{2}=-\frac{1}{2g_{1}g_{2}}[g_{2}^{\bullet \bullet }-\frac{%
g_{1}^{\bullet }g_{2}^{\bullet }}{2g_{1}}-\frac{(g_{2}^{\bullet })^{2}}{%
2g_{2}}+g_{1}^{^{\prime \prime }}-\frac{g_{1}^{^{\prime }}g_{2}^{^{\prime }}%
}{2g_{2}}-\frac{(g_{1}^{^{\prime }})^{2}}{2g_{1}}] &=&\lambda ,
\label{ricci1s} \\
R_{3}^{3}=R_{4}^{4}=-\frac{\beta }{2h_{3}h_{4}} &=&\lambda ,  \label{ricci2s}
\\
R_{3i^{\prime }}=-w_{i^{\prime }}\frac{\beta }{2h_{4}}-\frac{\alpha
_{i^{\prime }}}{2h_{4}} &=&0,  \label{ricci3s} \\
R_{4i^{\prime }}=-\frac{h_{4}}{2h_{3}}\left[ n_{i^{\prime }}^{\ast \ast
}+\gamma n_{i^{\prime }}^{\ast }\right] &=&0.  \label{ricci4s}
\end{eqnarray}%
The coefficients of equations (\ref{ricci1s}) - (\ref{ricci4s}) are given by
\begin{equation}
\alpha _{i}=\partial _{i}{h_{4}^{\ast }}-h_{4}^{\ast }\partial _{i}\ln \sqrt{%
|h_{3}h_{4}|},\qquad \beta =h_{4}^{\ast \ast }-h_{4}^{\ast }[\ln \sqrt{%
|h_{3}h_{4}|}]^{\ast },\qquad \gamma =\frac{3h_{4}^{\ast }}{2h_{4}}-\frac{%
h_{3}^{\ast }}{h_{3}}.  \label{abcs}
\end{equation}%
The various partial derivatives are denoted as $a^{\bullet }=\partial
a/\partial x^{1},a^{^{\prime }}=\partial a/\partial x^{2},a^{\ast }=\partial
a/\partial y^{3}.$ This system of equations can be solved by choosing one of
the ansatz functions (\textit{e.g.} $g_{1}\left( x^{i}\right) $ or $%
g_{2}\left( x^{i}\right) )$ and one of the ansatz functions (\textit{e.g.} $%
h_{3}\left( x^{i},y^{3}\right) $ or $h_{4}\left( x^{i},y^{3}\right) )$ to
take some arbitrary, but physically interesting form (see Theorem 3 in
Appendix B). Then, the other ansatz functions can be analytically determined
up to an integration in terms of this choice. In this way we can generate a
lost of different solutions, but we impose the condition that the initial,
arbitrary choice of the ansatz functions is ``physically interesting'' which
means that one wants to make this original choice so that the generated
final solution yield a well behaved metric.

In Ref. 19 %
%\cite{anhtv7} %
(see also previous subsection), we proved that for
\begin{eqnarray}
g_{1} &=&-1,\quad g_{2}=r^{2}\left( \xi \right) q\left( \xi \right) ,
\label{data10} \\
h_{3} &=&-\eta _{3}\left( \xi ,\varphi \right) r^{2}\left( \xi \right) \sin
^{2}\theta ,\quad  \notag \\
h_{4} &=&\eta _{4}\left( \xi ,\varphi \right) h_{4[0]}\left( \xi \right) =1-%
\frac{2\mu }{r}+\varepsilon \frac{\Phi _{4}\left( \xi ,\varphi \right) }{%
2\mu ^{2}},  \notag
\end{eqnarray}%
with coordinates $x^{1}=\xi =\int dr\sqrt{1-2m/r+\varepsilon /r^{2}}%
,x^{2}=\theta ,y^{3}=\varphi ,y^{4}=t$ (the $(r,\theta ,\varphi )$ being
usual radial coordinates), the ansatz (\ref{ansatz18}) is a vacuum solution
with $\lambda =0$ of the equations (\ref{eq17}) which defines a black
ellipsoid with mass $\mu ,$ eccentricity $\varepsilon $ and gravitational
polarizations $q\left( \xi \right) ,\eta _{3}\left( \xi ,\varphi \right) $
and $\Phi _{4}\left( \xi ,\varphi \right) .$ Such black holes are certain
deformations of the Schawarzschild metrics to static configurations with
ellipsoidal horizons which is possible if generic off--diagonal metrics and
anholonomic frames are considered. A complex generalization of this solution
is given by the values (\ref{data}). In this subsection we show that the
data (\ref{data10}) and/or (\ref{data}) can be extended as to generate exact
black ellipsoid solutions, defied correspondingly with respect to real or
complex N--frames, with nontrivial cosmological constant $\lambda =1/4$
which can be imbedded in string theory.

At the first \ step, we find a class of solutions with $g_{1}=-1$ and $\quad
g_{2}=g_{2}\left( \xi \right) $ solving the equation (\ref{ricci1s}), which
under such parametrizations transforms to
\begin{equation*}
g_{2}^{\bullet \bullet }-\frac{(g_{2}^{\bullet })^{2}}{2g_{2}}=2g_{2}\lambda
.
\end{equation*}%
With respect to the variable $Z=(g_{2})^{2}$ this equation is written as
\begin{equation*}
Z^{\bullet \bullet }+2\lambda Z=0
\end{equation*}%
which can be integrated in explicit form, $Z=Z_{[0]}\sin \left( \sqrt{%
2\lambda }\xi +\xi _{\lbrack 0]}\right) ,$ for some constants $Z_{[0]}$ and $%
\xi _{\lbrack 0]}$ which means that
\begin{equation}
g_{2}=-Z_{[0]}^{2}\sin ^{2}\left( \sqrt{2\lambda }\xi +\xi _{\lbrack
0]}\right)  \label{aux2p}
\end{equation}%
parametrize in 'real' string gravity a class of solution of (\ref{ricci1s})
for the signature $\left( -,-,-,+\right) .$ For $\lambda \rightarrow 0$ we
can approximate $g_{2}=r^{2}\left( \xi \right) q\left( \xi \right) =-\xi
^{2} $ and $Z_{[0]}^{2}=1$ which has compatibility with the data (\ref%
{data10}). The solution (\ref{aux2p}) with cosmological constant (of string
or non--string origin) induces oscillations in the ''horozontal'' part of
the metric written with respect to N--adapted frames.

The next step is to solve the equation (\ref{ricci2s}),%
\begin{equation*}
h_{4}^{\ast \ast }-h_{4}^{\ast }[\ln \sqrt{|h_{3}h_{4}|}]^{\ast }=-2\lambda
h_{3}h_{4}.
\end{equation*}%
For $\lambda =0$ a class of solution is given by any $\widehat{h}_{3}$ and $%
\widehat{h}_{4}$ related as
\begin{equation*}
\widehat{h}_{3}=\eta _{0}\left[ \left( \sqrt{|\hat{h}_{4}|}\right) ^{\ast }%
\right] ^{2}
\end{equation*}%
for a constant $\eta _{0}$ chosen to be negative in order to generate the
signature $\left( -,-,-,+\right) .$ For non--trivial $\lambda ,$ we may
search the solution as
\begin{equation}
h_{3}=\widehat{h}_{3}\left( \xi ,\varphi \right) ~f_{3}\left( \xi ,\varphi
\right) \mbox{ and }h_{4}=\widehat{h}_{4}\left( \xi ,\varphi \right) ,
\label{sol15}
\end{equation}%
which solves (\ref{ricci2s}) if $f_{3}=1$ for $\lambda =0$ and
\begin{equation*}
f_{3}=\frac{1}{4\lambda }\left[ \int \frac{\hat{h}_{3}\hat{h}_{4}}{\hat{h}%
_{4}^{\ast }}d\varphi \right] ^{-1}\mbox{ for }\lambda \neq 0.
\end{equation*}

Now it is easy to write down the solutions of equations (\ref{ricci3s})
(being a linear equation for $w_{i^{\prime }})$ and (\ref{ricci4s}) (after
two integrations of $n_{i^{\prime }}$ on $\varphi ),$%
\begin{equation}
w_{i^{\prime }}=\varepsilon \widehat{w}_{i^{\prime }}=-\alpha _{i^{\prime
}}/\beta ,  \label{aux3s}
\end{equation}%
were $\alpha _{i^{\prime }}$ and $\beta $ are computed by putting (\ref%
{sol15}) $\ $into corresponding values from (\ref{abcs}) (we chose the
initial conditions as $w_{i^{\prime }}\rightarrow 0$ for $\varepsilon
\rightarrow 0)$ and
\begin{equation*}
n_{1}=\varepsilon \widehat{n}_{1}\left( \xi ,\varphi \right)
\end{equation*}%
where the coefficients
\begin{eqnarray}
\widehat{n}_{1}\left( \xi ,\varphi \right) &=&n_{1[1]}\left( \xi \right)
+n_{1[2]}\left( \xi \right) \int d\varphi \ \eta _{3}\left( \xi ,\varphi
\right) /\left( \sqrt{|\eta _{4}\left( \xi ,\varphi \right) |}\right)
^{3},\eta _{4}^{\ast }\neq 0;  \label{aux4s} \\
&=&n_{1[1]}\left( \xi \right) +n_{1[2]}\left( \xi \right) \int d\varphi \
\eta _{3}\left( \xi ,\varphi \right) ,\eta _{4}^{\ast }=0;  \notag \\
&=&n_{1[1]}\left( \xi \right) +n_{1[2]}\left( \xi \right) \int d\varphi
/\left( \sqrt{|\eta _{4}\left( \xi ,\varphi \right) |}\right) ^{3},\eta
_{3}^{\ast }=0;  \notag
\end{eqnarray}%
being stated to be real or complex valued for a corresponding model of real
or complex gravity, with the functions $n_{k[1,2]}\left( \xi \right) $ to be
stated by boundary conditions.

We conclude that the set of data $g_{1}=-1,$ with non--trivial $g_{2}\left(
\xi \right) ,h_{3},h_{4},w_{i^{\prime }}$ and $n_{1}$ stated respectively by
(\ref{aux2p}), (\ref{sol15}), (\ref{aux3s}), (\ref{aux4s}) we can define a
black ellipsoid solution with explicit dependence on cosmological constant $%
\lambda ,$ i. e. a metric (\ref{ansatz18}). The stability of such static
black ellipsoids in (anti) De Sitter space can be proven exactly as it was
done in Ref. 19 %\cite{anhtv7}%
for the real case vanishing cosmological constant.

The analytic extension of black ellipsoid solutions with cosmological
constant can be performed similarly as in the previous subsection when for
the real/ complex solutions we are dealing with real/ complex values of $%
\widehat{n}_{1}\left( \xi ,\varphi \right) $ defining some components of
N--adapted frames. We note that the solution from string theory contains a
frame induced torsion with the components (\ref{dtors}) (in general, with
complex coefficients) computed for nontrivial $N_{i^{\prime }}^{3}=-\alpha
_{i^{\prime }}/\beta $ (see (\ref{aux3s})) and $N_{1}^{4}=\varepsilon
\widehat{n}_{1}\left( \xi ,\varphi \right) $ (see (\ref{aux4s})). This is an
explicit example illustrating that the anholonomic frame method is also
powerfull for generating exact solutions in models of gravity with
nontrivial torsion, induced by anholonomic frame transforms. For such
solutions we may elaborate corresponding analytic extension and Penrose
diagram formalisms if the constructions are considered with respect to
N--elongated vielbeins.

Finally, we analyze the structure of noncommutative symmetries associated to
the (anti) de Sitter black ellipsoid solutions. The metric (\ref{ansatz18})
with real and/or complex coefficients defining the corresponding solutions
and its analytic extensions also do not posses Killing symmetries being
deformed by anholonomic transforms. For this solution, we can associate
certain noncommutative symmetries following the same procedure as for the
Einstein real/ complex gravity but with additional nontrivial coefficients
of anholonomy and even with nonvanishing coefficients of the nonlinear
connection curvature, $\Omega _{12}^{3}=\delta _{1}N_{2}^{3}-\delta
_{2}N_{1}^{3}.$ Taking the data (\ref{aux3s}) and (\ref{aux4s}) and formulas
(\ref{anhol}), we compute the corresponding nontrivial anholonomy
coefficients
\begin{eqnarray}
w_{\ 31}^{[N]4} &=&-w_{\ 13}^{[N]4}=\partial n_{1}\left( \xi ,\varphi
\right) /\partial \varphi =n_{2}^{\ast }\left( \xi ,\varphi \right) ,
\label{auxxx2} \\
w_{\ 12}^{[N]4} &=&-w_{\ 21}^{[N]4}=\delta _{1}(\alpha _{2}/\beta )-\delta
_{2}(\alpha _{1}/\beta )  \notag
\end{eqnarray}%
for $\delta _{1}=\partial /\partial \xi -w_{1}\partial /\partial \varphi $
and $\delta _{2}=\partial /\partial \theta -w_{2}\partial /\partial \varphi
, $ with $n_{1}$ defined by (\ref{auxf4s}) and $\alpha _{1,2}$ and $\beta $
computed by using the formula (\ref{abcs}) for the solutions (\ref{sol15}).
Our exact solution, with nontrivial cosmological constant, is for 4D, like
in the previous subsection. So, for $n+m=4,$ the condition $k^{2}-1=n+m$ can
not satisfied by any integer numbers. We may similarly trivially extend the
dimensions like $n^{\prime }=6$ and $m^{\prime }=m=2$ and for $k=3$ to
consider the \ Lie group $SL\left( 3,\C\right) $ noncommutativity with
corresponding values of $Q_{\underline{\alpha }\underline{\beta }}^{%
\underline{\gamma }}$\ and structure constants $f_{~\underline{\alpha }%
\underline{\beta }}^{\underline{\gamma }},$ see (\ref{gr1}). An extension $%
w_{~\alpha \beta }^{[N]\gamma }\rightarrow W_{~\underline{\alpha }\underline{%
\beta }}^{\underline{\gamma }}$ may be performed by stating the N--deformed
''structure'' constants (\ref{anhb}), $W_{~\underline{\alpha }\underline{%
\beta }}^{\underline{\gamma }}=f_{~\underline{\alpha }\underline{\beta }}^{%
\underline{\gamma }}+w_{~\underline{\alpha }\underline{\beta }}^{[N]%
\underline{\gamma }},$ with nontrivial values of $w_{~\underline{\alpha }%
\underline{\beta }}^{[N]\underline{\gamma }}$ given by (\ref{auxxx2}). In a
similar form, we can consider anholonomic deformations of the $SU_{k}$
structure constants, see (\ref{anhcsu}).

\section{\quad Noncommutative Complex Wormholes}

The black ellipsoid solutions defined by real and certain complex metrics
elaborated in the previous section were for the 4D Einstein gravity, in
general, with nontrivial cosmological constant. In this section we construct
and analyze an exact 5D solution which can be also complexified by using
complex anholonomic transforms as well they can be provided with associated
noncommutative structure. \ For such configurations we can apply directly
the formulas stated in Appendix B. The metric ansatz (\ref{dmetric}) is
taken in the form%
\begin{eqnarray}
\delta s^{2} &=&g_{1}(dx^{1})^{2}+g_{2}(dx^{2})^{2}+g_{3}(dx^{3})^{2}+h_{4}({%
\delta }y^{4})^{2}+h_{5}(\delta y^{5})^{2},  \notag \\
{\delta }y^{4} &=&{d}y^{4}+w_{k^{\prime }}\left( x^{i^{\prime }},v\right)
dx^{k^{\prime }},{\delta }y^{5}={d}y^{5}+n_{k^{\prime }}\left( x^{i^{\prime
}},v\right) dx^{k^{\prime }};\ i^{\prime },k^{\prime }=1,2,3,  \label{ans20}
\end{eqnarray}%
where
\begin{eqnarray}
g_{1} &=&1,\quad g_{2}=g_{2}(r),\quad g_{3}=-a(r),  \label{anz6a} \\
h_{4} &=&\hat{h}_{4}=\widehat{\eta }_{4}\left( r,\theta ,\varphi \right)
h_{4[0]}(r),\quad h_{5}=\hat{h}_{5}=\widehat{\eta }_{5}\left( r,\theta
,\varphi \right) h_{5[0]}(r,\theta )  \notag
\end{eqnarray}%
for the parametrization of coordinate of type
\begin{equation}
x^{1}=t,x^{2}=r,x^{3}=\theta ,y^{4}=v=\varphi ,y^{5}=p=\chi  \label{coord5}
\end{equation}%
where $t$ is the time coordinate, $\left( r,\theta ,\varphi \right) $ are
spherical coordinates, $\chi $ is the 5th coordinate; $\varphi $ is the
anholonomic coordinate; for this ansatz there is not considered the
dependence of metric coefficients on the second anholonomic coordinate $\chi
.$ Following similar approximations as in subsection \ref{belsg} for
deriving the equations (\ref{eq17}), we can write the gravity equations with
cosmological constant as a system of 5D Einstein equations with constant
diagonal source (the related details on computing the Ricci tensors with
anholonomic variables and possible sources are given in Appendix B):
\begin{eqnarray}
\frac{1}{2}R_{1}^{1}=R_{2}^{2}=R_{3}^{3}=-\frac{1}{2g_{2}g_{3}}%
[g_{3}^{\bullet \bullet }-\frac{g_{2}^{\bullet }g_{3}^{\bullet }}{2g_{2}}-%
\frac{(g_{3}^{\bullet })^{2}}{2g_{3}}+g_{2}^{^{\prime \prime }}-\frac{%
g_{2}^{^{\prime }}g_{3}^{^{\prime }}}{2g_{3}}-\frac{(g_{2}^{^{\prime }})^{2}%
}{2g_{2}}] &=&\lambda ,  \label{ricci7a} \\
R_{4}^{4}=R_{5}^{5}=-\frac{\beta }{2h_{4}h_{5}} &=&\lambda ,  \label{ricci8a}
\\
R_{4i^{\prime }}=-w_{i^{\prime }}\frac{\beta }{2h_{5}}-\frac{\alpha
_{i^{\prime }}}{2h_{5}} &=&0,  \label{ricci9a} \\
R_{5i^{\prime }}=-\frac{h_{5}}{2h_{4}}\left[ n_{i^{\prime }}^{\ast \ast
}+\gamma n_{i^{\prime }}^{\ast }\right] &=&0,  \label{ricci10a}
\end{eqnarray}%
where $i^{\prime }=1,2,3.$ The coefficients of the equations are given by
\begin{equation}
\alpha _{i^{\prime }}=\partial _{i}{h_{5}^{\ast }}-h_{5}^{\ast }\partial
_{i^{\prime }}\ln \sqrt{|h_{4}h_{5}|},\qquad \beta =h_{5}^{\ast \ast
}-h_{5}^{\ast }[\ln \sqrt{|h_{4}h_{5}|}]^{\ast },\qquad \gamma =\frac{%
3h_{5}^{\ast }}{2h_{5}}-\frac{h_{4}^{\ast }}{h_{4}}.  \label{abc1}
\end{equation}%
The various partial derivatives are denoted as $a^{\bullet }=\partial
a/\partial x^{2},a^{^{\prime }}=\partial a/\partial x^{3},a^{\ast }=\partial
a/\partial v.$

The system of equations (\ref{ricci7a})--(\ref{ricci10a}) can be solved by
choosing one of the ansatz functions (\textit{e.g.} $h_{4}\left(
x^{i^{\prime }},v\right) $ or $h_{5}\left( x^{i^{\prime }},v\right) )$ to
take some arbitrary, but physically interesting form. Then the other ansatz
functions can be analytically determined up to an integration in terms of
this choice. In this way one can generate many solutions, but the
requirement that the initial, arbitrary choice of the ansatz functions be
``physically interesting'' means that one wants to make this original choice
so that the final solution generated in this way yield a well behaved
solution. To satisfy this requirement we start from well known solutions of
Einstein's equations and then use the above procedure to deform this
solutions in a number of ways as to include it in a string theory, for
instance, as a gravity model with cosmological constant.

The data%
\begin{eqnarray}
g_{1} &=&1,~\hat{g}_{2}=-1,~g_{3}=-a(r),  \label{data6a} \\
~h_{4[0]}(r) &=&-r_{0}^{2}e^{2\psi (r)},~\eta _{4}=1/\kappa _{r}^{2}\left(
r,\theta ,\varphi \right) ,~h_{5[0]}=-a\left( r\right) \sin ^{2}\theta
,~\eta _{5}=1,  \notag \\
w_{1} &=&\widehat{w}_{1}=\omega \left( r\right) ,~w_{2}=\widehat{w}%
_{2}=0,w_{3}=~\widehat{w}_{3}=n\cos \theta /\kappa _{n}^{2}\left( r,\theta
,\varphi \right) ,  \notag \\
n_{1} &=&\widehat{n}_{1}=0,~n_{2,3}=\widehat{n}_{2,3}=n_{2,3[1]}\left(
r,\theta \right) \int \ln |\kappa _{r}^{2}\left( r,\theta ,\varphi \right)
|d\varphi  \notag
\end{eqnarray}%
for some constants $r_{0}$ $\ $\ and $n$ and arbitrary functions $a(r),\psi
(r)$ and arbitrary vacuum gravitational polarizations $\kappa _{r}\left(
r,\theta ,\varphi \right) $ and $\kappa _{n}\left( r,\theta ,\varphi \right)
$ define an exact vacuum 5D solution of Kaluza--Klein gravity$^{14}$ \
%\cite{anhtv2}%
describing a locally anisotropic wormhole with elliptic gravitational vacuum
polarization of charges,
\begin{equation*}
\frac{q_{0}^{2}}{4a\left( 0\right) \kappa _{r}^{2}}+\frac{Q_{0}^{2}}{%
4a\left( 0\right) \kappa _{n}^{2}}=1,
\end{equation*}%
where $q_{0}=2\sqrt{a\left( 0\right) }\sin \alpha _{0}$ and $Q_{0}=2\sqrt{%
a\left( 0\right) }\cos \alpha _{0}$ are respectively the electric and
magnetic charges and $2\sqrt{a\left( 0\right) }\kappa _{r}$ and $2\sqrt{%
a\left( 0\right) }\kappa _{n}$ are ellipse's axes.

The first aim in this section is to prove that following the ansatz (\ref%
{ans20}) we can construct locally anisotropic wormhole metrics in (anti) de
Sitter gravity, in general complexified by a certain class of anhlonomic
frame transforms as solutions of the system of equations (\ref{ricci7a}) -- (%
\ref{ricci10a}) with redefined coordinates as in (\ref{coord5}). For
simplicity, we select such solutions when only the coefficients $n_{i}$ can
be real or complex valued functions. Having the vacuum data (\ref{data6a}),
we may try to generalize the solution for a nontrivial cosmological constant
by supposing that the new solutions may be represented as%
\begin{equation}
h_{4}=\widehat{h}_{4}\left( x^{i^{\prime }},v\right) ~q_{4}\left(
x^{i^{\prime }},v\right) \mbox{ and }h_{5}=\widehat{h}_{5}\left(
x^{i^{\prime }},v\right) ,  \label{shift3}
\end{equation}%
with $\widehat{h}_{4,5}$ taken as in (\ref{anz6a}) which solves (\ref%
{ricci8a}) if $q_{4}=1$ for $\lambda =0$ and
\begin{equation*}
q_{4}=\frac{1}{4\lambda }\left[ \int \frac{\hat{h}_{5}\left( r,\theta
,\varphi \right) \hat{h}_{4}\left( r,\theta ,\varphi \right) }{\hat{h}%
_{5}^{\ast }\left( r,\theta ,\varphi \right) }d\varphi \right] ^{-1}%
\mbox{
for }\lambda \neq 0.
\end{equation*}%
This $q_{4}$ can be considered as an additional polarization to $\eta _{4}$
induced by the cosmological constant $\lambda .$ We state $g_{2}=-1$ but
\begin{equation*}
g_{3}=-\sin ^{2}\left( \sqrt{2\lambda }\theta +\xi _{\lbrack 0]}\right) ,
\end{equation*}%
defining a solution of (\ref{ricci7a}) with signature $\left(
+,-,-,-,-\right) $ being different from the solution (\ref{aux2}). A
non--trivial $q_{4}$ results in modification of coefficients (\ref{abc1}),
\begin{eqnarray*}
\alpha _{i^{\prime }} &=&\hat{\alpha}_{i^{\prime }}+\alpha _{i^{\prime
}}^{[q]},~\beta =\hat{\beta}+\beta ^{\lbrack q]},~\gamma =\hat{\gamma}%
+\gamma ^{\lbrack q]}, \\
\hat{\alpha}_{i^{\prime }} &=&\partial _{i}{\hat{h}_{5}^{\ast }}-\hat{h}%
_{5}^{\ast }\partial _{i^{\prime }}\ln \sqrt{|\hat{h}_{4}\hat{h}_{5}|}%
,\qquad \hat{\beta}=\hat{h}_{5}^{\ast \ast }-\hat{h}_{5}^{\ast }[\ln \sqrt{|%
\hat{h}_{4}\hat{h}_{5}|}]^{\ast },\qquad \hat{\gamma}=\frac{3\hat{h}%
_{5}^{\ast }}{2\hat{h}_{5}}-\frac{\hat{h}_{4}^{\ast }}{\hat{h}_{4}} \\
\alpha _{i^{\prime }}^{[q]} &=&-h_{5}^{\ast }\partial _{i^{\prime }}\ln
\sqrt{|q_{4}|},\qquad \beta ^{\lbrack q]}=-h_{5}^{\ast }[\ln \sqrt{|q_{4}|}%
]^{\ast },\qquad \gamma ^{\lbrack q]}=-\frac{q_{4}^{\ast }}{q_{4}},
\end{eqnarray*}%
which following formulas (\ref{ricci9a}) and (\ref{ricci10a}) result in
additional terms to the N--connection coefficients, i. e.
\begin{equation}
w_{i^{\prime }}=\widehat{w}_{i^{\prime }}+w_{i^{\prime }}^{[q]}~\mbox{ and }%
n_{i^{\prime }}=\widehat{n}_{i^{\prime }}+n_{i^{\prime }}^{[q]},
\label{ncon05}
\end{equation}%
with $w_{i^{\prime }}^{[q]}$ and $n_{i^{\prime }}^{[q]}$ computed by using
respectively $\alpha _{i^{\prime }}^{[q]},\beta ^{\lbrack q]}$ and $\gamma
^{\lbrack q]}.$

The simplest way to generate complex solutions is to consider that $\widehat{%
n}_{i^{\prime }}$ from the data (\ref{data6a}) and (\ref{ncon05}) can be
complex valued functions, for instance, with complex valued coefficients $%
n_{2,3[1]}\left( r,\theta \right) $ resulting from integration. In this case
the metric (\ref{ans20}) has real coefficients describing wormhole solutions
with polarized constants but such metric coefficients are defined with
respect to anholonomic frames being N--elongated by some real and complex
functions.

Having nontrivial values of (\ref{ncon05}), we can associate certain
noncommutative symmetries following the same procedure as for real/ complex
black ellipsoids. The wormhole cases are described by a more general set of
nontrivial coefficients of anholonomy $w_{~\underline{\alpha }\underline{%
\beta }}^{[N]\underline{\gamma }},$ computed by using formulas (\ref{anhol})
and (\ref{anh1}) (for simplicity, we omit such cumbersome expressions), and
a nontrivial nonlinear connection curvature, $\ $in our case $\Omega
_{i^{\prime }k^{\prime }}^{a}=\delta _{i^{\prime }}N_{k^{\prime
}}^{a}-\delta _{k^{\prime }}N_{i^{\prime }}^{a}$ with $N_{k^{\prime
}}^{4}=w_{k^{\prime }}$ and $N_{k^{\prime }}^{5}=n_{k^{\prime }}.$ Such
coefficients depend on variables $\left( r,\theta ,\varphi \right) ,$ in
general, being complex valued functions. We have to extend trivially the
dimensions. We have to extend the dimensions like $n=5\rightarrow n^{\prime
}=6$ and $m^{\prime }=m=2$ and for $k=3$ if we wont to associate a \ Lie
group $SL\left( 3,\C\right) $ like noncommutativity with the corresponding
values of $Q_{\underline{\alpha }\underline{\beta }}^{\underline{\gamma }}$\
and structure constants $f_{~\underline{\alpha }\underline{\beta }}^{%
\underline{\gamma }},$ see (\ref{gr1}). An extension $w_{~\alpha \beta
}^{[N]\gamma }\rightarrow W_{~\underline{\alpha }\underline{\beta }}^{%
\underline{\gamma }}$ may be similarly performed by introducing N--deformed
''structure'' constants (\ref{anhb}), $W_{~\underline{\alpha }\underline{%
\beta }}^{\underline{\gamma }}=f_{~\underline{\alpha }\underline{\beta }}^{%
\underline{\gamma }}+w_{~\underline{\alpha }\underline{\beta }}^{[N]%
\underline{\gamma }},$ with nontrivial values of $w_{~\underline{\alpha }%
\underline{\beta }}^{[N]\underline{\gamma }}$ defined by (\ref{ncon05}).

\section{\quad Noncommutative Symmetries and Gauge Gravity}

We start with a discussing of the results of Refs. 6,7\ %\cite{v1,ch1}%
concerning noncommutative gauge models of gravity:

The basic idea of the Ref. 6\ %
%\cite{v1}%
was to use a geometrical result$^{21,22}$ \ %\cite{pd1,pd2}%
that the Einstein gravity can be equivalently represented as a gauge theory
with a specific connection in the bundle of affine frames. Such gauge
theories are with nonsemisimple structure gauge groups, i. e. with
degenerated metrics in the total spaces. Using an auxiliary symmetric form
for the typical fiber, any such model can be transformed into a variational
one. There is an alternative way to construct geometrically a usual
Yang--Mills theory by applying a corresponding set of absolute derivations
and dualities defined by the Hodge operator. \ For both such approaches,
there is a projection formalism reducing the geometric field equations on
the base space to be exactly the Einstein equations from the general
relativity theory.

For more general purposes, it was suggested to consider also extensions to a
nonlinear realization with the (anti) de Sitter gauge structural group$%
^{23}. $ \ %
% \cite{ts}.%
The constructions with nonlinear group realizations are very important
because they prescribe a consistent approach of distinguishing the frame
indices and coordinate indices subjected to different rules of
transformation. This approach to gauge gravity (of course, after a
corresponding generalizations of the Seiberg--Witten map) may include, in
general, quadratic on curvature and torsion terms (as it is stated in Ref.
6)\ %
%\cite{v1})%
being correlated to the results on gravity on noncommutative D-branes$^{53}.$%
\ %\cite{ard}.%
At the first step, it was very important to suggest an idea how to include
the general relativity into a gauge model being more explicitly developed in
noncommutative form$^{1,2,45}$\ %
%\cite{con1,con2,sw}%
(see recent developments in Refs. 54-60).%
%\cite{jsswa,msswa,sak,aasv,nr,gcors,card}).%

\subsection{Nonlinear gauge models for the (anti) de Sitter group}

There were elaborated some alternative approaches to the noncommutative
gauge gravity models in Refs. 7\ %
%\cite{ch1}%
(by deforming the Einstein gravity based on gauging the commutative
inhomogeneous Lorentz group $ISO\left( 3,1\right) $ using the
Seiberg--Witten map) and 60\ %
% \cite{card} %
(by considering some simplest noncommutative deformations of the gauge
theory $U(2,2)$ and of the Lorentz algebra $SO\left( 1,3\right) ).$ Such
theories reduce to the general relativity if certain constraints and
brocking symmetries are imposed. Perhaps, only some experimental data would
emphasize a priority of a theory of noncommutative gravity with a proper
prescription how the vielbeins and connection from 'commutative' gravity
have to be combined into components of a linear/nonlinear realizations of a
\ noncommutative gauge potentials defined by corresponding Seiberg--Witten
maps. At the present state of elaboration of noncommutative geometry and
physics, we have to analyze the physical consequences of different classes
of models of noncommutative gravity.

We introduce vielbein decompositions of (in general) complex metrics (\ref%
{offdig})
\begin{eqnarray*}
\widehat{g}_{\alpha \beta }(u) &=&e_{\alpha }^{\ \alpha ^{\prime }}\left(
u\right) e_{\beta }^{\ \beta ^{\prime }}\left( u\right) \eta _{\alpha
^{\prime }\beta ^{\prime }}, \\
e_{\alpha }^{\ \alpha ^{\prime }}e_{\ \alpha ^{\prime }}^{\beta } &=&\delta
_{\alpha }^{\beta }\mbox{ and }e_{\alpha }^{\ \alpha ^{\prime }}e_{\ \beta
^{\prime }}^{\alpha }=\delta _{\beta ^{\prime }}^{\alpha ^{\prime }}
\end{eqnarray*}%
where $\eta _{\alpha ^{\prime }\beta ^{\prime }}$ is a constant diagonal
matrix (for real spacetimes we can consider it as the flat Minkowski metric,
for instance, $\eta _{\alpha ^{\prime }\beta ^{\prime }}=diag\left(
-1,+1,...,+1\right) $) and $\delta _{\alpha }^{\beta }$ and $\delta _{\beta
^{\prime }}^{\alpha ^{\prime }}$ are Kronecker's delta symbols. The
vielbiens with an associated N--connection structure \ $N_{i}^{a}\left(
x^{j},y^{a}\right) ,$ being real or complex valued functions, have a special
parametrization%
\begin{equation}
e_{\alpha }^{\ \alpha ^{\prime }}(u)=\left[
\begin{array}{cc}
e_{i}^{\ i^{\prime }}\left( x^{j}\right) & N_{i}^{c}\left(
x^{j},y^{a}\right) \ e_{c}^{\ b^{\prime }}\left( x^{j},y^{a}\right) \\
0 & e_{e}^{\ e^{\prime }}\left( x^{j},y^{a}\right)%
\end{array}%
\right]  \label{viel1}
\end{equation}%
and
\begin{equation}
e_{\ \alpha ^{\prime }}^{\alpha \ }(u)=\left[
\begin{array}{cc}
e_{\ i^{\prime }}^{i}\left( x^{j}\right) & -N_{i}^{c}\left(
x^{j},y^{a}\right) \ e_{\ i^{\prime }}^{i\ }\left( x^{j}\right) \\
0 & e_{\ c^{\prime }}^{c}\left( x^{j},y^{a}\right)%
\end{array}%
\right]  \label{viel2}
\end{equation}%
with $e_{i}^{\ i^{\prime }}\left( x^{j}\right) $ and $e_{c}^{\ b^{\prime
}}\left( x^{j},y^{a}\right) $ generating the coefficients of metric (\ref%
{dmetric}) with the coefficients defined with respect to anholonmic frames,
\begin{equation}
g_{ij}\left( x^{j}\right) =e_{i}^{\ i^{\prime }}\left( x^{j}\right) e_{j}^{\
j^{\prime }}\left( x^{j}\right) \eta _{i^{\prime }j^{\prime }}\mbox{ and }%
h_{ab}\left( x^{j},y^{c}\right) =e_{a}^{\ a^{\prime }}\left(
x^{j},y^{c}\right) e_{b}^{\ b^{\prime }}\left( x^{j},y^{c}\right) \eta
_{a^{\prime }b^{\prime }}.  \label{metr01}
\end{equation}%
By using vielbeins and metrics of type (\ref{viel1}) and (\ref{viel2}) and,
respectively, (\ref{metr01}), we can model in a unified manner various types
of (pseudo) Riemannian, Einstein--Cartan, Riemann--Finsler and vector/
covector bundle nonlinear connection commutative and noncommutative
geometries in effective gauge and string theories (it depends on the
parametrization of $e_{i}^{\ i^{\prime }},e_{c}^{\ b^{\prime }}$ and $%
N_{i}^{c}$ on coordinates and anholonomy relations, see details in Refs.
24-28). %
%\cite{v41,v42,v43,v44,v45}).%

We consider the de Sitter space $\Sigma ^{4}$ as a hypersurface defined by
the equations $\eta _{AB}u^{A}u^{B}=-l^{2}$ in the four dimensional flat
space enabled with diagonal metric $\eta _{AB},\eta _{AA}=\pm 1$ (in this
section $A,B,C,...=1,2,...,5),$ where $\{u^{A}\}$ are global Cartesian
coordinates in $\R^{5};l>0$ is the curvature of de Sitter space (for
simplicity, we consider here only the de Sitter case; the anti--de Sitter
configuration is to be stated by a hypersurface $\eta
_{AB}u^{A}u^{B}=l^{2}). $ The de Sitter group $S_{\left( \eta \right)
}=SO_{\left( \eta \right) }\left( 5\right) $ is the isometry group of $%
\Sigma ^{5}$--space with $6$ generators of Lie algebra ${\mathit{s}o}%
_{\left( \eta \right) }\left( 5\right) $ satisfying the commutation
relations
\begin{eqnarray}
\left[ M_{AB},M_{CD}\right] &=&\eta _{AC}M_{BD}-\eta _{BC}M_{AD}-\eta
_{AD}M_{BC}+\eta _{BD}M_{AC}.  \label{dsc} \\
&&  \notag
\end{eqnarray}

We can decompose the capital indices $A,B,...$ as $A=\left( \alpha ^{\prime
},5\right) ,B=\left( \beta ^{\prime },5\right) ,...,$ and the metric $\eta
_{AB}$ as $\eta _{AB}=\left( \eta _{\alpha ^{\prime }\beta ^{\prime }},\eta
_{55}\right) .$ The operators (\ref{dsc}) $M_{AB}$ can be decomposed as $%
M_{\alpha ^{\prime }\beta ^{\prime }}=\mathcal{F}_{\alpha ^{\prime }\beta
^{\prime }}$ and $P_{\alpha ^{\prime }}=l^{-1}M_{5\alpha ^{\prime }}$
written as
\begin{eqnarray}
\left[ \mathcal{F}_{\alpha ^{\prime }\beta ^{\prime }},\mathcal{F}_{\gamma
^{\prime }\delta ^{\prime }}\right] &=&\eta _{\alpha ^{\prime }\gamma
^{\prime }}\mathcal{F}_{\beta ^{\prime }\delta ^{\prime }}-\eta _{\beta
^{\prime }\gamma ^{\prime }}\mathcal{F}_{\alpha ^{\prime }\delta ^{\prime
}}+\eta _{\beta ^{\prime }\delta ^{\prime }}\mathcal{F}_{\alpha ^{\prime
}\gamma ^{\prime }}-\eta _{\alpha ^{\prime }\delta ^{\prime }}\mathcal{F}%
_{\beta ^{\prime }\gamma ^{\prime }},  \notag \\
\left[ P_{\alpha ^{\prime }},P_{\beta ^{\prime }}\right] &=&-l^{-2}\mathcal{F%
}_{\alpha ^{\prime }\beta ^{\prime }},  \label{dsca} \\
\left[ P_{\alpha ^{\prime }},\mathcal{F}_{\beta ^{\prime }\gamma ^{\prime }}%
\right] &=&\eta _{\alpha ^{\prime }\beta ^{\prime }}P_{\underline{\gamma }%
}-\eta _{\alpha ^{\prime }\gamma ^{\prime }}P_{\beta ^{\prime }},  \notag
\end{eqnarray}%
where the Lie algebra ${\mathit{s}o}_{\left( \eta \right) }\left( 5\right) $
is split into a direct sum, ${\mathit{s}o}_{\left( \eta \right) }\left(
5\right) ={\mathit{s}o}_{\left( \eta \right) }(4)\oplus V_{4}$ with $V_{4}$
being the vector space stretched on vectors $P_{\underline{\alpha }}.$ We
remark that $\Sigma ^{4}=S_{\left( \eta \right) }/L_{\left( \eta \right) },$
where $L_{\left( \eta \right) }=SO_{\left( \eta \right) }\left( 4\right) .$
For $\eta _{AB}=diag\left( -1,+1,+1,+1\right) $ and $S_{10}=SO\left(
1,4\right) ,L_{6}=SO\left( 1,3\right) $ is the group of Lorentz rotations.

The generators $I^{\underline{a}}$ and structure constants $f_{~\underline{t}%
}^{\underline{s}\underline{p}}$ of the de Sitter Lie group can be
paramet\-riz\-ed in a form distinguishing the de Sitter generators and
commutations (\ref{dsca}). The action of the group $S_{\left( \eta \right) }$
may be realized by using $4\times 4$ matrices with a parametrization
distinguishing the subgroup $L_{\left( \eta \right) }:$%
\begin{equation}
B=bB_{L},  \label{parametriz}
\end{equation}%
where%
\begin{equation*}
B_{L}=\left(
\begin{array}{cc}
L & 0 \\
0 & 1%
\end{array}%
\right) ,
\end{equation*}%
$L\in L_{\left( \eta \right) }$ is the de Sitter bust matrix transforming
the vector $\left( 0,0,...,\rho \right) \in {\R}^{5}$ into the arbitrary
point $\left( V^{1},V^{2},...,V^{5}\right) \in \Sigma _{\rho }^{5}\subset
\mathcal{R}^{5}$ with curvature $\rho ,$ $(V_{A}V^{A}=-\rho ^{2},V^{A}=\tau
^{A}\rho ),$ and the matrix $b$ is expressed
\begin{equation*}
b=\left(
\begin{array}{cc}
\delta _{\quad \beta ^{\prime }}^{\alpha ^{\prime }}+\frac{\tau ^{\alpha
^{\prime }}\tau _{\beta ^{\prime }}}{\left( 1+\tau ^{5}\right) } & \tau
^{\alpha ^{\prime }} \\
\tau _{\beta ^{\prime }} & \tau ^{5}%
\end{array}%
\right) .
\end{equation*}

The de Sitter gauge field is associated with a ${\mathit{s}o}_{\left( \eta
\right) }\left( 5\right) $--valued connection 1--form
\begin{equation}
\widetilde{\Omega }=\left(
\begin{array}{cc}
\omega _{\quad \beta ^{\prime }}^{\alpha ^{\prime }} & \widetilde{\theta }%
^{\alpha ^{\prime }} \\
\widetilde{\theta }_{\beta ^{\prime }} & 0%
\end{array}%
\right) ,  \label{dspot}
\end{equation}%
where $\omega _{\quad \beta ^{\prime }}^{\alpha ^{\prime }}\in so(4)_{\left(
\eta \right) },$ $\widetilde{\theta }^{\alpha ^{\prime }}\in \mathcal{R}^{4},%
\widetilde{\theta }_{\beta ^{\prime }}\in \eta _{\beta ^{\prime }\alpha
^{\prime }}\widetilde{\theta }^{\alpha ^{\prime }}.$

The actions of $S_{\left( \eta \right) }$ mix the components of the matrix $%
\omega _{\quad \beta ^{\prime }}^{\alpha ^{\prime }}$ and $\widetilde{\theta
}^{\alpha ^{\prime }}$ fields in (\ref{dspot}). Because the introduced
para\-met\-ri\-za\-ti\-on is invariant on action on $SO_{\left( \eta \right)
}\left( 4\right) $ group, we cannot identify $\omega _{\quad \beta ^{\prime
}}^{\alpha ^{\prime }}$ and $\widetilde{\theta }^{\alpha ^{\prime }},$
respectively, with the connection $\Gamma ^{\lbrack c]}$ and the 1--form $%
e^{\alpha }$ defined by a N--connection structure like in (\ref{ndif}) with
the coefficients chosen as in (\ref{viel1}) and (\ref{viel2}). To avoid this
difficulty we \ can consider nonlinear gauge realizations of the de Sitter
group $S_{\left( \eta \right) }$ by introducing the nonlinear gauge field
\begin{equation}
\Gamma =b^{-1}{\widetilde{\Omega }}b+b^{-1}db=\left(
\begin{array}{cc}
\Gamma _{~\beta ^{\prime }}^{\alpha ^{\prime }} & \theta ^{\alpha ^{\prime }}
\\
\theta _{\beta ^{\prime }} & 0%
\end{array}%
\right) ,  \label{npot}
\end{equation}%
where
\begin{eqnarray}
\Gamma _{\quad \beta ^{\prime }}^{\alpha ^{\prime }} &=&\omega _{\quad \beta
^{\prime }}^{\alpha ^{\prime }}-\left( \tau ^{\alpha ^{\prime }}D\tau
_{\beta ^{\prime }}-\tau _{\beta ^{\prime }}D\tau ^{\alpha ^{\prime
}}\right) /\left( 1+\tau ^{5}\right) ,  \notag \\
\theta ^{\alpha ^{\prime }} &=&\tau ^{5}\widetilde{\theta }^{\alpha ^{\prime
}}+D\tau ^{\alpha ^{\prime }}-\tau ^{\alpha ^{\prime }}\left( d\tau ^{5}+%
\widetilde{\theta }_{\gamma ^{\prime }}\tau ^{\gamma ^{\prime }}\right)
/\left( 1+\tau ^{5}\right) ,  \notag \\
D\tau ^{\alpha ^{\prime }} &=&d\tau ^{\alpha ^{\prime }}+\omega _{\quad
\beta ^{\prime }}^{\alpha ^{\prime }}\tau ^{\beta ^{\prime }}.  \notag
\end{eqnarray}

The action of the group $S\left( \eta \right) $ is nonlinear, yielding the
transformation rules
\begin{equation*}
\Gamma ^{\prime }=L^{\prime }\Gamma \left( L^{\prime }\right)
^{-1}+L^{\prime }d\left( L^{\prime }\right) ^{-1},~\theta ^{\prime }=L\theta
,
\end{equation*}%
where the nonlinear matrix--valued function
\begin{equation*}
L^{\prime }=L^{\prime }\left( \tau ^{\alpha },b,B_{T}\right)
\end{equation*}%
is defined from $B_{b}=b^{\prime }B_{L^{\prime }}$ (see the parametrization (%
\ref{parametriz})). The de Sitter 'nonlinear' algebra is defined by
generators (\ref{dsca}) and nonlinear gauge transforms of type (\ref{npot}).

\subsection{\ De Sitter Nonlinear Gauge Gravity and General Relativity}

We generalize the constructions from Refs. 23,6\ %
% \cite{ts,v1} %
to the case when the de Sitter nonlinear gauge gravitational connection (\ref%
{npot}) is defined by the viebeins (\ref{viel1}) and (\ref{viel2}) and the
linear connection (\ref{dcon}) $\Gamma _{\quad \beta \mu }^{[c]\alpha
}=\{\Gamma _{\quad \beta \mu }^{\alpha }\},$
\begin{equation}
\Gamma =\left(
\begin{array}{cc}
\Gamma _{\quad \beta ^{\prime }}^{\alpha ^{\prime }} & l_{0}^{-1}e^{\alpha
^{\prime }} \\
l_{0}^{-1}e_{\beta ^{\prime }} & 0%
\end{array}%
\right)  \label{conds}
\end{equation}%
where
\begin{equation}
\Gamma _{\quad \beta ^{\prime }}^{\alpha ^{\prime }}=\Gamma _{\quad \beta
^{\prime }\mu }^{\alpha ^{\prime }}\delta u^{\mu },  \label{condsc}
\end{equation}%
for
\begin{eqnarray}
\Gamma _{\quad \beta ^{\prime }\mu }^{\alpha ^{\prime }} &=&e_{\alpha
}^{~\alpha ^{\prime }}e_{\quad \beta ^{\prime }}^{\beta }\Gamma _{\quad
\beta \mu }^{\alpha }+e_{\alpha }^{~\alpha ^{\prime }}\delta _{\mu }e_{\quad
\beta ^{\prime }}^{\alpha },  \label{coef3} \\
e^{\alpha ^{\prime }} &=&e_{\mu }^{~\alpha ^{\prime }}\delta u^{\mu }  \notag
\end{eqnarray}%
and $l_{0}$ being a dimensional constant.

The matrix components of the curvature of the connection (\ref{conds}),
\begin{equation*}
\mathcal{R}^{(\Gamma )}=d\Gamma +\Gamma \wedge \Gamma ,
\end{equation*}%
can be written
\begin{equation}
\mathcal{R}^{(\Gamma )}=\left(
\begin{array}{cc}
\mathcal{R}_{\quad \beta ^{\prime }}^{\alpha ^{\prime }}+l_{0}^{-1}\pi
_{\beta ^{\prime }}^{\alpha ^{\prime }} & l_{0}^{-1}T^{\alpha ^{\prime }} \\
l_{0}^{-1}T^{\beta ^{\prime }} & 0%
\end{array}%
\right) ,  \label{curvs}
\end{equation}%
where
\begin{equation*}
\pi _{\beta ^{\prime }}^{\alpha ^{\prime }}=e^{\alpha ^{\prime }}\wedge
e_{\beta ^{\prime }},~\mathcal{R}_{\quad \beta ^{\prime }}^{\alpha ^{\prime
}}=\frac{1}{2}\mathcal{R}_{\quad \beta ^{\prime }\mu \nu }^{\alpha ^{\prime
}}\delta u^{\mu }\wedge \delta u^{\nu },
\end{equation*}%
and
\begin{equation*}
\mathcal{R}_{\quad \beta ^{\prime }\mu \nu }^{\alpha ^{\prime }}=e_{~\beta
^{\prime }}^{\beta }e_{\alpha }^{\quad \alpha ^{\prime }}R_{\quad \beta
_{\mu \nu }}^{\alpha }.
\end{equation*}%
with the coefficients $R_{\quad \beta {\mu \nu }}^{\alpha }$ defined with
h--v--invariant components, see (\ref{dcurvatures}) in Appendix A.

The de Sitter gauge group is semisimple: we are able to construct a
variational gauge gravitational theory with the Lagrangian
\begin{equation}
L=L_{\left( g\right) }+L_{\left( m\right) }  \label{lagrangc}
\end{equation}%
where the gauge gravitational Lagrangian is defined
\begin{equation*}
L_{\left( g\right) }=\frac{1}{4\pi }Tr\left( \mathcal{R}^{(\Gamma )}\wedge
\ast _{G}\mathcal{R}^{(\Gamma )}\right) =\mathcal{L}_{\left( G\right)
}\left| g\right| ^{1/2}\delta ^{4}u,
\end{equation*}%
for
\begin{equation}
\mathcal{L}_{\left( g\right) }=\frac{1}{2l^{2}}T_{\quad \mu \nu }^{\alpha
^{\prime }}T_{\alpha ^{\prime }}^{\quad \mu \nu }+\frac{1}{8\lambda }%
\mathcal{R}_{\quad \beta ^{\prime }\mu \nu }^{\alpha ^{\prime }}\mathcal{R}%
_{\quad \alpha ^{\prime }}^{\beta ^{\prime }\quad \mu \nu }{}-\frac{1}{l^{2}}%
\left( {\overleftarrow{R}}\left( \Gamma \right) -2\lambda _{1}\right) ,
\notag
\end{equation}
with $\delta ^{4}u$ being the volume element, $\left| g\right| $ is the
determinant computed the metric coefficients (\ref{dmetric}) stated with
respect to N--elongated frames, the curvature scalar ${\overleftarrow{R}}%
\left( \Gamma \right) $ is computed as in (\ref{dscalar}), $T_{\quad \mu \nu
}^{\alpha ^{\prime }}=e_{\quad \alpha }^{\alpha ^{\prime }}T_{\quad \mu \nu
}^{\alpha }$ (the gravitational constant $l^{2}$ satisfies the relations $%
l^{2}=2l_{0}^{2}\lambda ,\lambda _{1}=-3/l_{0}),\quad Tr$ denotes the trace
on $\alpha ^{\prime },\beta ^{\prime }$ indices. The matter field Lagrangian
from (\ref{lagrangc}) is defined
\begin{equation*}
L_{\left( m\right) }=-\frac{1}{2}Tr\left( \Gamma \wedge \ast _{g}\mathcal{I}%
\right) =\mathcal{L}_{\left( m\right) }\left| g\right| ^{1/2}\delta ^{n}u,
\end{equation*}%
with the Hodge operator derived by $\left| g\right| $ and $\left| h\right| $
where
\begin{equation*}
\mathcal{L}_{\left( m\right) }=\frac{1}{2}\Gamma _{\quad \beta ^{\prime }\mu
}^{\alpha ^{\prime }}S_{\quad \alpha }^{\beta ^{\prime }\quad \mu }-t_{\quad
\alpha ^{\prime }}^{\mu }l_{\quad \mu }^{\alpha ^{\prime }}.
\end{equation*}%
The matter field source $\mathcal{J}$ is obtained as a variational
derivation of $\mathcal{L}_{\left( m\right) }$ on $\Gamma $ and is
paramet\-riz\-ed in the form
\begin{equation*}
\mathcal{J}=\left(
\begin{array}{cc}
S_{\quad \underline{\beta }}^{\alpha ^{\prime }} & -l_{0}\tau ^{\alpha
^{\prime }} \\
-l_{0}\tau _{\beta ^{\prime }} & 0%
\end{array}%
\right)
\end{equation*}%
with $\tau ^{\alpha ^{\prime }}=\tau _{\quad \mu }^{\alpha ^{\prime }}\delta
u^{\mu }$ and $S_{\quad \beta ^{\prime }}^{\alpha ^{\prime }}=S_{\quad \beta
^{\prime }\mu }^{\alpha ^{\prime }}\delta u^{\mu }$ being respectively the
canonical tensors of energy--momentum and spin density.

Varying the action
\begin{equation*}
S=\int \delta ^{4}u\left( \mathcal{L}_{\left( g\right) }+\mathcal{L}_{\left(
m\right) }\right)
\end{equation*}%
on the $\Gamma $--variables (\ref{conds}), we obtain the
gau\-ge--gra\-vi\-ta\-ti\-on\-al field equations:%
\begin{equation}
d\left( \ast \mathcal{R}^{(\Gamma )}\right) +\Gamma \wedge \left( \ast
\mathcal{R}^{(\Gamma )}\right) -\left( \ast \mathcal{R}^{(\Gamma )}\right)
\wedge \Gamma =-\lambda \left( \ast \mathcal{J}\right) ,  \label{eqs}
\end{equation}%
were the Hodge operator $\ast $ is used. This equations can be alternatively
derived in geometric form by applying the absolute derivation and dual
operators.

Distinguishing the variations on $\Gamma $ and $e$--variables, we rewrite (%
\ref{eqs})
\begin{eqnarray*}
\widehat{\mathcal{D}}\left( \ast \mathcal{R}^{(\Gamma )}\right) +\frac{%
2\lambda }{l^{2}}(\widehat{\mathcal{D}}\left( \ast \pi \right) +e\wedge
\left( \ast T^{T}\right) -\left( \ast T\right) \wedge e^{T}) &=&-\lambda
\left( \ast S\right) , \\
\widehat{\mathcal{D}}\left( \ast T\right) -\left( \ast \mathcal{R}^{(\Gamma
)}\right) \wedge e-\frac{2\lambda }{l^{2}}\left( \ast \pi \right) \wedge e
&=&\frac{l^{2}}{2}\left( \ast t+\frac{1}{\lambda }\ast \varsigma \right) ,
\end{eqnarray*}%
$e^{T}$ being the transposition of \ $e,$ where
\begin{eqnarray}
T^{t} &=&\{T_{\alpha ^{\prime }}=\eta _{\alpha ^{\prime }\beta ^{\prime
}}T^{\beta ^{\prime }},~T^{\beta ^{\prime }}=\frac{1}{2}T_{\quad \mu \nu
}^{\beta ^{\prime }}\delta u^{\mu }\wedge \delta u^{\nu }\},  \notag \\
e^{T} &=&\{e_{\alpha ^{\prime }}=\eta _{\alpha ^{\prime }\beta ^{\prime
}}e^{\beta ^{\prime }},~e^{\beta ^{\prime }}=e_{\quad \mu }^{\beta ^{\prime
}}\delta u^{\mu }\},\qquad \widehat{\mathcal{D}}=\delta +\widehat{\Gamma },
\notag
\end{eqnarray}%
($\widehat{\Gamma }$ acts as $\Gamma _{\quad \beta ^{\prime }\mu }^{\alpha
^{\prime }}$ on indices $\gamma ^{\prime },\delta ^{\prime },...$ and as $%
\Gamma _{\quad \beta \mu }^{\alpha }$ on indices $\gamma ,\delta ,...).$ The
value $\varsigma $ defines the energy--momentum tensor of the gauge
gravitational field $\widehat{\Gamma }:$%
\begin{equation*}
\varsigma _{\mu \nu }\left( \widehat{\Gamma }\right) =\frac{1}{2}Tr\left(
\mathcal{R}_{\mu \alpha }\mathcal{R}_{\quad \nu }^{\alpha }-\frac{1}{4}%
\mathcal{R}_{\alpha \beta }\mathcal{R}^{\alpha \beta }G_{\mu \nu }\right) .
\end{equation*}

Equations (\ref{eqs}) make up the complete system of variational field
equations for nonlinear de Sitter gauge gravity. We note that we can obtain
a nonvariational Poincare gauge gravitational theory if we consider the
contraction of the gauge potential (\ref{conds}) to a potential $\Gamma
^{\lbrack P]}$ with values in the Poincare Lie algebra
\begin{equation}
\Gamma =\left(
\begin{array}{cc}
\Gamma _{\quad \beta ^{\prime }}^{\alpha ^{\prime }} & l_{0}^{-1}e^{\alpha
^{\prime }} \\
l_{0}^{-1}e_{\beta ^{\prime }} & 0%
\end{array}%
\right) \rightarrow \Gamma ^{\lbrack P]}=\left(
\begin{array}{cc}
\Gamma _{\quad \beta ^{\prime }}^{\alpha ^{\prime }} & l_{0}^{-1}e^{\alpha
^{\prime }} \\
0 & 0%
\end{array}%
\right) .  \label{poinc}
\end{equation}%
A similar gauge potential was considered in the formalism of linear and
affine frame bundles on curved spacetimes by Popov and Dikhin $^{21,22}.$ \
%
%\cite{pd1,pd2}.%
They considered the gauge potential (\ref{poinc}) to be just the Cartan
connection form in the affine gauge like gravity and proved that the
Yang--Mills equations of their theory are equivalent, after projection on
the base, to the Einstein equations.

Let us give an example how an exact vacuum solution of the Einstein
equations, with associated noncommutative symmetry, can be included as to
define an exact solution in gauge gravity. Using the data (\ref{data})
defining a 4D black ellipsoid solution, we write the nontrivial vielbein
coefficients (\ref{viel1}) as
\begin{equation}
e_{2}^{\ 2^{\prime }}=1,e_{3}^{\ 3^{\prime }}=\sqrt{|g_{3}|},e_{4}^{\
4^{\prime }}=\sqrt{|h_{4}|},e_{5}^{\ 5^{\prime }}=\sqrt{|h_{5}|}%
,N_{2}^{5}=n_{2}  \label{framel}
\end{equation}%
for the diagonal Minkowski metric $\eta _{\alpha ^{\prime }\beta ^{\prime
}}=\left( -1,-1,-1,1\right) $ with the tetrad and coordinate indices running
respectively the values $\alpha ^{\prime },\beta ^{\prime },...=2,3,4,5$ and
$\alpha ,\beta ,...=2,3,4,5.$ The connection coefficients $\Gamma _{\quad
\beta ^{\prime }\mu }^{\alpha ^{\prime }},$ see formula (\ref{coef3}), are
computed by using the values $e_{\alpha }^{\ \alpha ^{\prime }}$ and (\ref%
{dcon}) and used for definition of the potential $\Gamma ^{\lbrack P]}$ (\ref%
{poinc}) which defines a gauge gravity model with the Yang--Mills equations (%
\ref{eqs}) being completely equivalent to the Einstein equations even the
frames are anholonomic (see details in Refs. 21,22,24-28). %
%\cite{pd1,pd2,v41,v42,v43,v44,v45}).% Taking complex valued
N--coefficients, for instance, a complex $N_{2}^{5}=n_{2}$ we can construct
both complex Einstein and gauge gravity vacuum configurations which are
stable and define anholonomically deformed black hole solutions with
associated noncommutative symmetries.

Finally, we emphasize that in a similar manner, by extending the dimensions
of spaces and gauge groups and introducing the cosmological constant, we can
include the solutions for the (anti) de Sitter black ellipsoids and
wormohles, with real or complex anholonomic structures (constructed
respectively in sections IV C and V), into a gauge gravity theory (Einstein
and Poincare like, or as a degenerated configuration in\ the nonlinear
(anti) de Sitter gravity).

\section{\quad Noncommutative Gauge Deformations of Gravity}

The noncommutative gravity theories are confrunted with the problem of
definition of noncommutative variants of pseudo--Eucliedean and
pseudo--Riemannian metrics. This is connected with another problem when the
generation of noncommutative metric structures via the Moyal product and the
Seiberg--Witten map$^{45}$ \ %
% \cite{sw}%
results in complex and noncommutative metrics for, in general, nonstable
and/or unphysical gravitational vacua. In order to avoid the mentioned
difficulties, we elaborated a model of noncommutative gauge gravity starting
from a variant of gauge gravity being equivalent to the Einstein gravity and
emphasizing in a such approach the vielbein (frame) and connection
structures, but not the metric configuration (see Ref. 6 and 61).\ %
%\cite{vnonc}).%
The metric for such theories is induced by an anholonomic (in general) frame
transform.

For explicit constructions, we follow the method of restricted enveloping
algebras$^{62}$ %
%\cite{js}%
and construct gauge gravitational theories by stating corresponding
structures with semisimple or nonsemisimple Lie algebras and their
extensions. We use power series of generators for the affine and nonlinearly
realized de Sitter gauge groups and compute the coefficient functions of all
the higher powers of the generators of the gauge group which are functions
of the coefficients of the first power. Such constructions are based on the
Seiberg--Witten map and on the formalism of $\ast $--product formulation of
the algebra$^{45,63-69}$\ %
%\cite{sw,w1,w2,wig,moy,bffls,konts,zot}%
when for functional objects, being functions of commuting variables, there
are associated some algebraic noncommutative properties encoded in the $\ast
$--product. Here we note that an approach to the gauge theory on
noncommutative spaces was introduced geometrically$^{70}$\ %
%\cite{mssw}%
by defining the covariant coordinates without speaking about derivatives.
This formalism was also developed for quantum planes$^{71,72}.$ \ %
%\cite{wz1,jw}.

In this section, we shall prove the existence for noncommutative spaces of
gauge models of gravity which agrees with usual gauge gravity theories being
equivalent, or extending, the general relativity theory in the limit of
commuting spaces. We shall show how it is possible to adapt mutually the
Seiberg--Witten map and anholonomic frame transforms in order to generate
solutions of the gauge gravity preserving noncommutative symmetries even in
the classical limit of commutative Einstein gravity.

\subsection{Enveloping algebras for gauge gravity connections}

We define the gauge fields on a noncommutative space as elements of an
algebra $\mathcal{A}_{u}$ that form a representation of the generator $I$%
--algebra for the de Sitter gauge group and the noncommutative space is
modelled as the associative algebra of $\C.$\ This algebra is freely
generated by the coordinates modulo ideal $\mathcal{R}$ generated by the
relations (one accepts formal power series)\ $\mathcal{A}_{u}=%
\C[[{\hat
u}^1,...,{\hat u}^N]]/\mathcal{R}.$ A variational gauge gravitational theory
can be formulated by using a minimal extension of the affine structural
group ${\mathcal{A}f}_{3+1}\left( {\R}\right) $ to the de Sitter gauge group
$S_{10}=SO\left( 4+1\right) $ acting on ${\R}^{4+1}$ space.

Let now us consider a noncommutative space (see Appendix \ref{asncalg} for a
brief outline of necessary concepts). The gauge fields are elements of the
algebra $\widehat{\psi }\in \mathcal{A}_{I}^{(dS)}$ that form the nonlinear
representation of the de Sitter algebra ${\mathit{s}o}_{\left( \eta \right)
}\left( 5\right) $ (the whole algebra is denoted $\mathcal{A}_{z}^{(dS)}).$
The elements transform
\begin{equation*}
\delta \widehat{\psi }=i\widehat{\gamma }\widehat{\psi },\widehat{\psi }\in
\mathcal{A}_{u},\widehat{\gamma }\in \mathcal{A}_{z}^{(dS)},
\end{equation*}%
under a nonlinear de Sitter transformation. The action of the generators (%
\ref{dsca}) on $\widehat{\psi }$ is defined as the resulting element will
form a nonlinear representation of $\mathcal{A}_{I}^{(dS)}$ and, in
consequence, $\delta \widehat{\psi }\in \mathcal{A}_{u}$ despite $\widehat{%
\gamma }\in \mathcal{A}_{z}^{(dS)}.$ We emphasize that for any
representation the object $\widehat{\gamma }$ takes values in enveloping de
Sitter algebra but not in a Lie algebra as would be for commuting spaces.

We introduce a connection $\widehat{\Gamma }^{\nu }\in \mathcal{A}%
_{z}^{(dS)} $ in order to define covariant coordinates,%
\begin{equation*}
\widehat{U}^{\nu }=\widehat{u}^{v}+\widehat{\Gamma }^{\nu }.
\end{equation*}%
The values $\widehat{U}^{\nu }\widehat{\psi }$ transform covariantly, i. e. $%
\delta \widehat{U}^{\nu }\widehat{\psi }=i\widehat{\gamma }\widehat{U}^{\nu }%
\widehat{\psi },$ if and only if the connection $\widehat{\Gamma }^{\nu }$
satisfies the transformation law of the enveloping nonlinear realized de
Sitter algebra,%
\begin{equation*}
\delta \widehat{\Gamma }^{\nu }\widehat{\psi }=-i[\widehat{u}^{v},\widehat{%
\gamma }]+i[\widehat{\gamma },\widehat{\Gamma }^{\nu }],
\end{equation*}%
where $\delta \widehat{\Gamma }^{\nu }\in \mathcal{A}_{z}^{(dS)}.$

The enveloping algebra--valued connection has infinitely many component
fields. Nevertheless, all component fields can be induced from a Lie
algebra--valued connection by a Seiberg--Witten map$^{45,62}$\ %
%(\cite{sw,js}%
and, for $SO(n)$ and $Sp(n),$ see Ref. 79. %\cite{bsst}.%
Here, we show that similar constructions can be performed for nonlinear
realizations of de Sitter algebra when the transformation of the connection
is considered%
\begin{equation*}
\delta \widehat{\Gamma }^{\nu }=-i[u^{\nu },^{\ast }~\widehat{\gamma }]+i[%
\widehat{\gamma },^{\ast }~\widehat{\Gamma }^{\nu }].
\end{equation*}%
We treat in more detail the canonical case with the star product (\ref{csp1a}%
). The first term in the variation $\delta \widehat{\Gamma }^{\nu }$ gives
\begin{equation*}
-i[u^{\nu },^{\ast }~\widehat{\gamma }]=\theta ^{\nu \mu }\frac{\partial }{%
\partial u^{\mu }}\gamma .
\end{equation*}%
Assuming that the variation of $\widehat{\Gamma }^{\nu }=\theta ^{\nu \mu
}Q_{\mu }$ starts with a linear term in $\theta ,$ we have%
\begin{equation*}
\delta \widehat{\Gamma }^{\nu }=\theta ^{\nu \mu }\delta Q_{\mu },~\delta
Q_{\mu }=\frac{\partial }{\partial u^{\mu }}\gamma +i[\widehat{\gamma }%
,^{\ast }~Q_{\mu }].
\end{equation*}%
We expand the star product (\ref{csp1a}) in $\theta $ but not in $g_{a}$ and
find up to first order in $\theta $ that \ \
\begin{equation}
\gamma =\gamma _{\underline{a}}^{1}I^{\underline{a}}+\gamma _{\underline{a}%
\underline{b}}^{1}I^{\underline{a}}I^{\underline{b}}+...,Q_{\mu }=q_{\mu ,%
\underline{a}}^{1}I^{\underline{a}}+q_{\mu ,\underline{a}\underline{b}%
}^{2}I^{\underline{a}}I^{\underline{b}}+...  \label{seriesa}
\end{equation}%
where $\gamma _{\underline{a}}^{1}$ and $q_{\mu ,\underline{a}}^{1}$ are of
order zero in $\theta $ and $\gamma _{\underline{a}\underline{b}}^{1}$ and $%
q_{\mu ,\underline{a}\underline{b}}^{2}$ are of second order in $\theta .$
The expansion in $I^{\underline{b}}$ leads to an expansion in $g_{a}$ of the
$\ast $--product because the higher order $I^{\underline{b}}$--derivatives
vanish. For de Sitter case, we take the generators $I^{\underline{b}}$ (\ref%
{dsca}), see commutators (\ref{commutators1}), with the corresponding de
Sitter structure constants $f_{~\underline{d}}^{\underline{b}\underline{c}%
}\simeq f_{~\underline{\beta }}^{\underline{\alpha }\underline{\beta }}$ (in
our further identifications with spacetime objects like frames and
connections we shall use Greek indices). The result of calculation of
variations of (\ref{seriesa}), by using $g_{a}$ to the order given in (\ref%
{gdecomp}), is%
\begin{eqnarray}
\delta q_{\mu ,\underline{a}}^{1} &=&\frac{\partial \gamma _{\underline{a}%
}^{1}}{\partial u^{\mu }}-f_{~\underline{a}}^{\underline{b}\underline{c}%
}\gamma _{\underline{b}}^{1}q_{\mu ,\underline{c}}^{1},\ \delta Q_{\tau
}=\theta ^{\mu \nu }\partial _{\mu }\gamma _{\underline{a}}^{1}\partial
_{\nu }q_{\tau ,\underline{b}}^{1}I^{\underline{a}}I^{\underline{b}}+...,
\notag \\
\delta q_{\mu ,\underline{a}\underline{b}}^{2} &=&\partial _{\mu }\gamma _{%
\underline{a}\underline{b}}^{2}-\theta ^{\nu \tau }\partial _{\nu }\gamma _{%
\underline{a}}^{1}\partial _{\tau }q_{\mu ,\underline{b}}^{1}-2f_{~%
\underline{a}}^{\underline{b}\underline{c}}\{\gamma _{\underline{b}%
}^{1}q_{\mu ,\underline{c}\underline{d}}^{2}+\gamma _{\underline{b}%
\underline{d}}^{2}q_{\mu ,\underline{c}}^{1}\}.  \notag
\end{eqnarray}

Let us introduce the objects $\varepsilon ,$ taking the values in de Sitter
Lie algebra and $W_{\mu },$ taking values in the enveloping de Sitter
algebra, i. e.
\begin{equation*}
\varepsilon =\gamma _{\underline{a}}^{1}I^{\underline{a}}\mbox{
and
}W_{\mu }=q_{\mu ,\underline{a}\underline{b}}^{2}I^{\underline{a}}I^{%
\underline{b}},
\end{equation*}%
with the variation $\delta W_{\mu }$ satisfying the equation
\begin{equation*}
\delta W_{\mu }=\partial _{\mu }(\gamma _{\underline{a}\underline{b}}^{2}I^{%
\underline{a}}I^{\underline{b}})-\frac{1}{2}\theta ^{\tau \lambda
}\{\partial _{\tau }\varepsilon ,\partial _{\lambda }q_{\mu
}\}+{}i[\varepsilon ,W_{\mu }]+i[(\gamma _{\underline{a}\underline{b}}^{2}I^{%
\underline{a}}I^{\underline{b}}),q_{\nu }].
\end{equation*}%
This equation can be solved$^{70,45}$ \ %
%\cite{mssw,sw}
in the form%
\begin{equation}
\gamma _{\underline{a}\underline{b}}^{2}=\frac{1}{2}\theta ^{\nu \mu
}(\partial _{\nu }\gamma _{\underline{a}}^{1})q_{\mu ,\underline{b}%
}^{1},~q_{\mu ,\underline{a}\underline{b}}^{2}=-\frac{1}{2}\theta ^{\nu \tau
}q_{\nu ,\underline{a}}^{1}\left( \partial _{\tau }q_{\mu ,\underline{b}%
}^{1}+R_{\tau \mu ,\underline{b}}^{1}\right) .  \notag
\end{equation}%
The values
\begin{equation*}
R_{\tau \mu ,\underline{b}}^{1}=\partial _{\tau }q_{\mu ,\underline{b}%
}^{1}-\partial _{\mu }q_{\tau ,\underline{b}}^{1}+f_{~\underline{d}}^{%
\underline{e}\underline{c}}q_{\tau ,\underline{e}}^{1}q_{\mu ,\underline{e}%
}^{1}
\end{equation*}%
could be identified with the coefficients $\mathcal{R}_{\quad \underline{%
\beta }\mu \nu }^{\underline{\alpha }}$ of de Sitter nonlinear gauge gravity
curvature (see formula (\ref{curvs})) if in the commutative limit $q_{\mu ,%
\underline{b}}^{1}\simeq \left(
\begin{array}{cc}
\Gamma _{\quad \underline{\beta }}^{\underline{\alpha }} & l_{0}^{-1}\chi ^{%
\underline{\alpha }} \\
l_{0}^{-1}\chi _{\underline{\beta }} & 0%
\end{array}%
\right) $ (see (\ref{conds})).

We note that the below presented procedure can be generalized to all the
higher powers of $\theta .$ As an example, we compute the first order
corrections to the gravitational curvature:

\subsection{Noncommutative Covariant\ Gauge Gravity Dynamics}

The constructions from the previous subsection can be summarized by a
conclusion that the de Sitter algebra valued object $\varepsilon =\gamma _{%
\underline{a}}^{1}\left( u\right) I^{\underline{a}}$ determines all the
terms in the enveloping algebra%
\begin{equation*}
\gamma =\gamma _{\underline{a}}^{1}I^{\underline{a}}+\frac{1}{4}\theta ^{\nu
\mu }\partial _{\nu }\gamma _{\underline{a}}^{1}\ q_{\mu ,\underline{b}%
}^{1}\left( I^{\underline{a}}I^{\underline{b}}+I^{\underline{b}}I^{%
\underline{a}}\right) +...
\end{equation*}%
and the gauge transformations are defined by $\gamma _{\underline{a}%
}^{1}\left( u\right) $ and $q_{\mu ,\underline{b}}^{1}(u),$ when
\begin{equation*}
\delta _{\gamma ^{1}}\psi =i\gamma \left( \gamma ^{1},q_{\mu }^{1}\right)
\ast \psi .
\end{equation*}

Applying the formula (\ref{csp1a}) we calculate%
\begin{eqnarray*}
\lbrack \gamma ,^{\ast }\zeta ] &=&i\gamma _{\underline{a}}^{1}\zeta _{%
\underline{b}}^{1}f_{~\underline{c}}^{\underline{a}\underline{b}}I^{%
\underline{c}}+\frac{i}{2}\theta ^{\nu \mu }\{\partial _{v}\left( \gamma _{%
\underline{a}}^{1}\zeta _{\underline{b}}^{1}f_{~\underline{c}}^{\underline{a}%
\underline{b}}\right) q_{\mu ,\underline{c}} \\
&&+{}\left( \gamma _{\underline{a}}^{1}\partial _{v}\zeta _{\underline{b}%
}^{1}-\zeta _{\underline{a}}^{1}\partial _{v}\gamma _{\underline{b}%
}^{1}\right) q_{\mu ,\underline{b}}f_{~\underline{c}}^{\underline{a}%
\underline{b}}+2\partial _{v}\gamma _{\underline{a}}^{1}\partial _{\mu
}\zeta _{\underline{b}}^{1}\}I^{\underline{d}}I^{\underline{c}},
\end{eqnarray*}%
where we used the properties that, for the de Sitter enveloping algebras,
one holds the general formula for compositions of two transformations%
\begin{equation*}
\delta _{\gamma }\delta _{\varsigma }-\delta _{\varsigma }\delta _{\gamma
}=\delta _{i(\varsigma \ast \gamma -\gamma \ast \varsigma )}.
\end{equation*}%
This is also true for the restricted transformations defined by $\gamma
^{1}, $%
\begin{equation*}
\delta _{\gamma ^{1}}\delta _{\varsigma ^{1}}-\delta _{\varsigma ^{1}}\delta
_{\gamma ^{1}}=\delta _{i(\varsigma ^{1}\ast \gamma ^{1}-\gamma ^{1}\ast
\varsigma ^{1})}.
\end{equation*}

Such commutators could be used for definition of tensors
\begin{equation}
\widehat{S}^{\mu \nu }=[\widehat{U}^{\mu },\widehat{U}^{\nu }]-i\widehat{%
\theta }^{\mu \nu },  \label{tensor1}
\end{equation}%
where $\widehat{\theta }^{\mu \nu }$ is respectively stated for the
canonical, Lie and quantum plane structures. Under the general enveloping
algebra one holds the transform%
\begin{equation*}
\delta \widehat{S}^{\mu \nu }=i[\widehat{\gamma },\widehat{S}^{\mu \nu }].
\end{equation*}%
For instance, the canonical case is characterized by%
\begin{eqnarray}
S^{\mu \nu } &=&i\theta ^{\mu \tau }\partial _{\tau }\Gamma ^{\nu }-i\theta
^{\nu \tau }\partial _{\tau }\Gamma ^{\mu }+\Gamma ^{\mu }\ast \Gamma ^{\nu
}-\Gamma ^{\nu }\ast \Gamma ^{\mu }  \notag \\
&=&\theta ^{\mu \tau }\theta ^{\nu \lambda }\{\partial _{\tau }Q_{\lambda
}-\partial _{\lambda }Q_{\tau }+Q_{\tau }\ast Q_{\lambda }-Q_{\lambda }\ast
Q_{\tau }\}.  \notag
\end{eqnarray}%
We introduce the gravitational gauge strength (curvature)
\begin{equation}
R_{\tau \lambda }=\partial _{\tau }Q_{\lambda }-\partial _{\lambda }Q_{\tau
}+Q_{\tau }\ast Q_{\lambda }-Q_{\lambda }\ast Q_{\tau },  \label{qcurv}
\end{equation}%
which could be treated as a noncommutative extension of de Sitter nonlinear
gauge gravitational curvature (\ref{curvs}), and calculate
\begin{equation}
R_{\tau \lambda ,\underline{a}}=R_{\tau \lambda ,\underline{a}}^{1}+\theta
^{\mu \nu }\{R_{\tau \mu ,\underline{a}}^{1}R_{\lambda \nu ,\underline{b}%
}^{1}{}-\frac{1}{2}q_{\mu ,\underline{a}}^{1}\left[ (D_{\nu }R_{\tau \lambda
,\underline{b}}^{1})+\partial _{\nu }R_{\tau \lambda ,\underline{b}}^{1}%
\right] \}I^{\underline{b}},  \notag
\end{equation}%
where the gauge gravitation covariant derivative is introduced,%
\begin{equation*}
(D_{\nu }R_{\tau \lambda ,\underline{b}}^{1})=\partial _{\nu }R_{\tau
\lambda ,\underline{b}}^{1}+q_{\nu ,\underline{c}}R_{\tau \lambda ,%
\underline{d}}^{1}f_{~\underline{b}}^{\underline{c}\underline{d}}.
\end{equation*}%
Following the gauge transformation laws for $\gamma $ and $q^{1}$ we find
\begin{equation*}
\delta _{\gamma ^{1}}R_{\tau \lambda }^{1}=i\left[ \gamma ,^{\ast }R_{\tau
\lambda }^{1}\right]
\end{equation*}%
with the restricted form of $\gamma .$ Such formulas were proved in Ref. 45
%\cite{sw}%
for usual gauge (nongravitational) fields. Here we reconsidered them for the
gravitational gauge fields.

One can be formulated a gauge covariant gravitational dynamics of
noncommutative spaces following the nonlinear realization of de Sitter
algebra and the $\ast $--formalism and introducing derivatives in such a way
that one does not obtain new relations for the coordinates. In this case, a
Leibniz rule can be defined that
\begin{equation*}
\widehat{\partial }_{\mu }\widehat{u}^{\nu }=\delta _{\mu }^{\nu }+d_{\mu
\sigma }^{\nu \tau }\ \widehat{u}^{\sigma }\ \widehat{\partial }_{\tau }
\end{equation*}%
where the coefficients $d_{\mu \sigma }^{\nu \tau }=\delta _{\sigma }^{\nu
}\delta _{\mu }^{\tau }$ are chosen to have not new relations when $\widehat{%
\partial }_{\mu }$ acts again to the right hand side. One holds the $\ast $%
--derivative formulas
\begin{equation}
{\partial }_{\tau }\ast f=\frac{\partial }{\partial u^{\tau }}f+f\ast {%
\partial }_{\tau },~[{\partial }_{l},{{}^{\ast }}(f\ast g)]=([{\partial }%
_{l},{{}^{\ast }}f])\ast g+f\ast ([{\partial }_{l},{}^{\ast }g])  \notag
\end{equation}%
and the Stokes theorem%
\begin{equation*}
\int [\partial _{l},f]=\int d^{N}u[\partial _{l},^{\ast }f]=\int d^{N}u\frac{%
\partial }{\partial u^{l}}f=0,
\end{equation*}%
where, for the canonical structure, the integral is defined,%
\begin{equation*}
\int \widehat{f}=\int d^{N}uf\left( u^{1},...,u^{N}\right) .
\end{equation*}

An action can be introduced by using such integrals. For instance, for a
tensor of type (\ref{tensor1}), when%
\begin{equation*}
\delta \widehat{L}=i\left[ \widehat{\gamma },\widehat{L}\right] ,
\end{equation*}
we can define a gauge invariant action%
\begin{equation*}
W=\int d^Nu\ Tr\widehat{L},~\delta W=0,
\end{equation*}
were the trace has to be taken for the group generators.

For the nonlinear de Sitter gauge gravity a proper action is
\begin{equation*}
L=\frac{1}{4}R_{\tau \lambda }R^{\tau \lambda },
\end{equation*}%
where $R_{\tau \lambda }$ is defined by (\ref{qcurv}) (in the commutative
limit we shall obtain the connection (\ref{conds})). In this case the
dynamic of noncommutative space is entirely formulated in the framework of
quantum field theory of gauge fields. In general, we are dealing with
anisotropic gauge gravitational interactions. The method works for matter
fields as well to restrictions to the general relativity theory.

\subsection{Noncommutative Symmetries and Star Product Deformations}

The aim of this subsection is to prove that there are possible extensions of
exact solutions from the Einstein and gauge gravity possessing hidden
noncommutative symmetries without introducing new fields. For simplicity, we
present the formulas including decompositions up to the second order on
noncommutative parameter $\theta ^{\alpha \beta }$ for vielbeins,
connections and curvatures which can be arranged to result in different
models of noncommutative gravity. We give the data for the $SU\left(
1,n+m-1\right) $ and $SO\left( 1,n+m-1\right) $ gauge models containing, in
general, complex N--elongated frames, modelling some exact solutions, for
instance, those derived in the sections IV and V. All data can be considered
for extensions with nonlinear realizations into a bundle of affine/or de
Sitter frames (in this case, one generates noncommutative gauge theories of
type considered in Ref. 6\ %
%\cite{v1})%
or to impose certain constraints and broking of symmetries (in order to
construct other models$^{7,60}$). %
% \cite{ch1,card}).%

In previous sections we considered noncommutative geometric structures
introduced by frame anholonomic relations (\ref{anh}), or (\ref{anh1}). The
standard approaches to noncommutative geometry also contain certain
noncommutative relations for coordinates,
\begin{equation}
\lbrack u^{\alpha },u^{\beta }]=u^{\alpha }u^{\beta }-u^{\beta }u^{\alpha
}=i\theta ^{\alpha \beta }(u^{\gamma })  \label{coordnc}
\end{equation}%
were, in the simplest models, the commutator $[u^{\alpha },u^{\beta }]$ is
approximated to be constant, but there were elaborated approaches for
general manifolds with the noncommutative parameter $\theta ^{\alpha \beta }$
treated as functions on $u^{\gamma }$ in Ref. 68.\ %
% \cite{konts}.%
We define the star (Moyal) product to include possible N--elongated partial
derivatives (\ref{nder}) and \ a quantum constant $\hbar $,
\begin{eqnarray}
f\ast \varphi &=&f\varphi +\frac{\hbar }{2}B^{\overline{\alpha }\overline{%
\beta }}\left( \delta _{\overline{\alpha }}f\delta _{\overline{\beta }%
}\varphi +\delta _{\overline{\beta }}f\delta _{\overline{\alpha }}\varphi
\right) +\hbar ^{2}B^{\overline{\alpha }\overline{\beta }}B^{\overline{%
\gamma }\overline{\mu }}\left[ \delta _{(\overline{\alpha }}\delta _{%
\overline{\gamma })}f\right] \left[ \delta _{(\overline{\beta }}\delta _{%
\overline{\mu })}\varphi \right]  \notag \\
&&+\frac{2}{3}\hbar ^{2}B^{\overline{\alpha }\overline{\beta }}\delta _{%
\overline{\beta }}B^{\overline{\gamma }\overline{\mu }}\{~\left[ \delta _{(%
\overline{\alpha }}\delta _{\overline{\gamma })}f\right] \delta _{\overline{%
\mu }}\varphi +[\delta _{(\overline{\alpha }}\delta _{\overline{\gamma }%
)}\varphi ]\delta _{\overline{\mu }}f\}+O\left( \hbar ^{3}\right) ,
\label{form01}
\end{eqnarray}%
where, \ for instance, $\delta _{(\mu }\delta _{\nu )}=(1/2)(\delta _{\mu
}\delta _{\nu }+\delta _{\nu }\delta _{\mu })$
\begin{equation}
B^{\overline{\alpha }\overline{\beta }}=\frac{\theta ^{\alpha \beta }}{2}%
\left( \delta _{\alpha }u^{\overline{\alpha }}\delta _{\beta }u^{\overline{%
\beta }}+\delta _{\beta }u^{\overline{\alpha }}\delta _{\alpha }u^{\overline{%
\beta }}\right) +O\left( \hbar ^{3}\right)  \label{form02}
\end{equation}%
is defined for new coordinates $u^{\overline{\alpha }}=u^{\overline{\alpha }%
}\left( u^{\alpha }\right) $ inducing a suitable Poisson bi--vector field $%
B^{\overline{\alpha }\overline{\beta }}\left( \hbar \right) $ being related
to a quantum diagram formalism (we shall not consider details concerning
geometric quantization in this paper by investigating only classicassical
deformations related to any anholonomic frame and coordinate (\ref{coordnc})
noncommutativity origin). The formulas (\ref{form01}) and (\ref{form02})
transform into the usual ones with partial derivatives $\partial _{\alpha }$
and $\partial _{\overline{\alpha }}$ from Refs. 68,7\ %
%\cite{konts,ch1} %
considered for vanishing anholonmy coefficients. We can define a star
product being invariant under diffeomorphism transforms, $\ast \rightarrow
\ast ^{\lbrack -]},$ adapted to the N--connection structure ( in a vector
bundle provided with N--connection configuration, we use the label $[-]$ in
order to emphasize the dependence on coordinates $u^{\overline{\alpha }}$
with 'overlined' indices), by introducing the transforms%
\begin{eqnarray*}
f^{[-]}\left( \hbar \right) &=&\Theta f\left( \hbar \right) , \\
f^{[-]}\ast ^{\lbrack -]}\varphi ^{\lbrack -]} &=&\Theta \left( \Theta
^{-1}f^{[-]}\ast \Theta ^{-1}\right) \varphi ^{\lbrack -]}
\end{eqnarray*}%
for $\Theta =1+\sum_{[k=1]}\hbar ^{k}\Theta _{\lbrack k]},$ for simplicity,
computed up to the squared order on $\hbar ,$%
\begin{equation*}
\Theta =1-2\hbar ^{2}\theta ^{\mu \nu }\theta ^{\rho \sigma }\left\{ \left[
\delta _{(\mu }\delta _{\nu )}u^{\overline{\alpha }}\right] \left[ \delta
_{(\rho }\delta _{\sigma )}u^{\overline{\beta }}\right] \delta _{(\overline{%
\alpha }}\delta _{\overline{\beta })}+[\delta _{(\mu }\delta _{\rho )}u^{%
\overline{\alpha }}](\delta _{\nu }u^{\overline{\beta }})(\delta _{\sigma
}u^{\overline{\gamma }})\left[ \delta _{(\overline{\alpha }}\delta _{%
\overline{\beta }}\delta _{\overline{\gamma })}\right] \right\} +O\left(
\hbar ^{4}\right) ,
\end{equation*}%
where $\delta _{(\overline{\alpha }}\delta _{\overline{\beta }}\delta _{%
\overline{\gamma })}=$ $(1/3!)(\delta _{\overline{\alpha }}\delta _{%
\overline{\beta }}\delta _{\overline{\gamma }}+$ all \textsl{symmetric
permutations). }In our further constructions we shall omit the constant $%
\hbar $ considering that $\theta \sim \hbar $ is a small value by writing
the necessary terms in the approximation $O\left( \theta ^{3}\right) $ or $%
O\left( \theta ^{4}\right) .$

We consider a noncommutative gauge theory on a space with N--connection
structure stated by the gauge fields $\widehat{A}_{\mu }=\left( \widehat{A}%
_{i},\widehat{A}_{a}\right) $ when ''hats'' on symbols will be used for the
objects defined on spaces with coordinate noncommutativity. In general, the
gauge model can be with different types of structure groups like $SL\left( k,%
\C\right) ,$ $SU_{k},$ $U_{k},SO(k-1,1)$ and their nonlinear realizations.
For instance, for the $U\left( n+m\right) $ gauge fields there are satisfied
the conditions $\widehat{A}_{\mu }^{+}=-\widehat{A}_{\mu },$where $"+"$ is
the Hermitian conjugation. It is useful to present the basic geometric
constructions for a unitary structural group containing the $SO\left(
4,1\right) $ as a particular case if we wont to consider noncommutative
extensions of 4D exact solutions.

The noncommutative gauge transforms of potentials are defined by using the
star product%
\begin{equation*}
\widehat{A}_{\mu }^{[\varphi ]}=\widehat{\varphi }\ast \widehat{A}_{\mu }%
\widehat{\varphi }_{[\ast ]}^{-1}-\widehat{\varphi }\ast \delta _{\mu }%
\widehat{\varphi }_{[\ast ]}^{-1}
\end{equation*}%
where the N--elongated partial derivatives (\ref{nder}) are used $\widehat{%
\varphi }\ast \widehat{\varphi }_{[\ast ]}^{-1}=1=$ $\widehat{\varphi }%
_{[\ast ]}^{-1}\ast \widehat{\varphi }.$ The matrix coefficients of fields
will be distinguished by ''overlined'' indices, for instance, $\widehat{A}%
_{\mu }=\{\widehat{A}_{\mu }^{\underline{\alpha }\underline{\beta }}\},$ and
for commutative values, $A_{\mu }=\{A_{\mu }^{\underline{\alpha }\underline{%
\beta }}\}.$ Such fields are subjected to the conditions%
\begin{equation*}
(\widehat{A}_{\mu }^{\underline{\alpha }\underline{\beta }})^{+}\left(
u,\theta \right) =-\widehat{A}_{\mu }^{\underline{\beta }\underline{\alpha }%
}\left( u,\theta \right) \mbox{ and }\widehat{A}_{\mu }^{\underline{\alpha }%
\underline{\beta }}\left( u,-\theta \right) =-\widehat{A}_{\mu }^{\underline{%
\beta }\underline{\alpha }}\left( u,\theta \right) .
\end{equation*}%
There is a basic assumption$^{45}$ \ %
% \cite{sw}%
that the noncommutative fields are related to the commutative fields by the
Seiberg--Witten map in a manner that there are not new degrees of freedom
being satisfied the equation%
\begin{equation}
\widehat{A}_{\mu }^{\underline{\alpha }\underline{\beta }}(A)+\Delta _{%
\widehat{\lambda }}\widehat{A}_{\mu }^{\underline{\alpha }\underline{\beta }%
}(A)=\widehat{A}_{\mu }^{\underline{\alpha }\underline{\beta }}(A+\Delta _{%
\widehat{\lambda }}A)  \label{sw1}
\end{equation}%
where $\widehat{A}_{\mu }^{\underline{\alpha }\underline{\beta }}(A)$
denotes a functional dependence on commutative field $A_{\mu }^{\underline{%
\alpha }\underline{\beta }},\widehat{\varphi }=\exp \widehat{\lambda }$ and
the infinitesimal deformations $\widehat{A}_{\mu }^{\underline{\alpha }%
\underline{\beta }}(A)$ and of $A_{\mu }^{\underline{\alpha }\underline{%
\beta }}$ are given respectively by
\begin{equation*}
\Delta _{\widehat{\lambda }}\widehat{A}_{\mu }^{\underline{\alpha }%
\underline{\beta }}=\delta _{\mu }\widehat{\lambda }^{\underline{\alpha }%
\underline{\beta }}+\widehat{A}_{\mu }^{\underline{\alpha }\underline{\gamma
}}\ast \widehat{\lambda }^{\underline{\gamma }\underline{\beta }}-\widehat{%
\lambda }^{\underline{\alpha }\underline{\gamma }}\ast \widehat{A}_{\mu }^{%
\underline{\gamma }\underline{\beta }}
\end{equation*}%
and
\begin{equation*}
\Delta _{\lambda }A_{\mu }^{\underline{\alpha }\underline{\beta }}=\delta
_{\mu }\lambda ^{\underline{\alpha }\underline{\beta }}+A_{\mu }^{\underline{%
\alpha }\underline{\gamma }}\ast \lambda ^{\underline{\gamma }\underline{%
\beta }}-\lambda ^{\underline{\alpha }\underline{\gamma }}\ast A_{\mu }^{%
\underline{\gamma }\underline{\beta }}
\end{equation*}%
where instead of partial derivatives $\partial _{\mu }$ we use the
N--elongated ones, $\delta _{\mu }$ and sum on index $\underline{\gamma }.$

Solutions of the Seiberg--Witten equations for models of gauge gravity are
considered, for instance, in Refs. 6,7\ %
% \cite{v1,ch1} %
(there are discussed procedures of deriving expressions on $\theta $ to all
orders). Here we present only the first order on $\theta $ for the
coefficients $\widehat{\lambda }^{\underline{\alpha }\underline{\beta }}$
and the first and second orders for $\widehat{A}_{\mu }^{\underline{\alpha }%
\underline{\beta }}$ including anholonomy relations and not depending on
model considerations,%
\begin{equation*}
\widehat{\lambda }^{\underline{\alpha }\underline{\beta }}=\lambda ^{%
\underline{\alpha }\underline{\beta }}+\frac{i}{4}\theta ^{\nu \tau
}\{(\delta _{\nu }\lambda ^{\underline{\alpha }\underline{\gamma }})A_{\mu
}^{\underline{\gamma }\underline{\beta }}+A_{\mu }^{\underline{\alpha }%
\underline{\gamma }}(\delta _{\nu }\lambda ^{\underline{\gamma }\underline{%
\beta }})\}+O\left( \theta ^{2}\right)
\end{equation*}%
and
\begin{eqnarray}
\widehat{A}_{\mu }^{\underline{\alpha }\underline{\beta }} &=&A_{\mu }^{%
\underline{\alpha }\underline{\beta }}-\frac{i}{4}\theta ^{\nu \tau
}\{A_{\mu }^{\underline{\alpha }\underline{\gamma }}\left( \delta _{\tau
}A_{\nu }^{\underline{\gamma }\underline{\beta }}+R_{\quad \tau \nu }^{%
\underline{\gamma }\underline{\beta }}\right) +\left( \delta _{\tau }A_{\mu
}^{\underline{\alpha }\underline{\gamma }}+R_{\quad \tau \mu }^{\underline{%
\alpha }\underline{\gamma }}\right) A_{\nu }^{\underline{\gamma }\underline{%
\beta }}\}+  \label{nnccon1} \\
&&\frac{1}{32}\theta ^{\nu \tau }\theta ^{\rho \sigma }\{[2A_{\rho }^{%
\underline{\alpha }\underline{\gamma }}(R_{\quad \sigma \nu }^{\underline{%
\gamma }\underline{\varepsilon }}R_{\quad \mu \tau }^{\underline{e}%
\underline{\beta }}+R_{\quad \mu \tau }^{\underline{\gamma }\underline{%
\varepsilon }}R_{\quad \sigma \nu }^{\underline{\varepsilon }\underline{%
\beta }})+2(R_{\quad \sigma \nu }^{\underline{\alpha }\underline{\varepsilon
}}R_{\quad \mu \tau }^{\underline{\varepsilon }\underline{\gamma }}+R_{\quad
\mu \tau }^{\underline{\alpha }\underline{\varepsilon }}R_{\quad \sigma \nu
}^{\underline{\varepsilon }\underline{\gamma }})A_{\rho }^{\underline{\gamma
}\underline{\beta }}]  \notag \\
&&-[A_{\nu }^{\underline{\alpha }\underline{\gamma }}\left( D_{\tau
}R_{\quad \sigma \mu }^{\underline{\gamma }\underline{\beta }}+\delta _{\tau
}R_{\quad \sigma \mu }^{\underline{\gamma }\underline{\beta }}\right)
+\left( D_{\tau }R_{\quad \sigma \mu }^{\underline{\alpha }\underline{\gamma
}}+\delta _{\tau }R_{\quad \sigma \mu }^{\underline{\alpha }\underline{%
\gamma }}\right) A_{\nu }^{\underline{\gamma }\underline{\beta }}]-  \notag
\\
&&\delta _{\sigma }[A_{\nu }^{\underline{\alpha }\underline{\gamma }}\left(
\delta _{\tau }A_{\mu }^{\underline{\gamma }\underline{\beta }}+R_{\quad
\tau \mu }^{\underline{\gamma }\underline{\beta }}\right) +\left( \delta
_{\tau }A_{\mu }^{\underline{\alpha }\underline{\gamma }}+R_{\quad \tau \mu
}^{\underline{\alpha }\underline{\gamma }}\right) A_{\nu }^{\underline{%
\gamma }\underline{\beta }}]+  \notag
\end{eqnarray}%
\begin{eqnarray*}
&&[(\delta _{\nu }A_{\rho }^{\underline{\alpha }\underline{\gamma }})\left(
2\delta _{(\tau }\delta _{\sigma )}A_{\mu }^{\underline{\gamma }\underline{%
\beta }}+\delta _{\tau }R_{\quad \sigma \mu }^{\underline{\gamma }\underline{%
\beta }}\right) +\left( 2\delta _{(\tau }\delta _{\sigma )}A_{\mu }^{%
\underline{\alpha }\underline{\gamma }}+\delta _{\tau }R_{\quad \sigma \mu
}^{\underline{\alpha }\underline{\gamma }}\right) (\delta _{\nu }A_{\rho }^{%
\underline{\gamma }\underline{\beta }})]- \\
&&[A_{\nu }^{\underline{\alpha }\underline{\varepsilon }}\left( \delta
_{\tau }A_{\rho }^{\underline{\varepsilon }\underline{\gamma }}+R_{\quad
\tau \rho }^{\underline{\varepsilon }\underline{\gamma }}\right) +\left(
\delta _{\tau }A_{\rho }^{\underline{\alpha }\underline{\varepsilon }%
}+R_{\quad \tau \rho }^{\underline{\alpha }\underline{\varepsilon }}\right)
A_{\nu }^{\underline{\varepsilon }\underline{\gamma }}]\left( \delta
_{\sigma }A_{\mu }^{\underline{\gamma }\underline{\beta }}+R_{\quad \sigma
\mu }^{\underline{\gamma }\underline{\beta }}\right) - \\
&&\left( \delta _{\sigma }A_{\mu }^{\underline{\alpha }\underline{\gamma }%
}+R_{\quad \sigma \mu }^{\underline{\alpha }\underline{\gamma }}\right)
[A_{\nu }^{\underline{\gamma }\underline{\varepsilon }}\left( \delta _{\tau
}A_{\rho }^{\underline{\varepsilon }\underline{\beta }}+R_{\quad \tau \rho
}^{\underline{\varepsilon }\underline{\beta }}\right) +\left( \delta _{\tau
}A_{\rho }^{\underline{\gamma }\underline{\varepsilon }}+R_{\quad \tau \rho
}^{\underline{\gamma }\underline{\varepsilon }}\right) A_{\nu }^{\underline{%
\varepsilon }\underline{\beta }}]+O\left( \theta ^{3}\right) ,
\end{eqnarray*}%
where the curvature is defined $R_{\quad \tau \nu }^{\underline{\alpha }%
\underline{\beta }}=e_{\alpha }^{\underline{\alpha }}e^{\underline{\beta }%
\beta }R_{\beta \ \tau \nu }^{\ \alpha }$ with $R_{\beta \ \tau \nu }^{\
\alpha }$ computed as in Appendix A, see formula (\ref{curvature}), when $%
\Gamma \rightarrow A,$ and for the gauge model of gravity, see (\ref{curvs})
and (\ref{qcurv}). By using the star product, we can write symbolically the
solution (\ref{nnccon1}) in general form,%
\begin{equation*}
\Delta \widehat{A}_{\mu }^{\underline{\alpha }\underline{\beta }}\left(
\theta \right) =-\frac{i}{4}\theta ^{\nu \tau }\left[ \widehat{A}_{\mu }^{%
\underline{\alpha }\underline{\gamma }}\ast \left( \delta _{\tau }\widehat{A}%
_{\nu }^{\underline{\gamma }\underline{\beta }}+\widehat{R}_{\quad \tau \nu
}^{\underline{\gamma }\underline{\beta }}\right) +\left( \delta _{\tau }%
\widehat{A}_{\mu }^{\underline{\alpha }\underline{\gamma }}+\widehat{R}%
_{\quad \tau \mu }^{\underline{\alpha }\underline{\gamma }}\right) \ast
\widehat{A}_{\nu }^{\underline{\gamma }\underline{\beta }}\right]
\end{equation*}%
where $\widehat{R}_{\quad \tau \nu }^{\underline{\gamma }\underline{\beta }}$
is defined by the same formulas as $R_{\quad \tau \nu }^{\underline{\alpha }%
\underline{\beta }}$ but with the star products, $\ $\ like $AA\rightarrow
A\ast A.$

There is a problem how to determine the dependence of the noncommutative
vielbeins $\widehat{e}_{\alpha }^{\underline{\alpha }}$ on commutative ones $%
e_{\alpha }^{\underline{\alpha }}.$ If we consider the frame fields to be
included into a (anti) de Sitter gauge gravity model with the connection (%
\ref{conds}), the vielbein components should be treated as certain
coefficients of the gauge potential with specific nonlinear transforms for
which the results of Ref. 6.\ %
%\cite{v1} hold.
The main difference (considered in this work) is that the frames are in
general with anholonomy induced by a N--connection field. In order to derive
in a such model of the Einstein gravity we have to analyze the reduction (%
\ref{poinc}) to a Poincare gauge gravity. An explicit calculus of the
curvature of such gauge potential (see details in Refs. 21,22,24-28),\ %
%\cite{pd1,pd2,v41,v42,v43,v44,v45})),%
show that the coefficients of the curvature of the connection (\ref{poinc}),
obtained as a reduction from the $SO\left( 4,1\right) $ gauge group is given
by the coefficients (\ref{curvs}) with vanishing torsion and constraints of
type $\widehat{A}_{\nu }^{\underline{\gamma }\underline{5}}=\epsilon
\widehat{e}_{\nu }^{\underline{\gamma }}$ and $\widehat{A}_{\nu }^{%
\underline{5}\underline{5}}=\epsilon \widehat{\phi }_{\nu }$ with $\widehat{R%
}_{\quad \tau \nu }^{\underline{5}\underline{5}}\sim \epsilon $ vanishing in
the limit $\epsilon \rightarrow 0$ like in Ref. 7\ %
%\cite{ch1}%
(we obtain the same formulas for the vielbein and curvature components
derived for the inhomogenious Lorentz group but generalized to N--elongated
derivatives and with distinguishing into h--v--components). The result for $%
\widehat{e}_{\mu }^{\underline{\mu }}$ in the limit $\epsilon \rightarrow 0$
generalized to the case of canonical connections (\ref{dcon}) defining the
covariant derivatives $D_{\tau }$ and corresponing curvatures (\ref%
{curvature}) is%
\begin{eqnarray}
\widehat{e}_{\mu }^{\underline{\mu }} &=&e_{\mu }^{\underline{\mu }}-\frac{i%
}{4}\theta ^{\nu \tau }\left[ A_{\nu }^{\underline{\mu }\underline{\gamma }%
}\delta _{\tau }e_{\mu }^{\underline{\gamma }}+\left( \delta _{\tau }A_{\mu
}^{\underline{\mu }\underline{\gamma }}+R_{\quad \tau \mu }^{\underline{\mu }%
\underline{\gamma }}\right) e_{\nu }^{\underline{\gamma }}\right] +
\label{qfr} \\
&&\frac{1}{32}\theta ^{\nu \tau }\theta ^{\beta \sigma }\{2(R_{\quad \sigma
\nu }^{\underline{\mu }\underline{\varepsilon }}R_{\quad \mu \tau }^{%
\underline{\varepsilon }\underline{\gamma }}+R_{\quad \mu \tau }^{\underline{%
\mu }\underline{\varepsilon }}R_{\quad \sigma \nu }^{\underline{\varepsilon }%
\underline{\gamma }})e_{\beta }^{\underline{\gamma }}-A_{\beta }^{\underline{%
\mu }\underline{\gamma }}\left( D_{\tau }R_{\quad \sigma \mu }^{\underline{%
\gamma }\underline{\beta }}+\delta _{\tau }R_{\quad \sigma \mu }^{\underline{%
\gamma }\underline{\beta }}\right) e_{\beta }^{\underline{\beta }}-  \notag
\\
&&[A_{\nu }^{\underline{\mu }\underline{\gamma }}\left( D_{\tau }R_{\quad
\sigma \mu }^{\underline{\gamma }\underline{\beta }}+\delta _{\tau }R_{\quad
\sigma \mu }^{\underline{\gamma }\underline{\beta }}\right) +\left( D_{\tau
}R_{\quad \sigma \mu }^{\underline{\mu }\underline{\gamma }}+\delta _{\tau
}R_{\quad \sigma \mu }^{\underline{\mu }\underline{\gamma }}\right) A_{\nu
}^{\underline{\gamma }\underline{\beta }}]e_{\beta }^{\underline{\beta }}-
\notag
\end{eqnarray}%
\begin{eqnarray*}
&&e_{\beta }^{\underline{\beta }}\delta _{\sigma }\left[ A_{\nu }^{%
\underline{\mu }\underline{\gamma }}\left( \delta _{\tau }A_{\mu }^{%
\underline{\gamma }\underline{\beta }}+R_{\quad \tau \mu }^{\underline{%
\gamma }\underline{\beta }}\right) +\left( \delta _{\tau }A_{\mu }^{%
\underline{\mu }\underline{\gamma }}+R_{\quad \tau \mu }^{\underline{\mu }%
\underline{\gamma }}\right) A_{\nu }^{\underline{\gamma }\underline{\beta }}%
\right] +2\left( \delta _{\nu }A_{\beta }^{\underline{\mu }\underline{\gamma
}}\right) \delta _{(\tau }\delta _{\sigma )}e_{\mu }^{\underline{\gamma }}-
\\
&&A_{\beta }^{\underline{\mu }\underline{\gamma }}\delta _{\sigma }\left[
A_{\nu }^{\underline{\gamma }\underline{\beta }}\delta _{\tau }e_{\mu }^{%
\underline{\beta }}+\left( \delta _{\tau }A_{\mu }^{\underline{\gamma }%
\underline{\beta }}+R_{\quad \tau \mu }^{\underline{\gamma }\underline{\beta
}}\right) e_{\nu }^{\underline{\beta }}\right] -\left( \delta _{\nu
}e_{\beta }^{\underline{\gamma }}\right) \delta _{\tau }\left( \delta
_{\sigma }A_{\mu }^{\underline{\mu }\underline{\gamma }}+R_{\quad \sigma \mu
}^{\underline{\mu }\underline{\gamma }}\right) -
\end{eqnarray*}%
\begin{eqnarray*}
&&\left[ A_{\nu }^{\underline{\mu }\underline{\gamma }}\left( \delta _{\tau
}A_{\beta }^{\underline{\gamma }\underline{\beta }}+R_{\quad \tau \beta }^{%
\underline{\gamma }\underline{\beta }}\right) +\left( \delta _{\tau
}A_{\beta }^{\underline{\mu }\underline{\gamma }}+R_{\quad \tau \beta }^{%
\underline{\mu }\underline{\gamma }}\right) A_{\nu }^{\underline{\gamma }%
\underline{\beta }}\right] \delta _{\sigma }e_{\mu }^{\underline{\beta }}- \\
&&\left( \delta _{\sigma }A_{\mu }^{\underline{\mu }\underline{\gamma }%
}+R_{\quad \sigma \mu }^{\underline{\mu }\underline{\gamma }}\right) \left[
A_{\mu }^{\underline{\gamma }\underline{\beta }}\left( \delta _{\nu
}e_{\beta }^{\underline{\beta }}\right) +e_{\nu }^{\underline{\beta }}\left(
\delta _{\sigma }A_{\mu }^{\underline{\gamma }\underline{\beta }}+R_{\quad
\sigma \mu }^{\underline{\gamma }\underline{\beta }}\right) \right]
\}+O\left( \theta ^{3}\right) .
\end{eqnarray*}

Having the decompositions (\ref{qfr}), we can define the inverse vielbein $%
\widehat{e}_{\ast \underline{\mu }}^{\mu }$ from the equation
\begin{equation*}
\widehat{e}_{\ast \underline{\mu }}^{\mu }\ast \widehat{e}_{\mu }^{%
\underline{\nu }}=\delta _{\underline{\mu }}^{\underline{\nu }}
\end{equation*}%
and consequently compute $\theta $--deformations of connections, curvatures,
torsions and any type of actions and field equations (for simplicity, we
omit such cumbersome formulas being certain analogies to those computed in
Ref. 7 %
%\cite{ch1}%
but with additional N--deformations).

The main result of this section consists in formulation of a procedure
allowing to map exact solutions of a 'commutative' gravity model into a
corresponding 'noncommutative' model without introducing new fields. For
instance, we can take the data (\ref{data}) and (\ref{framel}) and construct
the $\theta $--deformation of the exact solution defining a static black
ellipsoid (a similar prescription works in transforming both real and
complex wormholes from Section V). The analysis presented in subsections IV
B and IV C illustrates a possibility to preserve the black ellipsoid
stability for a certain class of extensions of solutions to
noncommutative/complex gravity with complexified N--frames. In other turn,
if we consider arbitrary noncommutative relations for coordinates ($\theta $%
--noncommutativity) (\ref{coordnc}) the resulting $\theta $--deformation of
stable solution will be, in general, unstable because arbitrary
decompositions of type (\ref{qfr}) will induce arbitrary complex terms in
the metric, connection and curvature coefficients, i. e. will result in
complex terms in the ''inverse Schrodinger problem'' and related instability
(see Refs. 18,19). %
%\cite{anhtv6,anhtv7}).%
Perhaps, a certain class of stable $\theta $--deformed solutions can be
defined if we constrain the $\theta $--noncommutativity (\ref{coordnc}) to
be dual to the so--called anholnomic frame noncommutativity (\ref{anh}), or (%
\ref{anh1}), by connecting the nontrivial values of $\theta ^{\alpha \beta }$
to certain complex $N_{i}^{a}$ resulting in stable noncommutative
configurations, or (in the simplest case) to say that the noncommutative
extensions are modelled only by N--fields $N_{i}^{a}\sim \hbar \theta .$ The
resulting noncommutative extensions could be defined as to preserve
stability at least in the first order of $(\hbar \theta )$--terms.

\section{\quad Discussion and Conclusions}

With this paper we begin the investigation of spacetimes with anholonomic
noncommutative symmetry. The exact solutions we find are parametrized by
generic off--diagonal metric ansatz and anholonomic frame (vielbein)
structures with associated nonlinear connections (defining nontrivial
anholonomy relations and inducing natural matrix noncommutative differential
geometries). Their noncommutative symmetries are derived from exact
solutions of the field equations in the Einstein gravity theory and their
extra dimension and gauge like generalizations.

We analyzed the geometric and physical properties of new classes of exact
solutions with 'hidden' noncommutativity describing specific
non--perturbative vacuum and non--vacuum gravitational configurations. Such
spacetimes with generic off--diagonal metrics are very different, for
instance, from those possessing Killing symmetries. We also addressed to a
particular class of solutions with noncommutative symmetries defining static
black ellipsoid spacetimes which are not prohibited by the uniqueness black
hole theorems (proved for metrics with Killing symmetries and satisfying
asymptotic flat conditions) because the generic anholonomic noncommutative
configurations are very different from the Killing ones.

Let us comment on the difference between our approach and the former
elaborated ones: In the so--called Connes--Lot models$^{1,2},$\ %
% \cite{con1,con2},%
the gravitational models with Euclidean signature were elaborated from a
spectral analysis of Dirac operators connected to the noncommutative
geometry. This type of noncommutative geometry was constructed by replacing
the algebra of smooth function on a manifold with a more general associative
but noncommutative algebra. The fundamental matter field interactions and
Riemannian gravity were effectively derived from a corresponding spectral
calculus. In an alternative approach, the noncommutative geometry, as a low
energy noncommutativity of coordinates, can be obtained in string theory
because of presence of the so--called $B$--fields. A number of models of
gravity were proposed in order to satisfy certain noncommutativity relations
for coordinates and frames (of Lie group, or quantum group type, or by
computing some effective actions from string/ brane theory and
noncommutative gauge generalizations of gauge, Kaluza--Klein and Einstein
gravity), see Refs. 3-5,8.\ %
%\cite{ncg1,ncg2,ncg3,madore}.%
All mentioned noncommutative theories suppose that that the noncommutative
geometry transforms into a commutative one in some limits to the Einstein
theory and its extra dimension generalizations. In our approach we argue
that the existence of hidden noncommutative structures suggests a natural
way for constructing noncommutative models of gravitational interactions.

Our strategy explained in section \ref{ncgg} is quite different from the
previous attempts to construct the noncommutative gravity theory. We give a
proof and analyze some explicit examples illustrating that that there are
some specific hidden noncommutative geometric structures even in the
classical Einstein and gauge gravity models. This fact can be of fundamental
importance in constructing more general models of noncommutative gravity
with complex and nonsymmetric metrics.

Of course, there are two different notions of noncommutativity: The first
one is related to the spacetime deformations via Seiberg--Witten transforms
with noncommutative coordinates and the second one is associated to
noncommutative algebra modelled by anholonomic vielbeins. In general, the
result of such deformations and frame maps can not be distinguished exactly
on a resulting complex spacetime because there is a ''mixture'' of
coordinate, gauge and frame transforms in the case of noncommutative
geometry. Nevertheless, there are certain type of gravitational
configurations possessing Lie type (noncommutative) symmetries which
''survive'' in the limit of commutative coordinates and real valued metrics.
We say that a such type of solutions of the gravitational field equations
posses hidden noncommutative symmetries and describe a generic off--diagonal
class of metrics and anholonomic frame transforms. It is a very difficult
task to get exact solutions of the deformed gravity. In this work, we
succeeded to do this by generating such gravitational configurations which
are adapted both to the Seiberg--Witten type deformations and to the
vielbein transforms. In the section VII C we proved that there are possible
extensions (on deformation parameters) of exact solutions from the Einstein
and gauge gravity possessing hidden noncommutative symmetries without
introducing new fields.

In this work, from a number of results following from application of
Seiberg--Witten maps and related anholonomic vielbein transforms, we
selected only those which allow us to define classes of solutions as
noncommutative generalizations of some commutative ones of special physical
interest. Such metrics with hidden noncommutative symmetry are described by
a general off--diagonal ansatz for the metric and vielbein coefficients. The
solutions can be extended to complex metrics by allowing that some subsets
of vielbeins coefficients (with associated nonlinear connection structure)
may be complex valued. With respect to adapted frames, such metrics have
real coefficients describing vacuum black ellipsoid or wormhole
configurations (there were elaborated procedures of their analytical
extensions and proofs of stability). The new types of metrics may be
considered as certain exact solutions in complex gravity which have to be
considered if some noncommutative relations for coordinates are introduced.
\ Such configurations may play an important role in the further
understanding of vacua of noncommutative gauge and gravity theories and
investigation of their quantum variants.

The anholonomic noncommutative symmetry of exact solutions of four
dimensional (4D) vacuum Einstein equations positively does not violate the
local (real) Lorentz symmetry. This symmetry may be preserved in a specific
form even by anholonomic complex vielbein transforms mapping the 4D real
Einstein's metrics into certain complex ones for noncommutative gravity.
Such frames may be defined as the generated new solutions will be a formal
analogy with their real (diagonal) 'pedigrees', to be stable with well
defined geodesic and horizon properties, like it was concluded for black
ellipsoids solutions in general relativity$^{18,19}.$\ %
%\cite{anhtv6,anhtv7}.%
By complex frame transforms with noncommutative symmetries we demonstrated
that we may deform the horizon of the Schwarzshild solution to a static
ellipsoid configurations as well to induce an effective electric charge of
'complex noncommutative' origin.

We compare the generated off--diagonal ellipsoidal (in general, complex)
metrics possessing anholonomic noncommutative symmetries with those
describing the distorted diagonal black hole solutions (see the vacuum case
in Refs. 80,81\ %\cite{ms1,gh}%
and an extension to the case of non--vanishing electric fields in Ref. 82).
%
%\cite{fk}).%
In the complex ellipsoidal cases the spacetime distorsion is caused by some
anisotropic off--diagonal terms being non--trivial in some regions but in
the case of ''pure diagonal'' distorsions one treats such effects following
the fact that the vacuum Einstein equations are not satisfied in some
regions because of presence of matter. Alternatively, the complex ellipsoid
solutions may be described as in a 'real' world with real metric
coefficients but defined with respect to complex frames.

Here we emphasize that the off--diagonal gravity may model some
gravity--matter like interactions (for instance, in the Kaluza--Klein theory
by emphasizing some very particular metric's configurations and topological
compactifications) but, in general, the off--diagonal vacuum gravitational
dynamics can not be associated to any effective matter dynamics in a
holonomic gravitational background. So, we may consider that the anholonomic
ellipsoidal deformations of the Schwarzschild metric defined by real and/or
complex anholonomic frame transforms generate some kind of anisotropic
off--diagonal distorsions modelled by certain vacuum gravitational fields
with the distorsion parameters (equivalently, vacuum gravitational
polarizations) depending both on radial, angular and extra dimension
coordinates. For complex valued nonlinear connection coefficients, we obtain
a very specific complex spacetime distorsion instead of matter type
distorsion. We note that both classes of off--diagonal anisotropic and
''pure'' diagonal distorsions (like in Refs. 80,81) %
%\cite{ms1,gh})%
result in solutions which in general are not asymptotically flat. However,
it is possible to find asymptotically flat extensions, as it was shown in
this paper and in Refs. 18,19, %
%\cite{anhtv6,anhtv7},%
even for ellipsoidal configurations by introducing the corresponding
off--diagonal terms. The asymptotic conditions for the diagonal distorsions
are discussed in Ref. 82 %
% \cite{fk} %
where it was suggested that to satisfy such conditions one has to include
some additional matter fields in the exterior portion of spacetime. For
ellipsoidal real/complex metrics, we should consider that the off--diagonal
metric terms have a corresponding behavior as to result fare away from the
horizon in the Minkowski metric.

The deformation parameter $\varepsilon $ effectively seems to put an
''electric charge'' on the black hole which is of gravitational
off--diagonal/anholonomic origin. For complex metrics such ''electric
charges'' may be induced by complex values of off--diagonal metric/
anholonomic frame coefficients. It can describe effectively both positive
and negative gravitational polarizations (even some repulsive gravitational
effects). The polarization may have very specific nonlinearities induced by
complex gravity terms. This is not surprising because the coefficients of an
anisotropic black hole are similar to those of the Reissner--Nordstrom
solution only with respect to corresponding anholonomic complex/ real frames
which are subjected to some constraints (anholonomy conditions).

Applying the method of anholonomic frame transforms elaborated and developed
in Refs. 13-19,24-31, %
% \cite{anhtv1,anhtv2,anhtv3,anhtv4,anhtv5,anhtv6,anhtv7,v41,v42,v43,v44,v45,v5,vmethod1,vmethod2},%
we can construct off--diagonal ellipsoidal extensions of the already
diagonally disturbed Schwarzschild metric (see the metric (3.7) from Ref.
82). %%
Such anholonomic deformations would contain in the diagonal limit
configurations with $\varepsilon \rightarrow 0$ but $\eta _{3}\neq 1$ (see (%
\ref{data10}) and/or (\ref{data}) and (\ref{aux4s})) for such configurations
the function $\eta _{3}$ has to be related in the corresponding limits with
the values $\overline{\gamma }_{D},\overline{\psi }_{D}$ and $A$ from Ref.
82). %
%\cite{fk}.%
We remark that there are different classes of ellipsoidal deformations of
the Schwarschild metric which result in real or complex vacuum
configuration. The conditions $\varepsilon \rightarrow 0$ and $q,\eta _{3}=1$
and $\lambda \rightarrow 0$ select just the limit of the usual radial
Schwarschild asymptotics without any (also possible) additional diagonal
distorsions. For nontrivial values of $q,\eta _{3}$ and $\eta _{4}$ we may
obtain diagonal distorsions.

In the case of ellipsoidal metrics with the Schwarzschild asymptotics, the
ellipsoidal character could result in some observational effects in the
vicinity of the horizon. For instance, scattering of particles on a static
ellipsoid can be computed. We can can also model anisotropic matter
accretion effects on an ellipsoidal black hole put in the center of a
galactic being of ellipsoidal, toroidal or another configuration. Even in 4D
the nonshperic topology of horizons seem to be prohibited in the general
relativity theory$^{52,83-86,20}$ \ %
%\cite{haw,tors1,gall,cwald,jvent,heus}%
following some general differential geometry and cenzorship theorems, such
objects can be induced from extra dimensions and can exist in theories with
cosmological constant $^{87-89},$ % \cite{lem1,lem2,lem3},%
nontrivial torsion or induced by anholonomic frames$^{13-19,30,31}. $\
%\cite{anhtv1,anhtv2,anhtv3,anhtv4,anhtv5,anhtv6,anhtv7,vmethod1,vmethod2}.
We can consider black torus/ellipsoid solutions as a background for
potential tests for existence of extra dimensions and of general relativity.
A point of further investigations could be the anisotropic ellipsoidal
collapse when both the matter and spacetime are of ellipsoidal generic
off--diagonal symmetry (former theoretical and computational investigations
were performed only for rotoids with anisotropic matter and particular
classes of perturbations of the Schwarzshild solutions$^{90}$). %
% \cite{st}). %
It is interesting to elaborate collapse scenaria with respect to compexified
anholonomic frames. For very small eccentricities, we may not have any
observable effects like perihelion shift or light bending if we restrict our
investigations only to the Schwarzshild--Newton asymptotics but some kind of
dissipation can be considered for complex metrics.

We also present some comments on mechanics and thermodynamics of ellipsoidal
black holes. For the static black ellipsoids/tori with flat asymptotics, we
can compute the area of the ellipsoidal horizon, associate an entropy and
develop a corresponding black ellipsoid thermodynamics. But this is a rough
approximation because, in general, we are dealing with off--diagonal metrics
depending anisotropically on two/three coordinates. Such solutions are with
anholonomically deformed Killing horizons, or with anholonomic
noncommutative symmetries, and should be described by a thermodynamics (in
general, both non-equilibrium and irreversible) of black ellipsoids/tori
self--consistently embedded into an off--diagonal anisotropic gravitational
vacuum with potential dissipation described by some complex metric and frame
coefficients. This forms a ground for numerous new conceptual issues to be
developed and related to anisotropic black holes and the anisotropic
kinetics and thermodynamics$^{29}$\ %\cite{v5}%
as well to a framework of isolated anisotropic horizons$^{91-93}$,\ %
%\cite{asht1,asht2,asht3},%
defined in a locally anistoropic/ noncommutative / complex background with
wormhole real and/or complex configurations which is a matter of our further
investigations.

Finally, we note that we elaborated a general formalism of generating
noncommutative solutions starting from exact vacuum solutions with
anholonomic noncommutativity, but we do not know how to extend our solutions
via star (Moyal) product as to preserve their stability because of induced
general complex terms in the metrics. For some particular duality relations
between the coordinate and frame noncommutativity it seems possible to get
stability at least in the first approximation of noncommutative deformation
parameter but an arbitrary noncommutative coordinate relation results in
less defined physical configurations. A better understanding of the physical
relevance of the anholonomic noncommutative configurations completed also to
general coordinate noncommutativity is an interesting open question which we
leave for the future.

\subsection*{Acknowledgements}

~~ The work is partially supported by a NATO/ Portugal fellowship and by a
sabbatical research grant of the Ministry of Education and Science of Spain.

\appendix

\section{\quad Einstein Equations and N--Connections}

For convenience, we present in this Appendix a selection of necessary
results from Refs. 13-19,30,31: %
%\cite{anhtv1,anhtv2,anhtv3,anhtv4,anhtv5,anhtv6,anhtv7,vmethod1,vmethod2}:%

The curvature tensor of a connection $\Gamma ^{\lbrack c]}$ with
h--v--components (\ref{dcon}) induced by anholonomic frames (\ref{ndif}) and
(\ref{nder}) with associated N--connection structure is defined $R(\delta
_{\tau },\delta _{\gamma })\delta _{\beta }=R_{\beta \ \gamma \tau }^{\
\alpha }\delta _{\alpha }$ where
\begin{equation}
R_{\beta \ \gamma \tau }^{\ \alpha }=\delta _{\tau }\Gamma _{\ \beta \gamma
}^{\alpha }-\delta _{\gamma }\Gamma _{\ \beta \tau }^{\alpha }+\Gamma _{\
\beta \gamma }^{\varphi }\Gamma _{\ \varphi \tau }^{\alpha }-\Gamma _{\
\beta \tau }^{\varphi }\Gamma _{\ \varphi \gamma }^{\alpha }+\Gamma _{\
\beta \varphi }^{\alpha }w_{\ \gamma \tau }^{\varphi },  \label{curvature}
\end{equation}%
splits into irreducible h--v--components $R_{\beta \ \gamma \tau }^{\ \alpha
}=%
\{R_{h.jk}^{.i},R_{b.jk}^{.a},P_{j.ka}^{.i},P_{b.ka}^{.c},S_{j.bc}^{.i},S_{b.cd}^{.a}\},
$\ with
\begin{eqnarray}
R_{h.jk}^{.i} &=&\delta _{k}L_{.hj}^{i}-\delta
_{j}L_{.hk}^{i}+L_{.hj}^{m}L_{mk}^{i}-L_{.hk}^{m}L_{mj}^{i}-C_{.ha}^{i}%
\Omega _{.jk}^{a},  \label{dcurvatures} \\
R_{b.jk}^{.a} &=&\delta _{k}L_{.bj}^{a}-\delta
_{j}L_{.bk}^{a}+L_{.bj}^{c}L_{.ck}^{a}-L_{.bk}^{c}L_{.cj}^{a}-C_{.bc}^{a}%
\Omega _{.jk}^{c},  \notag \\
P_{j.ka}^{.i} &=&\partial _{a}L_{.jk}^{i}+C_{.jb}^{i}T_{.ka}^{b}-(\delta
_{k}C_{.ja}^{i}+L_{.lk}^{i}C_{.ja}^{l}-L_{.jk}^{l}C_{.la}^{i}-L_{.ak}^{c}C_{.jc}^{i}),
\notag \\
P_{b.ka}^{.c} &=&\partial _{a}L_{.bk}^{c}+C_{.bd}^{c}T_{.ka}^{d}-(\delta
_{k}C_{.ba}^{c}+L_{.dk}^{c\
}C_{.ba}^{d}-L_{.bk}^{d}C_{.da}^{c}-L_{.ak}^{d}C_{.bd}^{c}),  \notag \\
S_{j.bc}^{.i} &=&\partial _{c}C_{.jb}^{i}-\partial
_{b}C_{.jc}^{i}+C_{.jb}^{h}C_{.hc}^{i}-C_{.jc}^{h}C_{hb}^{i},  \notag \\
S_{b.cd}^{.a} &=&\partial _{d}C_{.bc}^{a}-\partial
_{c}C_{.bd}^{a}+C_{.bc}^{e}C_{.ed}^{a}-C_{.bd}^{e}C_{.ec}^{a},  \notag
\end{eqnarray}%
where we omitted the label $[c]$ in formulas. The Ricci tensor $R_{\beta
\gamma }=R_{\beta ~\gamma \alpha }^{~\alpha }$ has the \ irreducible
h--v--components
\begin{eqnarray}
R_{ij} &=&R_{i.jk}^{.k},\quad R_{ia}=-^{2}P_{ia}=-P_{i.ka}^{.k},
\label{dricci} \\
R_{ai} &=&^{1}P_{ai}=P_{a.ib}^{.b},\quad R_{ab}=S_{a.bc}^{.c}.  \notag
\end{eqnarray}%
This tensor is non-symmetric, $^{1}P_{ai}\neq ~^{2}P_{ia}$ (this could be
with respect to anholonomic frames of reference). The scalar curvature of
the metric d--connection, $\overleftarrow{R}=g^{\beta \gamma }R_{\beta
\gamma },$ is computed
\begin{equation}
{\overleftarrow{R}}=G^{\alpha \beta }R_{\alpha \beta }=\widehat{R}+S,
\label{dscalar}
\end{equation}%
where $\widehat{R}=g^{ij}R_{ij}$ and $S=h^{ab}S_{ab}.$

By substituting (\ref{dricci}) and (\ref{dscalar}) into the 5D Einstein
equations
\begin{equation*}
R_{\alpha \beta }-\frac{1}{2}g_{\alpha \beta }R=\kappa \Upsilon _{\alpha
\beta },
\end{equation*}%
where $\kappa $ and $\Upsilon _{\alpha \beta }$ are respectively the
coupling constant and the energy--momentum tensor, we obtain the
h-v-decomposition of the Einstein equations
\begin{eqnarray}
R_{ij}-\frac{1}{2}\left( \widehat{R}+S\right) g_{ij} &=&\kappa \Upsilon
_{ij},  \label{einsteq2} \\
S_{ab}-\frac{1}{2}\left( \widehat{R}+S\right) h_{ab} &=&\kappa \Upsilon
_{ab},  \notag \\
^{1}P_{ai}=\kappa \Upsilon _{ai},\ ^{2}P_{ia} &=&\kappa \Upsilon _{ia}.
\notag
\end{eqnarray}

The vacuum 5D gravitational field equations, in invariant h--v--components,
are written
\begin{equation}
R_{ij}=0,S_{ab}=0,^{1}P_{ai}=0,\ ^{2}P_{ia}=0.  \label{einsteq3}
\end{equation}

The main `trick' of the anholonomic frames method of integrating the
Einstein equations in general relativity and various (super) string and
higher / lower dimension gravitational theories consist in a procedure of
definition of such coefficients $N_{j}^{a}$ such that the block matrices $%
g_{ij}$ and $h_{ab}$ are diagonalized. This substantially simplifies
computations but we have to apply N--elongated partial derivatives.

\section{\quad Main Theorems for 5D}

We restrict our considerations to a five dimensional (in brief, 5D) \
spacetime provided with a generic off--diagonal (pseudo) Riemannian metric
and labeled by local coordinates $u^{\alpha }=(x^{i},y^{4}=v,y^{5}),$ for $%
i=1,2,3.$ We state the condition when exact solutions of the Einstein
equations depending on holonomic variables $x^{i}$ and on one anholonomic
(equivalently, anisotropic) variable $y^{4}=v$ can be constructed in
explicit form. Every coordinate from a set $u^{\alpha }$ can may be time
like, 3D space like, or extra dimensional. For simplicity, the partial
derivatives will be denoted like $a^{\times }=\partial a/\partial
x^{1},a^{\bullet }=\partial a/\partial x^{2},a^{^{\prime }}=\partial
a/\partial x^{3},a^{\ast }=\partial a/\partial v.$

The 5D quadratic line element is chosen
\begin{equation}
ds^{2}=g_{\alpha \beta }\left( x^{i},v\right) du^{\alpha }du^{\beta }
\label{metric}
\end{equation}%
when the metric components $g_{\alpha \beta }$ are parametrized with respect
to the coordinate dual basis by an off--diagonal matrix (ansatz) {\
%%\footnotesize
\begin{equation}
\left[
\begin{array}{ccccc}
g_{1}+w_{1}^{\ 2}h_{4}+n_{1}^{\ 2}h_{5} & w_{1}w_{2}h_{4}+n_{1}n_{2}h_{5} &
w_{1}w_{3}h_{4}+n_{1}n_{3}h_{5} & w_{1}h_{4} & n_{1}h_{5} \\
w_{1}w_{2}h_{4}+n_{1}n_{2}h_{5} & g_{2}+w_{2}^{\ 2}h_{4}+n_{2}^{\ 2}h_{5} &
w_{2}w_{3}h_{4}+n_{2}n_{3}h_{5} & w_{2}h_{4} & n_{2}h_{5} \\
w_{1}w_{3}h_{4}+n_{1}n_{3}h_{5} & w_{2}w_{3}h_{4}+n_{2}n_{3}h_{5} &
g_{3}+w_{3}^{\ 2}h_{4}+n_{3}^{\ 2}h_{5} & w_{3}h_{4} & n_{3}h_{5} \\
w_{1}h_{4} & w_{2}h_{4} & w_{3}h_{4} & h_{4} & 0 \\
n_{1}h_{5} & n_{2}h_{5} & n_{3}h_{5} & 0 & h_{5}%
\end{array}%
\right] ,  \label{ansatz}
\end{equation}%
} with the coefficients being some necessary smoothly class functions of
type
\begin{eqnarray}
g_{1} &=&\pm 1,g_{2,3}=g_{2,3}(x^{2},x^{3}),h_{4,5}=h_{4,5}(x^{i},v),  \notag
\\
w_{i} &=&w_{i}(x^{i},v),n_{i}=n_{i}(x^{i},v),  \notag
\end{eqnarray}%
where the $N$--coefficients from (\ref{nder}) and (\ref{ndif}) are
parametrized $N_{i}^{4}=w_{i}$ and $N_{i}^{5}=n_{i}.$

By straightforward calculation, we can prove$^{30,31}:$

\begin{theorem}
The nontrivial components of the 5D vacuum Einstein equations, $R_{\alpha
}^{\beta }=0$ (see (\ref{einsteq3}) in the Appendix A) for the metric (\ref%
{dmetric}) defined by the ansatz (\ref{ansatz}), computed with respect to
anholonomic frames (\ref{ndif}) and (\ref{nder}) can be written in the form:
\begin{eqnarray}
R_{2}^{2}=R_{3}^{3}=-\frac{1}{2g_{2}g_{3}}[g_{3}^{\bullet \bullet }-\frac{%
g_{2}^{\bullet }g_{3}^{\bullet }}{2g_{2}}-\frac{(g_{3}^{\bullet })^{2}}{%
2g_{3}}+g_{2}^{^{\prime \prime }}-\frac{g_{2}^{^{\prime }}g_{3}^{^{\prime }}%
}{2g_{3}}-\frac{(g_{2}^{^{\prime }})^{2}}{2g_{2}}] &=&0,  \label{ricci1a} \\
S_{4}^{4}=S_{5}^{5}=-\frac{\beta }{2h_{4}h_{5}} &=&0,  \label{ricci2a} \\
R_{4i}=-w_{i}\frac{\beta }{2h_{5}}-\frac{\alpha _{i}}{2h_{5}} &=&0,
\label{ricci3a} \\
R_{5i}=-\frac{h_{5}}{2h_{4}}\left[ n_{i}^{\ast \ast }+\gamma n_{i}^{\ast }%
\right] &=&0,  \label{ricci4a}
\end{eqnarray}%
where
\begin{equation}
\alpha _{i}=\partial _{i}{h_{5}^{\ast }}-h_{5}^{\ast }\partial _{i}\ln \sqrt{%
|h_{4}h_{5}|},\beta =h_{5}^{\ast \ast }-h_{5}^{\ast }[\ln \sqrt{|h_{4}h_{5}|}%
]^{\ast },\gamma =3h_{5}^{\ast }/2h_{5}-h_{4}^{\ast }/h_{4}.  \label{abc}
\end{equation}
\end{theorem}

Following this theorem, 1) we can define a function $g_{2}(x^{2},x^{3})$ for
a given $g_{3}(x^{2},x^{3}),$ or inversely, to define a function $%
g_{2}(x^{2},x^{3})$ for a given $g_{3}(x^{2},x^{3}),$ from equation (\ref%
{ricci1a}); 2) we can define a function $h_{4}(x^{1},x^{2},x^{3},v)$ for a
given $h_{5}(x^{1},x^{2},x^{3},v),$ or inversely, to define a function $%
h_{5}(x^{1},x^{2},x^{3},v)$ for a given $h_{4}(x^{1},x^{2},x^{3},v),$ from
equation (\ref{ricci2a}); 3-4) having the values of $h_{4}$ and $h_{5},$ we
can compute the coefficients (\ref{abc}) which allow to solve the algebraic
equations (\ref{ricci3a}) and to integrate two times on $v$ the equations (%
\ref{ricci4a}) which allow to find respectively the coefficients $%
w_{i}(x^{k},v)$ and $n_{i}(x^{k},v).$

We can generalize the construction by introducing a conformal factor $\Omega
(x^{i},v)$ and additional deformations of the metric via coefficients $\zeta
_{\hat{\imath}}(x^{i},v)$ (here, the indices with 'hat' take values like $%
\hat{{i}}=1,2,3,5),$ i. e. for metrics of type
\begin{equation}
ds^{2}=\Omega ^{2}(x^{i},v)\hat{{g}}_{\alpha \beta }\left( x^{i},v\right)
du^{\alpha }du^{\beta },  \label{cmetric}
\end{equation}%
were the coefficients $\hat{{g}}_{\alpha \beta }$ are parametrized by the
ansatz {\scriptsize
\begin{equation}
\left[
\begin{array}{ccccc}
g_{1}+(w_{1}^{\ 2}+\zeta _{1}^{\ 2})h_{4}+n_{1}^{\ 2}h_{5} &
(w_{1}w_{2}+\zeta _{1}\zeta _{2})h_{4}+n_{1}n_{2}h_{5} & (w_{1}w_{3}+\zeta
_{1}\zeta _{3})h_{4}+n_{1}n_{3}h_{5} & (w_{1}+\zeta _{1})h_{4} & n_{1}h_{5}
\\
(w_{1}w_{2}+\zeta _{1}\zeta _{2})h_{4}+n_{1}n_{2}h_{5} & g_{2}+(w_{2}^{\
2}+\zeta _{2}^{\ 2})h_{4}+n_{2}^{\ 2}h_{5} & (w_{2}w_{3}++\zeta _{2}\zeta
_{3})h_{4}+n_{2}n_{3}h_{5} & (w_{2}+\zeta _{2})h_{4} & n_{2}h_{5} \\
(w_{1}w_{3}+\zeta _{1}\zeta _{3})h_{4}+n_{1}n_{3}h_{5} & (w_{2}w_{3}+\zeta
_{2}\zeta _{3})h_{4}+n_{2}n_{3}h_{5} & g_{3}+(w_{3}^{\ 2}+\zeta _{3}^{\
2})h_{4}+n_{3}^{\ 2}h_{5} & (w_{3}+\zeta _{3})h_{4} & n_{3}h_{5} \\
(w_{1}+\zeta _{1})h_{4} & (w_{2}+\zeta _{2})h_{4} & (w_{3}+\zeta _{3})h_{4}
& h_{4} & 0 \\
n_{1}h_{5} & n_{2}h_{5} & n_{3}h_{5} & 0 & h_{5}+\zeta _{5}h_{4}%
\end{array}%
\right] .  \label{ansatzc}
\end{equation}%
}Such 5D metrics have a second order anisotropy$^{32,37,38}$
%\cite{vnp,ma1,m2}%
when the $N$--coefficients are paramet\-riz\-ed in the first order
anisotropy like $N_{i}^{4}=w_{i}$ and $N_{i}^{5}=n_{i}$ (with three
anholonomic, $x^{i},$ and two anholonomic, $y^{4}$ and $y^{5},$ coordinates)
and in the second order anisotropy (on the second 'shell', \ with four
anholonomic, $(x^{i},y^{5}),$ and one anholonomic,$y^{4},$ coordinates) with
$N_{\hat{{i}}}^{5}=\zeta _{\hat{{i}}},$ in this work we state, for
simplicity, $\zeta _{{5}}=0.$ For trivial values $\Omega =1$ and $\zeta _{%
\hat{\imath}}=0,$ the squared line interval (\ref{cmetric}) transforms into (%
\ref{metric}).

The Theorem 1 can be extended as to include the generalization to the second
ansatz (\ref{cmetric}):

\begin{theorem}
The nontrivial components of the 5D vacuum Einstein equations, $R_{\alpha
}^{\beta }=0,$ (see (\ref{einsteq3}) in the Appendix A) for the metric (\ref%
{cmetric}) consist from the system (\ref{ricci1a})--(\ref{ricci4a}) with the
additional conditions that
\begin{equation}
\hat{{\delta }}_{i}h_{4}=0\mbox{\ and\  }\hat{{\delta }}_{i}\Omega =0
\label{conf1}
\end{equation}%
for $\hat{{\delta }}_{i}=\partial _{i}-\left( w_{i}+\zeta _{i}\right)
\partial _{4}+n_{i}\partial _{5}$ when the values $\zeta _{\widetilde{i}%
}=\left( \zeta _{{i}},\zeta _{{5}}=0\right) $ are to be found as to be a
solution of (\ref{conf1}); for instance, if
\begin{equation}
\Omega ^{q_{1}/q_{2}}=h_{4}~(q_{1}\mbox{ and }q_{2}\mbox{ are
integers}),  \label{confq}
\end{equation}%
$\zeta _{{i}}$ satisfy the equations \
\begin{equation}
\partial _{i}\Omega -(w_{i}+\zeta _{{i}})\Omega ^{\ast }=0.  \label{confeq}
\end{equation}
\end{theorem}

The proof of Theorem 2 consists from a straightforward calculation of the
components of the Ricci tensor (\ref{dricci}) as it is outlined in the
Appendix A. The simplest way is to use the calculus for Theorem 1 and then
to compute deformations of the canonical connection (\ref{dcon}). \ Such
deformations induce corresponding deformations of the Ricci tensor (\ref%
{dricci}). \ The condition that we have the same values of the Ricci tensor
for the (\ref{ansatz}) and (\ref{ansatzc}) results in equations (\ref{conf1}%
) and (\ref{confeq}) which are compatible, for instance, if $\Omega
^{q_{1}/q_{2}}=h_{4}.$\ There are also another possibilities to satisfy the
condition (\ref{conf1}), for instance, if $\Omega =\Omega _{1}$ $\Omega
_{2}, $ we can consider that $h_{4}=\Omega _{1}^{q_{1}/q_{2}}$ $\Omega
_{2}^{q_{3}/q_{4}}$ $\ $for some integers $q_{1},q_{2},q_{3}$ and $q_{4}.$

A very surprising result is that we are able to construct exact solutions of
the 5D vacuum Einstein equations for both types of the ansatz (\ref{ansatz})
and (\ref{ansatzc}):

\begin{theorem}
The system of second order nonlinear partial differential equations (\ref%
{ricci1a})--(\ref{ricci4a}) and (\ref{confeq}) can be solved in general form
if there are given some values of functions $g_{2}(x^{2},x^{3})$ (or $%
g_{3}(x^{2},x^{3})),h_{4}\left( x^{i},v\right) $ (or $h_{5}\left(
x^{i},v\right) )$ and $\Omega \left( x^{i},v\right) :$

\begin{itemize}
\item The general solution of equation (\ref{ricci1a}) can be written in the
form
\begin{equation}
\varpi =g_{[0]}\exp [a_{2}\widetilde{x}^{2}\left( x^{2},x^{3}\right) +a_{3}%
\widetilde{x}^{3}\left( x^{2},x^{3}\right) ],  \label{solricci1a}
\end{equation}%
were $g_{[0]},a_{2}$ and $a_{3}$ are some constants and the functions $%
\widetilde{x}^{2,3}\left( x^{2},x^{3}\right) $ define any coordinate
transforms $x^{2,3}\rightarrow \widetilde{x}^{2,3}$ for which the 2D line
element becomes conformally flat, i. e.
\begin{equation}
g_{2}(x^{2},x^{3})(dx^{2})^{2}+g_{3}(x^{2},x^{3})(dx^{3})^{2}\rightarrow
\varpi \left[ (d\widetilde{x}^{2})^{2}+\epsilon (d\widetilde{x}^{3})^{2}%
\right] .  \label{con10}
\end{equation}

\item The equation (\ref{ricci2a}) relates two functions $h_{4}\left(
x^{i},v\right) $ and $h_{5}\left( x^{i},v\right) $ following two
possibilities:

a) to compute
\begin{eqnarray}
\sqrt{|h_{5}|} &=&h_{5[1]}\left( x^{i}\right) +h_{5[2]}\left( x^{i}\right)
\int \sqrt{|h_{4}\left( x^{i},v\right) |}dv,~h_{4}^{\ast }\left(
x^{i},v\right) \neq 0;  \notag \\
&=&h_{5[1]}\left( x^{i}\right) +h_{5[2]}\left( x^{i}\right) v,h_{4}^{\ast
}\left( x^{i},v\right) =0,  \label{p2}
\end{eqnarray}
for some functions $h_{5[1,2]}\left( x^{i}\right) $ stated by boundary
conditions;

b) or, inversely, to compute $h_{4}$ for a given $h_{5}\left( x^{i},v\right)
,h_{5}^{\ast }\neq 0,$%
\begin{equation}
\sqrt{|h_{4}|}=h_{[0]}\left( x^{i}\right) (\sqrt{|h_{5}\left( x^{i},v\right)
|})^{\ast },  \label{p1}
\end{equation}
with $h_{[0]}\left( x^{i}\right) $ given by boundary conditions.

\item The exact solutions of (\ref{ricci3a}) for $\beta \neq 0$ is
\begin{equation}
w_{k}=\partial _{k}\ln [\sqrt{|h_{4}h_{5}|}/|h_{5}^{\ast }|]/\partial
_{v}\ln [\sqrt{|h_{4}h_{5}|}/|h_{5}^{\ast }|],  \label{w}
\end{equation}
with $\partial _{v}=\partial /\partial v$ and $h_{5}^{\ast }\neq 0.$ If $%
h_{5}^{\ast }=0,$ or even $h_{5}^{\ast }\neq 0$ but $\beta =0,$ the
coefficients $w_{k}$ could be arbitrary functions on $\left( x^{i},v\right)
. $ \ For vacuum Einstein equations this is a degenerated case which imposes
the the compatibility conditions $\beta =\alpha _{i}=0,$ which are
satisfied, for instance, if the $h_{4}$ and $h_{5}$ are related as in the
formula (\ref{p1}) but with $h_{[0]}\left( x^{i}\right) =const.$

\item The exact solution of (\ref{ricci4a}) is
\begin{eqnarray}
n_{k} &=&n_{k[1]}\left( x^{i}\right) +n_{k[2]}\left( x^{i}\right) \int
[h_{4}/(\sqrt{|h_{5}|})^{3}]dv,~h_{5}^{\ast }\neq 0;  \notag \\
&=&n_{k[1]}\left( x^{i}\right) +n_{k[2]}\left( x^{i}\right) \int
h_{4}dv,\qquad ~h_{5}^{\ast }=0;  \label{n} \\
&=&n_{k[1]}\left( x^{i}\right) +n_{k[2]}\left( x^{i}\right) \int [1/(\sqrt{%
|h_{5}|})^{3}]dv,~h_{4}^{\ast }=0,  \notag
\end{eqnarray}
for some functions $n_{k[1,2]}\left( x^{i}\right) $ stated by boundary
conditions.

\item The exact solution of (\ref{confeq}) is given by some arbitrary
functions $\zeta _{i}=\zeta _{i}\left( x^{i},v\right) $ if \ both $\partial
_{i}\Omega =0$ and $\Omega ^{\ast }=0,$ we chose $\zeta _{i}=0$ for $\Omega
=const,$ and
\begin{eqnarray}
\zeta _{i} &=&-w_{i}+(\Omega ^{\ast })^{-1}\partial _{i}\Omega ,\quad \Omega
^{\ast }\neq 0,  \label{confsol} \\
&=&(\Omega ^{\ast })^{-1}\partial _{i}\Omega ,\quad \Omega ^{\ast }\neq 0,%
\mbox{ for vacuum solutions}.  \notag
\end{eqnarray}
\end{itemize}
\end{theorem}

\bigskip We note that a transform (\ref{con10}) is always possible for 2D
metrics and the explicit form of solutions depends on chosen system of 2D
coordinates and on the signature $\epsilon =\pm 1.$ In the simplest case the
equation (\ref{ricci1a}) is solved by arbitrary two functions $g_{2}(x^{3})$
and $g_{3}(x^{2}).$ The equation (\ref{ricci2a}) is satisfied by arbitrary
pairs of coefficients $h_{4}\left( x^{i},v\right) $ and $h_{5[0]}\left(
x^{i}\right) .$

The proof of Theorem 3, following from a direct integration of (\ref{ricci1a}%
)--(\ref{ricci4a}) and (\ref{confeq}) is given in the Appendix B of Ref. 30.
%
% \cite{vmethod1}.%

There are some important consequences of the Theorems 1--3:

\begin{corollary}
The non--trivial diagonal components of the Einstein tensor, $G_{\beta
}^{\alpha }=R_{\beta }^{\alpha }-\frac{1}{2}R\quad \delta _{\beta }^{\alpha
},$ for the metric (\ref{dmetric}), given with respect to N--frames, are
\begin{equation}
G_{1}^{1}=-\left( R_{2}^{2}+S_{4}^{4}\right)
,G_{2}^{2}=G_{3}^{3}=-S_{4}^{4},G_{4}^{4}=G_{5}^{5}=-R_{2}^{2}
\label{einstdiag}
\end{equation}%
imposing the condition that the dynamics is defined by two values $R_{2}^{2}$
and $S_{4}^{4}.$ The rest of non--diagonal components of the Ricci (Einstein
tensor) are compensated by fixing corresponding values of N--coefficients.
\end{corollary}

The formulas (\ref{einstdiag}) are obtained following the relations for the
Ricci tensor (\ref{ricci1a})--(\ref{ricci4a}).

\begin{corollary}
We can extend the system of 5D vacuum Einstein equations (\ref{ricci1a})--(%
\ref{ricci4a}) by introducing matter fields for which the coefficients of
the energy--momentum tensor $\Upsilon _{\alpha \beta }$ given with respect
to N-- frames satisfy the conditions
\begin{equation}
\Upsilon _{1}^{1}=\Upsilon _{2}^{2}+\Upsilon _{4}^{4},\Upsilon
_{2}^{2}=\Upsilon _{3}^{3},\Upsilon _{4}^{4}=\Upsilon _{5}^{5}.
\label{emcond}
\end{equation}
\end{corollary}

We note that, in general, the tensor $\Upsilon _{\alpha \beta }$ for the
non--vacuum Einstein equations,
\begin{equation*}
R_{\alpha \beta }-\frac{1}{2}g_{\alpha \beta }R=\kappa \Upsilon _{\alpha
\beta },
\end{equation*}%
is not symmetric because with respect to anholonomic frames there are
imposed constraints which makes non symmetric the Ricci and Einstein tensors
(the symmetry conditions may be defined explicitly only with respect to
holonomic, coordinate frames; for details see the Appendix A and the
formulas (\ref{einsteq2})).

For simplicity, in our investigations we can consider only diagonal matter
sources, given with respect to N--frames, satisfying the conditions
\begin{equation}
\kappa \Upsilon _{2}^{2}=\kappa \Upsilon _{3}^{3}=\Upsilon _{2},\kappa
\Upsilon _{4}^{4}=\kappa \Upsilon _{5}^{5}=\Upsilon _{4},\mbox{ and }%
\Upsilon _{1}=\Upsilon _{2}+\Upsilon _{4},  \label{diagemt}
\end{equation}%
where $\kappa $ is the gravitational coupling constant. In this case the
equations (\ref{ricci1a}) and (\ref{ricci2a}) are respectively generalized
to
\begin{equation}
R_{2}^{2}=R_{3}^{3}=-\frac{1}{2g_{2}g_{3}}[g_{3}^{\bullet \bullet }-\frac{%
g_{2}^{\bullet }g_{3}^{\bullet }}{2g_{2}}-\frac{(g_{3}^{\bullet })^{2}}{%
2g_{3}}+g_{2}^{^{\prime \prime }}-\frac{g_{2}^{^{\prime }}g_{3}^{^{\prime }}%
}{2g_{3}}-\frac{(g_{2}^{^{\prime }})^{2}}{2g_{2}}]=-\Upsilon _{4}
\label{ricci1b}
\end{equation}%
and
\begin{equation}
S_{4}^{4}=S_{5}^{5}=-\frac{\beta }{2h_{4}h_{5}}=-\Upsilon _{2}.
\label{ricci2b}
\end{equation}

\begin{corollary}
An arbitrary solution of the system of equations (\ref{ricci1a})--(\ref%
{ricci4a}) and (\ref{confeq}) is defined for a canonical connection (\ref%
{dcon}) containing, in general, non-trivial torsion coefficients. This can
be effectively applied in order to construct exact solutions, for instance,
in string gravity containing nontrivial torsion. We can select solutions
corresponding to the Levi--Civita connection (\ref{lcc}) for a generic
off--diagonal (pseudo) Riemannian metric if we impose the condition that the
coefficients $N_{i}^{4}=w_{i}(x^{k},v),$ $N_{i}^{5}=$ $n_{i}(x^{k},v)$ and $%
N_{\hat{{i}}}^{5}=\zeta _{\hat{{i}}}$ are fixed to result in a zero
N--curvature, $\Omega _{jk}^{a}=0,$ on all ''shells'' of anisotropy. Such
selections are possible by fixing corresponding boundary conditions and
selecting corresponding classes of functions like $n_{k[1,2]}\left(
x^{i}\right) ,$ obtained after a general integration, in formulas (\ref{w}),
(\ref{n}) and (\ref{confsol}).
\end{corollary}

The above presented results are for generic 5D off--diagonal metrics,
anholonomic transforms and nonlinear field equations. Reductions to a lower
dimensional theory are not trivial in such cases. We emphasize here some
specific points of this procedure (see details in Ref. 30).
%\cite{vmethod1}).

\section{\quad Reduction from 5D to 4D}

The simplest way to construct a $5D\rightarrow 4D$ reduction for the ansatz (%
\ref{ansatz}) and (\ref{ansatzc}) is to eliminate from formulas the variable
$x^{1}$ and to consider a 4D space (parametrized by local coordinates $%
\left( x^{2},x^{3},v,y^{5}\right) )$ being trivially embedded into 5D space
(parametrized by local coordinates $\left( x^{1},x^{2},x^{3},v,y^{5}\right) $
with $g_{11}=\pm 1,g_{1\widehat{\alpha }}=0,\widehat{\alpha }=2,3,4,5)$ with
further possible \ 4D conformal and anholonomic transforms depending only on
variables $\left( x^{2},x^{3},v\right) .$ We suppose that the 4D metric $g_{%
\widehat{\alpha }\widehat{\beta }}$ could be of arbitrary signature. In
order to emphasize that some coordinates are stated just for a such 4D space
we underline the Greek indices, $\widehat{\alpha },\widehat{\beta },...$ \
and the Latin indices from the middle of alphabet, $\widehat{i},\widehat{j}%
,...=2,3,$ where $u^{\widehat{\alpha }}=\left( x^{\widehat{i}},y^{a}\right)
=\left( x^{2},x^{3},y^{4},y^{5}\right) .$

In result, the analogs , Theorems 1-3 and Corollaries 1-3 can be
reformulated for 4D gravity with mixed holonomic--anholonomic variables. We
outline here the most important properties of a such reduction.

\begin{itemize}
\item The line element (\ref{metric}) with ansatz (\ref{ansatz}) and the
line element (\ref{metric}) with (\ref{ansatzc}) are respectively
transformed on 4D space to the values:

The first type 4D quadratic line element is taken
\begin{equation}
ds^{2}=g_{\widehat{\alpha }\widehat{\beta }}\left( x^{\widehat{i}},v\right)
du^{\widehat{\alpha }}du^{\widehat{\beta }}  \label{metric4}
\end{equation}%
with the metric coefficients $g_{\widehat{\alpha }\widehat{\beta }}$
parametrized

{%%\footnotesize
\begin{equation}
\left[
\begin{array}{cccc}
g_{2}+w_{2}^{\ 2}h_{4}+n_{2}^{\ 2}h_{5} & w_{2}w_{3}h_{4}+n_{2}n_{3}h_{5} &
w_{2}h_{4} & n_{2}h_{5} \\
w_{2}w_{3}h_{4}+n_{2}n_{3}h_{5} & g_{3}+w_{3}^{\ 2}h_{4}+n_{3}^{\ 2}h_{5} &
w_{3}h_{4} & n_{3}h_{5} \\
w_{2}h_{4} & w_{3}h_{4} & h_{4} & 0 \\
n_{2}h_{5} & n_{3}h_{5} & 0 & h_{5}%
\end{array}%
\right] ,  \label{ansatz4}
\end{equation}%
} where the coefficients are some necessary smoothly class functions of
type:
\begin{eqnarray}
g_{2,3} &=&g_{2,3}(x^{2},x^{3}),h_{4,5}=h_{4,5}(x^{\widehat{k}},v),  \notag
\\
w_{\widehat{i}} &=&w_{\widehat{i}}(x^{\widehat{k}},v),n_{\widehat{i}}=n_{%
\widehat{i}}(x^{\widehat{k}},v);~\widehat{i},\widehat{k}=2,3.  \notag
\end{eqnarray}

The anholonomically and conformally transformed 4D line element is
\begin{equation}
ds^{2}=\Omega ^{2}(x^{\widehat{i}},v)\hat{{g}}_{\widehat{\alpha }\widehat{%
\beta }}\left( x^{\widehat{i}},v\right) du^{\widehat{\alpha }}du^{\widehat{%
\beta }},  \label{cmetric4}
\end{equation}%
were the coefficients $\hat{{g}}_{\widehat{\alpha }\widehat{\beta }}$ are
parametrized by the ansatz {\scriptsize
\begin{equation}
\left[
\begin{array}{cccc}
g_{2}+(w_{2}^{\ 2}+\zeta _{2}^{\ 2})h_{4}+n_{2}^{\ 2}h_{5} &
(w_{2}w_{3}++\zeta _{2}\zeta _{3})h_{4}+n_{2}n_{3}h_{5} & (w_{2}+\zeta
_{2})h_{4} & n_{2}h_{5} \\
(w_{2}w_{3}+\zeta _{2}\zeta _{3})h_{4}+n_{2}n_{3}h_{5} &
g_{3}+(w_{3}^{\ 2}+\zeta _{3}^{\ 2})h_{4}+n_{3}^{\ 2}h_{5} &
(w_{3}+\zeta _{3})h_{4} &
n_{3}h_{5} \\
(w_{2}+\zeta _{2})h_{4} & (w_{3}+\zeta _{3})h_{4} & h_{4} & 0 \\
n_{2}h_{5} & n_{3}h_{5} & 0 & h_{5}+\zeta _{5}h_{4}%
\end{array}%
\right] .  \label{ansatzc4}
\end{equation}%
}where $\zeta _{\widehat{i}}=\zeta _{\widehat{i}}\left( x^{\widehat{k}%
},v\right) $ and we shall restrict our considerations for $\zeta _{5}=0.$

\item We have a quadratic line element (\ref{dmetric}) which can be written
\begin{equation}
\delta s^{2}=g_{2}(dx^{2})^{2}+g_{3}(dx^{3})^{2}+h_{4}(\delta
v)^{2}+h_{5}(\delta y^{5})^{2},  \label{dmetric4}
\end{equation}%
with respect to the anholonomic co--frame $\left( dx^{\widehat{i}},\delta
v,\delta y^{5}\right) ,$ where
\begin{equation}
\delta v=dv+w_{\widehat{i}}dx^{\widehat{i}}\mbox{ and }\delta
y^{5}=dy^{5}+n_{\widehat{i}}dx^{\widehat{i}}  \label{ddif4}
\end{equation}%
is the dual of $\left( \delta _{\widehat{i}},\partial _{4},\partial
_{5}\right) ,$ where
\begin{equation}
\delta _{\widehat{i}}=\partial _{\widehat{i}}+w_{\widehat{i}}\partial
_{4}+n_{\widehat{i}}\partial _{5}.  \label{dder4}
\end{equation}

\item In the conditions of the 4D variant of the Theorem 1 we have the same
equations (\ref{ricci1a})--(\ref{ricci4a}) were we must put $%
h_{4}=h_{4}\left( x^{\widehat{k}},v\right) $ and $h_{5}=h_{5}\left( x^{%
\widehat{k}},v\right) .$ As a consequence we have that $\alpha _{i}\left(
x^{k},v\right) \rightarrow \alpha _{\widehat{i}}\left( x^{\widehat{k}%
},v\right) ,\beta =\beta \left( x^{\widehat{k}},v\right) $ and $\gamma
=\gamma \left( x^{\widehat{k}},v\right) $ which result that $w_{\widehat{i}%
}=w_{\widehat{i}}\left( x^{\widehat{k}},v\right) $ and $n_{\widehat{i}}=n_{%
\widehat{i}}\left( x^{\widehat{k}},v\right) .$

\item The 4D line element with conformal factor (\ref{cmetric}) subjected to
an an anhlonomic map with $\zeta _{5}=0$ transforms into
\begin{equation}
\delta s^{2}=\Omega ^{2}(x^{\widehat{i}%
},v)[g_{2}(dx^{2})^{2}+g_{3}(dx^{3})^{2}+h_{4}(\hat{{\delta }}%
v)^{2}+h_{5}(\delta y^{5})^{2}],  \label{cdmetric4}
\end{equation}%
given with respect to the anholonomic co--frame $\left( dx^{\widehat{i}},%
\hat{{\delta }}v,\delta y^{5}\right) ,$ where
\begin{equation}
\delta v=dv+(w_{\widehat{i}}+\zeta _{\widehat{i}})dx^{\widehat{i}}%
\mbox{ and
}\delta y^{5}=dy^{5}+n_{\widehat{i}}dx^{\widehat{i}}  \label{ddif24}
\end{equation}%
is dual to the frame $\left( \hat{{\delta }}_{\widehat{i}},\partial _{4},%
\hat{{\partial }}_{5}\right) $ with
\begin{equation}
\hat{{\delta }}_{\widehat{i}}=\partial _{\widehat{i}}-(w_{\widehat{i}}+\zeta
_{\widehat{i}})\partial _{4}+n_{\widehat{i}}\partial _{5},\hat{{\partial }}%
_{5}=\partial _{5}.  \label{dder24}
\end{equation}

\item The formulas (\ref{conf1}) and (\ref{confeq}) from Theorem 2 must be
modified into a 4D form
\begin{equation}
\hat{{\delta }}_{\widehat{i}}h_{4}=0\mbox{\ and\  }\hat{{\delta }}_{\widehat{%
i}}\Omega =0  \label{conf14}
\end{equation}%
and the values $\zeta _{\widetilde{{i}}}=\left( \zeta \widehat{_{{i}}},\zeta
_{{5}}=0\right) $ are found as to be a unique solution of (\ref{conf1}); for
instance, if
\begin{equation*}
\Omega ^{q_{1}/q_{2}}=h_{4}~(q_{1}\mbox{ and }q_{2}\mbox{ are
integers}),
\end{equation*}%
$\zeta _{\widehat{{i}}}$ satisfy the equations \
\begin{equation}
\partial _{\widehat{i}}\Omega -(w_{\widehat{i}}+\zeta _{\widehat{{i}}%
})\Omega ^{\ast }=0.  \label{confeq4}
\end{equation}

\item One holds the same formulas (\ref{p2})-(\ref{n}) from the Theorem 3 on
the general form of exact solutions with that difference that their 4D
analogs are to be obtained by reductions of holonomic indices, $\widehat{i}%
\rightarrow i,$ and holonomic coordinates, $x^{i}\rightarrow x^{\widehat{i}%
}, $ i. e. in the 4D solutions there is not contained the variable $x^{1}.$

\item The formulae (\ref{einstdiag}) for the nontrivial coefficients of the
Einstein tensor in 4D stated by the Corollary 1 are \ written
\begin{equation}
G_{2}^{2}=G_{3}^{3}=-S_{4}^{4},G_{4}^{4}=G_{5}^{5}=-R_{2}^{2}.
\label{einstdiag4}
\end{equation}

\item For symmetries of the Einstein tensor (\ref{einstdiag4}), \ we can
introduce a matter field source with a diagonal energy momentum tensor,
like\ it is stated in the Corollary 2 by the conditions (\ref{emcond}),
which in 4D are transformed into
\begin{equation}
\Upsilon _{2}^{2}=\Upsilon _{3}^{3},\Upsilon _{4}^{4}=\Upsilon _{5}^{5}.
\label{emcond4}
\end{equation}
\end{itemize}

\section{\quad Star--Products, Enveloping Algebras and Noncommutative
Geometry}

\label{asncalg}For a noncommutative space the coordinates ${\hat{u}}^{i},$ $%
(i=1,...,N)$ satisfy some noncommutative relations
\begin{equation}
\lbrack {\hat{u}}^{i},{\hat{u}}^{j}]=\left\{
\begin{array}{rcl}
& i\theta ^{ij}, & \theta ^{ij}\in \C,\mbox{ canonical structure;
} \\
& if_{k}^{ij}{\hat{u}}^{k}, & f_{k}^{ij}\in \C,\mbox{ Lie
structure; } \\
& iC_{kl}^{ij}{\hat{u}}^{k}{\hat{u}}^{l}, & C_{kl}^{ij}\in \C,%
\mbox{ quantum
plane }%
\end{array}%
\right.  \label{ncr}
\end{equation}%
where $\C$ denotes the complex number field.

The noncommutative space is modelled as the associative algebra of $\C;$\
this algebra is freely generated by the coordinates modulo ideal $\mathcal{R}
$ generated by the relations (one accepts formal power series)\ $\mathcal{A}%
_{u}=\C[[{\hat u}^1,...,{\hat u}^N]]/\mathcal{R}.$ One restricts attention$%
^{94}$\ %\cite{jssw}%
to algebras having the (so--called, Poincare--Birkhoff--Witt) property that
any element of $\mathcal{A}_{u}$ is defined by its coefficient function and
vice versa,%
\begin{equation*}
\widehat{f}=\sum\limits_{L=0}^{\infty }f_{i_{1},...,i_{L}}:{\hat{u}}%
^{i_{1}}\ldots {\hat{u}}^{i_{L}}:\quad \mbox{ when
}\widehat{f}\sim \left\{ f_{i}\right\} ,
\end{equation*}%
where $:{\hat{u}}^{i_{1}}\ldots {\hat{u}}^{i_{L}}:$ denotes that the basis
elements satisfy some prescribed order (for instance, the normal order $%
i_{1}\leq i_{2}\leq \ldots \leq i_{L},$ or, another example, are totally
symmetric). The algebraic properties are all encoded in the so--called
diamond $(\diamond )$ product which is defined by
\begin{equation*}
\widehat{f}\widehat{g}=\widehat{h}~\sim ~\left\{ f_{i}\right\} \diamond
\left\{ g_{i}\right\} =\left\{ h_{i}\right\} .
\end{equation*}

In the mentioned approach to every function $f(u)=f(u^{1},\ldots ,u^{N})$ of
commuting variables $u^{1},\ldots ,u^{N}$ one associates an element of
algebra $\widehat{f}$ when the commuting variables are substituted by
anticommuting ones,
\begin{equation}
f(u)=\sum f_{i_{1}\ldots i_{L}}u^{1}\cdots u^{N}\rightarrow \widehat{f}%
=\sum\limits_{L=0}^{\infty }f_{i_{1},...,i_{L}}:{\hat{u}}^{i_{1}}\ldots {%
\hat{u}}^{i_{L}}:  \notag
\end{equation}%
when the $\diamond $--product leads to a bilinear $\ast $--product of
functions (see details in Ref. [70]) %\cite{mssw})%
\begin{equation*}
\left\{ f_{i}\right\} \diamond \left\{ g_{i}\right\} =\left\{ h_{i}\right\}
\sim \left( f\ast g\right) \left( u\right) =h\left( u\right) .
\end{equation*}

The $*$--product is defined respectively for the cases (\ref{ncr})
\begin{equation*}
f*g=\left\{
\begin{array}{rcl}
\exp [{\frac i2}{\frac \partial {\partial u^i}}{\theta }^{ij}\frac \partial
{\partial {u^{\prime }}^j}]f(u)g(u^{\prime }){|}_{u^{\prime }\to u}, &  &
\\
\exp [\frac i2u^kg_k(i\frac \partial {\partial u^{\prime }},i\frac \partial
{\partial u^{\prime \prime }})]f(u^{\prime })g(u^{\prime \prime }){|}%
_{u^{\prime \prime }\to u}^{u^{\prime }\to u}, &  &  \\
q^{{\frac 12}(-u^{\prime }{\frac \partial {\partial u^{\prime }}}v{\frac
\partial {\partial v}}+u{\frac \partial {\partial u}}v^{\prime }{\frac
\partial {\partial v^{\prime }}})}f(u,v)g(u^{\prime },v^{\prime }){|}%
_{v^{\prime }\to v}^{u^{\prime }\to u}, &  &
\end{array}
\right.
\end{equation*}
where there are considered values of type%
\begin{eqnarray}
&e^{ik_n\widehat{u}^n}& e^{ip_{nl}\widehat{u}^n} =e^{i\{k_n+p_n+\frac
12g_n\left( k,p\right) \}\widehat{u}^n,}  \label{gdecomp} \\
&g_n\left( k,p\right)& = -k_ip_jf_{\ n}^{ij}+\frac 16k_ip_j\left(
p_k-k_k\right) f_{\ m}^{ij}f_{\ n}^{mk}+...,  \notag \\
&e^Ae^B& = e^{A+B+\frac 12[A,B]+\frac 1{12}\left( [A,[A,B]]+[B,[B,A]]\right)
}+...  \notag
\end{eqnarray}
and for the coordinates on quantum (Manin) planes one holds the relation $%
uv=qvu.$

A non--Abelian gauge theory on a noncommutative space is given by two
algebraic structures, the algebra $\mathcal{A}_{u}$ and a non--Abelian Lie
algebra $\mathcal{A}_{I}$ of the gauge group with generators $%
I^{1},...,I^{S} $ and the relations
\begin{equation}
\lbrack I^{\underline{s}},I^{\underline{p}}]=if_{~\underline{t}}^{\underline{%
s}\underline{p}}I^{\underline{t}}.  \label{commutators1}
\end{equation}%
In this case both algebras are treated on the same footing and one denotes
the generating elements of the big algebra by $\widehat{u}^{i},$%
\begin{equation}
\widehat{z}^{\underline{i}}=\{\widehat{u}^{1},...,\widehat{u}%
^{N},I^{1},...,I^{S}\},\mathcal{A}_{z}=\C[[\widehat{u}^1,...,%
\widehat{u}^{N+S}]]/\mathcal{R}  \notag
\end{equation}%
and the $\ast $--product formalism is to be applied for the whole algebra $%
\mathcal{A}_{z}$ when there are considered functions of the commuting
variables $u^{i}\ (i,j,k,...=1,...,N)$ and $I^{\underline{s}}\
(s,p,...=1,...,S).$

For instance, in the case of a canonical structure for the space variables $%
u^{i}$ we have
\begin{eqnarray}
(F\ast G)(u) &=&\exp \left[ \frac{i}{2}\left( \theta ^{ij}\frac{\partial }{%
\partial u^{\prime i}}\frac{\partial }{\partial u^{\prime \prime j}}%
+t^{s}g_{s}\left( i\frac{\partial }{\partial t^{\prime }},i\frac{\partial }{%
\partial t^{\prime \prime }}\right) \right) \times F\left( u^{\prime
},t^{\prime }\right) G\left( u^{\prime \prime },t^{\prime \prime }\right)
\mid _{t^{\prime }\rightarrow t,t^{\prime \prime }\rightarrow t}^{u^{\prime
}\rightarrow u,u^{\prime \prime }\rightarrow u.}\right]  \notag \\
&&  \label{csp1a}
\end{eqnarray}%
This formalism was developed in Ref. 94 %
%\cite{jssw}%
for general Lie algebras. \vspace{5mm}

%\begin{thebibliography}{99}
{\footnotesize \noindent $^{1}$ %
%\bibitem{con1}
A. Connes and J. Lott, Nucl. Phys. B \ (Proc. Suppl) \textbf{18,}\ 29 (1990).%
\newline
\noindent $^{2}$ %
%\bibitem{con2}
A.\ Connes \textit{Noncommutative Geometry,}\ (Academic Press, 1994).\newline
\noindent $^{3}$ A. H. Chamseddine, G. Felder and J. Frohlich, Commun. Math.
Phys. \textbf{155,}\ 205 (1993).\newline
\noindent $^{4}$ D. Kastler, Commun. Math. Phys. \textbf{166,}\ 633 (1995).%
\newline
\noindent $^{5}$ %
%\bibitem{ncg3}
I. Vancea, Phys. Rev. Lett. \textbf{79,} 3121 (1997).\newline
\noindent $^{6}$ %
%\bibitem{v1}
S. Vacaru, Phys. Lett. B \textbf{498,}\ 74 (2001).\newline
\noindent $^{7}$ %
%\bibitem{ch1}
A. H. Chamsedine, Phys. Lett. B \textbf{504,}\ 33 (2001).\newline
\noindent $^{8}$ J. Madore, \textit{An Introduction to Noncommutative
Geometry and its Physical Applications,} LMS lecture note Series 257, 2nd
ed.\ (Cambridge University Press, 1999).\newline
\noindent $^{9}$ %
%\bibitem{cartan1}
E. Cartan, \textit{Les Espaces de Finsler,}\ (Hermann, Paris, 1934).\newline
\noindent $^{10}$ %
%\bibitem{cartan2}
E. Cartan, \textit{La Methode du Repere Mobile, la Theorie des Groupes
Continus et les Espaces Generalises,}\ (Herman, Paris, 1935) [Russian
translation:\ (Moscow Univ. Press, 1963).\newline
\noindent $^{11}$ %
%\bibitem{cartan3}
E. Cartan, \textit{Riemannian geometry in an orthogonal frame}, \'{E}lie
Cartan lectures at Sorbonne, 1926--27,\ (World Scientific Publishing Co.,
Inc., River Edge, NJ, 2001)\ [from Russian].\newline
\noindent $^{12}$ E. Cartan, \textit{On manifolds with an affine connection
and the theory of general relativity}. \ Monographs and Textbooks in
Physical Science, 1. (Bibliopolis, Naples, 1986) \ [from French].\newline
\noindent $^{13}$ %
%\bibitem{anhtv1}
S. Vacaru, JHEP \textbf{04,}\ 009\ (2001)\newline
\noindent $^{14}$ %
%\bibitem{anhtv2}
S. Vacaru and D. Singleton, J. Math. Phys. \textbf{43,}\ 2486 (2002).\newline
\noindent $^{15}$ %
%\bibitem{anhtv3}
S. Vacaru and O. Tintareanu-Mircea, Nucl. Phys. \textbf{B 626}\ 239 (2002).%
\newline
\noindent $^{16}$ %
%\bibitem{anhtv4}
Vacaru and D. Singleton, Class. Quant. Gravity, \textbf{19,}\ 2793 (2002).%
\newline
\noindent $^{17}$ %
%\bibitem{anhtv5}
S. Vacaru and H. Dehnen, Gen. Rel. Gravity, \textbf{35,}\ 209 (2003).\newline
\noindent $^{18}$ %
%\bibitem{anhtv6}
S.\ Vacaru, Int. J. Mod. Phys. D \textbf{12,}\ 461 (2003).\newline
\noindent $^{19}$ %
%\bibitem{anhtv7}
S.\ Vacaru, Int. J. Mod. Phys. D \textbf{12,}\ 479 (2003).\newline
\noindent $^{20}$ %
%\bibitem{heus}
M. Heusler, \textit{Black Hole Uniequeness Theorems,}\ Cambridge Lecture
Notes in Physics 6\ (Cambridge University Press, 1996).\newline
\noindent $^{21}$ %
%\bibitem{pd1}
D. A. Popov, Theor. Math. Phys. \textbf{24,}\ 347 (1975) (in Russian).%
\newline
\noindent $^{22}$ %
%\bibitem{pd2}
D. A. Popov and L. I. Dikhin, Dokl. Akad. Nauk SSSR \textbf{245,}\ 347
(1975) (in Russian).\newline
\noindent $^{23}$ %
%\bibitem{ts}
A.\ A. Tseytlin, Phys. Rev. \textbf{D 26,}\ 3327 (1982).\newline
\noindent $^{24}$ %
%\bibitem{v41}
S.\ Vacaru and Yu. Goncharenko, Int. J. Theor. Phys. \textbf{34,}\ 1955
(1995).\newline
\noindent $^{25}$ %
%\bibitem{v42}
S. Vacaru, Ann. Phys. (NY) \textbf{256,}\ 39 (1997).\newline
\noindent $^{26}$ %
%\bibitem{v43}
H. Dehnen and S. Vacaru, Gen. Rel. Grav. \textbf{35,}\ 807 (2003).\newline
\noindent $^{27}$ %
%\bibitem{v44}
S. Vacaru, \textit{Noncommutative Finsler Geometry, Gauge Fields and Gravity}%
,\ math--ph/ 0205023.\newline
\noindent $^{28}$ %
%\bibitem{v45}
S. Vacaru, \textit{(Non) Commutative Finsler Geometry from String/M--Theory}%
,\ hep--th/ 0211068.\newline
\noindent $^{29}$ %
%\bibitem{v5}
S. Vacaru, Ann. Phys. (NY) \textbf{290, } 83 (2001).\newline
\noindent $^{30}$ %
%\bibitem{vmethod1}
S. Vacaru, \textit{\ A New Method of Constructing Black Hole Solutions in
Einstein and 5D Dimension Gravity}, hep-th/0110250.\newline
\noindent $^{31}$ %
%\bibitem{vmethod2}
S. Vacaru, \textit{\ Black Tori in Einstein and 5D Gravity}, hep-th/0110284.
\newline
\noindent $^{32}$ %
%\bibitem{vnp}
S.\ Vacaru, Nucl.\ Phys. \textbf{B494,}\ 590 (1997).\newline
\noindent $^{33}$ %
%\bibitem{barthel}
W. Barthel, J. Reine Angew. Math. \textbf{212,}\ 120 (1963).\newline
\noindent $^{34}$ %
%\bibitem{kaw1}
A. Kawaguchi, Tensor, N. S. \textbf{2, } 123 (1952).\newline
\noindent $^{35}$ %
%\bibitem{kaw2}
A. Kawaguchi, Tensor, N. S. \textbf{2, } 165 (1956).\newline
\noindent $^{36}$ %
%\bibitem{kaw3}
A. Kawaguchi, Akad. Wetensch. Amsterdam, Proc. \textbf{40, } 596 (1937).%
\newline
\noindent $^{37}$ %
%\bibitem{ma1}
R. Miron and M. Anastasiei, \textit{\ The Geometry of Lagrange Spaces Theory
and Applications } (Kluwer, 1994).\newline
\noindent $^{38}$ %
%\bibitem{m2}
R. Miron, \textit{\ The Geometry of Higher Order Lagrange Spaces:\
Applications to Mechanics and Physics } (Kluwer, 1997).\newline
\noindent $^{39}$ %
%\bibitem{ste1}
J.\ Stewart, \textit{\ Advanced General Relativity }\ (Cambridge University
Press, 1991).\newline
\noindent $^{40}$ %
%\bibitem{mtw}
C.\ Misner, K. Thorne, J. Wheeler, \textit{\ Gravitation }\ (W. H. Freeman
and Company, NY, 1973).\newline
\noindent $^{41}$ %
%\bibitem{salst}
A. Salam and J. Strathdee, Ann. Phys. (NY) \textbf{141, } 316 (1982).\newline
\noindent $^{42}$ %
%\bibitem{prd}
R. Percacci and S. Randjbar--Daemi, J. Math. Phys. \textbf{24, } 807 (1983).
\newline
\noindent $^{43}$ %
%\bibitem{ow}
J. M. Overduin and P. S. Wesson, Phys. Rep. \textbf{283, } 303 (1997).%
\newline
\noindent $^{44}$ %
%\bibitem{dub}
M.\ Dubois--Violette, R. Kerner and J. Madore, J. Math. Phys. \textbf{\ 31, }
316 (1990).\newline
\noindent $^{45}$ %
%\bibitem{sw}
N. Seiberg and E. Witten, JHEP, \textbf{9909, } 032 (1999).\newline
\noindent $^{46}$ %
%\bibitem{dvkm1}
M. Dubois--Violette, R. Kerner and J. Madore, Phys. Lett. \textbf{B217,} 485
(1989).\newline
\noindent $^{47}$ %
%\bibitem{dvkm2}
M. Dubois--Violette, R. Kerner and J. Madore, Class. Quant. Grav. \textbf{\
7,} 1709 (1989).\newline
\noindent $^{48}$ %
%\bibitem{madm}
J. Madore and J. Mourad, Class. Quant. Grav. \textbf{\ 10, } 2157 (1993).%
\newline
\noindent $^{49}$ %
%\bibitem{mmm}
J. Madore, T. Masson and J. Mourad, Class. Quant. Grav. \textbf{\ 12, } 1249
(1995).\newline
\noindent $^{50}$ %
%\bibitem{gb}
J. C. Graves and D. R. Brill, Phys. Rev. \textbf{120,} 1507 (1960).\newline
\noindent $^{51}$ %
%\bibitem{carter}
B.\ Carter, Phys. Lett. \textbf{21, } 423 (1966).\newline
\noindent $^{52}$ %
%\bibitem{haw}
S. W. Hawking and G. F. R. Ellis, \textit{\ The Large Scale Structure of
Space--Time } (Cambridge University Press, 1973).\newline
\noindent $^{53}$ %
%\bibitem{ard}
F. Ardalan, H. Arfaei, M. R. Garousi and A. Ghodsi, Int. J. Mod. Phys.
\textbf{A 18,} 1051 (2003).\newline
\noindent $^{54}$ %
%\bibitem{jsswa}
B. Jurco, S. Schraml, P. Shupp and J. Wess, Eur. Phys. J. \textbf{C 14, }
367 (2000).\newline
\noindent $^{55}$ %
%\bibitem{msswa}
J. Madore, S. Schraml, P.\ Schupp and J.\ Wess, Eur. Phys. J. \textbf{C 16, }
161 (2000).\newline
\noindent $^{56}$ %
%\bibitem{sak}
V. Sakian, JHEP \textbf{0106 } 037 (2001).\newline
\noindent $^{57}$ %
%\bibitem{aasv}
M. C. B. Abdalla, M. A. De Anrade, M. A. Santos and I. V. Vancea, Phys.
Lett. \textbf{B 548,} 88 (2002).\newline
\noindent $^{58}$ %
%\bibitem{nr}
H.\ Nishino and S. Rajpoot, \textit{\ Noncommutative Nonlinear
Supersymmetry, } hep--th/ 0212329.\newline
\noindent $^{59}$ %
%\bibitem{gcors}
H. Garcia--Compean, O. Obregon, C. Ramirez and M.\ Sabido, Phys. Rev.
\textbf{\ D68 } 044015 (2003).\newline
\noindent $^{60}$ %
%\bibitem{card}
M. A. Cardella and D. Zanon, Class. Quant. Grav. \textbf{\ 20, } L95 (2003).%
\newline
\noindent $^{61}$ %
%\bibitem{vnonc}
S.I. Vacaru, I. A. Chiosa and Nadejda A. Vicol,  in: NATO Advanced Research
Workshop Proceedings ''Noncommutative Structures in Mathematics and
Physics'', eds S. Duplij and J. Wess, September 23-27, Kyiv, Ukraine (Kluwer
Academic Publishers, 2001), 229 -- 243, hep-th/0011221. \newline
\noindent $^{62}$ %
%\bibitem{js}
B. Jur\v{c}o and P. Schupp, Eur. Phys. J. \textbf{C 14, }367 (2000).\newline
\noindent $^{63}$ %
%\bibitem{w1}
H. Weyl, Z. Physik \textbf{46,} 1 (1927).\newline
\noindent $^{64}$ %
%\bibitem{w2}
H. Weyl, \textit{\ The theory of groups and quantum mechanics } (Dover,
New--York, 1931);\ translated from \textit{\ Gruppentheorie and
Quantenmechanik } (Hirzel Verlag, Leipzig, 1928).\newline
\noindent $^{65}$ %
%\bibitem{wig}
E. P. Wigner, Phys. Rev. \textbf{40, } 749 (1932).\newline
\noindent $^{66}$ %
%\bibitem{moy}
J. E. Moyal, Proc. Cambridge Phil. Soc. \textbf{45, } 99 (1949).\newline
\noindent $^{67}$ %
%\bibitem{bffls}
F. Bayen, M. Flato, C. Fronsdal, A. Lichnerowicz and D. Sternheimer, Ann.
Physics\ \textbf{111, } 61 (1978).\newline
\noindent $^{68}$ %\bibitem{konts}
M. Kontsevich, \textit{\ Deformation quantization of Poisson manifolds, I.}\
q--alg/9709040.\newline
\noindent $^{69}$ %
%\bibitem{zot}
A. Zotov, Mod. Phys. Lett. \textbf{\ A16, } 615 (2001).\newline
\noindent $^{70}$ %
%\bibitem{mssw}
J. Madore, S. Schraml, P. Schupp and J. Wess, Eur. Phys. J. \textbf{C 16,}
161 (2000).\newline
\noindent $^{71}$ %
%\bibitem{wz1}
J. Wess and B. Zumino, Nucl. Phys. Phys. Proc. Suppl. \textbf{18B, } 302
(1991).\newline
\noindent $^{72}$ %
%\bibitem{jw}
J. Wess, q-Deformed Heisenberg Algebras, in \textit{Proceeding of the 38
International Universit\"{a}tswochen f\"{u}r Kern-- und Teilchenphysik}, no.
543 in Lect. Notes in Phys., (Springer--Verlag, 2000) Schladming, January
1999, eds. H. Gusterer, H. Grosse and L. Pitner;\ math--ph / 9910013.\newline
\noindent $^{73}$ %
%\bibitem{vgauge1}
S. Vacaru, Gauge Like Treatment of Generalized Lagrange and Finsler Gravity,
Buletinul Academiei de Stiinte a Republicii Moldova, Fizica si Tehnica
[Izvestia Academii Nauk Respubliky Moldova, fizica i tehnika], 3, 31-34
(1996). \newline
\noindent $^{74}$ %
%\bibitem{vgauge2}
S. Vacaru, \textit{\ Gauge Gravity and Conservation Laws in Higher Order
Anisotropic Spaces, } hep-th/9810229.\newline
\noindent $^{75}$ %
%\bibitem{vgauge3}
S. Vacaru, Buletinul Academiei de Stiinte a Republicii Moldova, Fizica si
Tehnica [Izvestia Academii Nauk Respubliky Moldova, fizica i tehnika],
\textbf{3, } 26 (1996).\newline
\noindent $^{76}$ %
%\bibitem{ut1}
R. Utiyama, Phys. Rev. \textbf{101,} 1597 (1956).\newline
\noindent $^{77}$ %
%\bibitem{hgmn}
F. Hehl, J. D. McGrea, E. W. Mielke and Y. Ne'eman, Phys. Rep. \textbf{258,}
1 (1995).\newline
\noindent $^{78}$ %
%\bibitem{dh}
H. Dehnen and E. Hitzer, Int. J. Theor. Phys.\ \textbf{34,} 1981 (1995).%
\newline
\noindent $^{79}$ %
%\bibitem{bsst}
L. Bonora, M. Schnabl, M. M. Sheikh--Jabbari and A. Tomasiello, Nucl. Phys.
\textbf{\ B 589,} 461 (2000).\newline
\noindent $^{80}$ %
%\bibitem{ms1}
L. Mysak and G. Szekeres, Can. J. Phys. \textbf{44, } 617 (1966).\newline
\noindent $^{81}$ %
%\bibitem{gh}
R. Geroch and J. B. Hartle, J. Math. Phys. \textbf{23, } 680 (1982).\newline
\noindent $^{82}$ %
%\bibitem{fk}
S. Fairhurst and B. Krishnan, Int. J. Mod. Phys. \textbf{10, } 691 (2001).%
\newline
\noindent $^{83}$ %
%\bibitem{tors1}
J.\ L. Friedman, K. Schleich and D.\ M. Witt, Phys.\ Rev. Lett. \textbf{71, }
1486 (1993).\newline
\noindent $^{84}$ %
%\bibitem{gall}
G.\ Galloway, Commun. Math. Phys. \textbf{\ 151, } 53 (1993).\newline
\noindent $^{85}$ %
%\bibitem{cwald}
P.\ T. Chrusciel and R.\ M. Wald, Class. Quant. Grav. \textbf{\ 11, } L147
(1994).\newline
\noindent $^{86}$ %
%\bibitem{jvent}
T. Jacobson and S.\ Ventkataranami, Class. Quant. Grav. \textbf{\ 12, } 1055
(1995).\newline
\noindent $^{87}$ %
%\bibitem{lem1}
J. P. S. Lemos, Phys. Lett. \textbf{B 352, } 46 (1995).\newline
\noindent $^{88}$ %
%\bibitem{lem2}
J. P. S. Lemos and V. T. Zanchin, Phys. Rev. \textbf{D 54, } 3840 (1996).%
\newline
\noindent $^{89}$ %
%\bibitem{lem3}
P. M. Sa and J. P. S. Lemos, Phys.\ Lett. \textbf{\ B 423, } 49 (1998).%
\newline
\noindent $^{90}$ %
%\bibitem{st}
S. L. Shapiro and S.\ A. Teukolsky, Phys. Rev. Lett. \textbf{66,} 944 (1991).%
\newline
\noindent $^{91}$ %
%\bibitem{asht1}
A.\ Ashtekar, C. Beetle and S.\ Fairhurst, Class. Quant. Grav. \textbf{\ 17,
} 253 (2000).\newline
\noindent $^{92}$ %
%\bibitem{asht2}
A.\ Ashtekar, S.\ Fairhurst and B. Krishnan, Phys. Rev. \textbf{D 62 }
1040025 (2000).\newline
\noindent $^{93}$ %
%\bibitem{asht3}
A.\ Ashtekar, C. Beetle, O. Dreyer et all. Phys. Rev. Lett. \textbf{5, }
3564 (2000).\newline
\noindent $^{94}$ %
%\bibitem{jssw}
B. Jur\v{c}o, S. Schraml, P. Shupp and J. Wess, Eur. Phys. J. \textbf{C 17, }
521 (2000). %\end{thebibliography}
}

\end{document}